\documentclass[acmsmall]{acmart}
\usepackage{xspace}
\usepackage{multirow}
\usepackage{subfigure}
\usepackage[ruled,vlined,linesnumbered,lined, commentsnumbered]{algorithm2e}
\usepackage[noend]{algpseudocode}
\AtBeginDocument{%
  \providecommand\BibTeX{{%
    \normalfont B\kern-0.5em{\scshape i\kern-0.25em b}\kern-0.8em\TeX}}}

\newcommand{\modelname}{\textsf{FeSoG}\xspace}
\newtheorem{mydef}{Definition}
\setcopyright{acmcopyright}
\acmJournal{TIST}
\acmYear{2021} \acmVolume{1} \acmNumber{1} \acmArticle{1} \acmMonth{1} \acmPrice{15.00}\acmDOI{10.1145/3501815}

\begin{document}

%%
%% The "title" command has an optional parameter,
%% allowing the author to define a "short title" to be used in page headers.
\title{Federated Social Recommendation with Graph Neural Network}

%%
%% The "author" command and its associated commands are used to define
%% the authors and their affiliations.
%% Of note is the shared affiliation of the first two authors, and the
%% "authornote" and "authornotemark" commands
%% used to denote shared contribution to the research.
\author{Zhiwei Liu}
% \authornote{Both authors contributed equally to this research.}
\affiliation{%
  \institution{University of Illinois at Chicago}
%   \streetaddress{P.O. Box 1212}
  \city{Chicago}
  \state{IL}
  \country{USA}
  \postcode{60607}
}
\email{zliu213@uic.edu}
% \orcid{1234-5678-9012}

\author{Liangwei Yang}
% \authornotemark[1]
\email{lyang84@uic.edu}
\affiliation{%
  \institution{University of Illinois at Chicago}
%   \streetaddress{P.O. Box 1212}
  \city{Chicago}
  \state{IL}
  \country{USA}
  \postcode{60607}
}

\author{Ziwei Fan}
\affiliation{%
  \institution{University of Illinois at Chicago}
%   \streetaddress{P.O. Box 1212}
  \city{Chicago}
  \state{IL}
  \country{USA}
  \postcode{60607}
}
\email{zfan20@uic.edu}

\author{Hao Peng}
\authornote{This is the corresponding author.}
\affiliation{%
  \institution{Beihang University}
  \city{Beijing}
  \country{China}
  }
\email{penghao@act.buaa.edu.cn}

\author{Philip S. Yu}
\affiliation{%
  \institution{University of Illinois at Chicago}
%   \streetaddress{P.O. Box 1212}
  \city{Chicago}
  \state{IL}
  \country{USA}
  \postcode{60607}
}
\email{psyu@uic.edu}

%%
%% By default, the full list of authors will be used in the page
%% headers. Often, this list is too long, and will overlap
%% other information printed in the page headers. This command allows
%% the author to define a more concise list
%% of authors' names for this purpose.
\renewcommand{\shortauthors}{Liu and Yang, et al.}

%%
%% The abstract is a short summary of the work to be presented in the
%% article.
\begin{abstract}
  Recommender systems have become prosperous nowadays, designed to predict users' potential interests in items by learning embeddings. Recent developments of the Graph Neural Networks~(GNNs) also provide recommender systems with powerful backbones to learn embeddings from a user-item graph. However, only leveraging the user-item interactions suffers from the cold-start issue due to the difficulty in data collection. Hence, current endeavors propose fusing social information with user-item interactions to alleviate it, which is the social recommendation problem. Existing work employs GNNs to aggregate both social links and user-item interactions simultaneously. However, they all require centralized storage of the social links and item interactions of users, which leads to privacy concerns. Additionally, according to strict privacy protection under General Data Protection Regulation, centralized data storage may not be feasible in the future, urging a decentralized framework of social recommendation. 
  
  As a result, we design a federated learning recommender system for the social recommendation task, which is rather challenging because of its heterogeneity, personalization, and privacy protection requirements. To this end, we devise a novel framework \textbf{Fe}drated \textbf{So}cial recommendation with \textbf{G}raph neural network (\modelname). Firstly, \modelname adopts relational attention and aggregation to handle heterogeneity. Secondly, \modelname infers user embeddings using local data to retain personalization. Last but not least, the proposed model employs pseudo-labeling techniques with item sampling to protect the privacy and enhance training. Extensive experiments on three real-world datasets justify the effectiveness of \modelname in completing social recommendation and privacy protection. We are the first work proposing a federated learning framework for social recommendation to the best of our knowledge. 
\end{abstract}

%%
%% Keywords. The author(s) should pick words that accurately describe
%% the work being presented. Separate the keywords with commas.

\authorsaddresses{
Authors' addresses: 
Z. Liu, L. Yang, Z. Fan, and P. S. Yu, Department of Computer Science, University of Illinois at Chicago, Chicago, IL; email: \{zliu213, lyang84, zfan20, psyu\}@uic.edu;
H. Peng, School of Cyber Science and Technology, Beihang University, No. 37 Xue Yuan Road, Haidian District, Beijing, 100191, China; email: penghao@act.buaa.edu.cn.
}

%%
%% The code below is generated by the tool at http://dl.acm.org/ccs.cfm.
%% Please copy and paste the code instead of the example below.
%%


%%
%% This command processes the author and affiliation and title
%% information and builds the first part of the formatted document.
\maketitle

\section{Introduction}
The developments of Recommender Systems~(RSs)~\cite{rendle2009bpr,liu2020basconv,liu2021augmenting,fan2019graph,zhou2021intrinsic} become prosperous nowadays. 
A well-designed recommender system is able to predict users' potential interests in items. 
% The core of it is to learn user/item embeddings~\cite{rendle2009bpr,zheng2018spectral,liu2020deoscillated}, which are trained from historical user-item interactions. 
The core of it is to learn user/item embeddings~\cite{rendle2009bpr,zhou2021pure,liu2020deoscillated} by fitting historical user-item interactions.
Recently, the prosperity of graph neural networks~(GNNs)~\cite{dou2020enhancing,liu2020alleviating,cao2021knowledge,peng2021lime} provide powerful frameworks to learn node embeddings, which also motivates the community to design GNN-based RS models~\cite{wang19neural,liu2020basconv,liu2020deoscillated,fan2019graph,wang2021pre}. 
However, the cold-start issue~\cite{liu2021augmenting,zhou2021pure}, which is associated with users having few records, impairs the performance of learned embeddings. 

To cope with this, one can leverage the social information of users~\cite{fan2019graph,liu2019real,shen2012learning,mu2019graph}. 
In this way, we assume that users with social links also share similar item interests. 
Therefore, we could simultaneously aggregate social information and user-item interactions~\cite{yang2021consisrec,fan2019graph,fan2020graph,wu2018socialgcn} to alleviate the cold-start issue. SocialGCN~\cite{wu2018socialgcn,wu2019neural} employs the Graph Convolutional Network~(GCN) to enhance user embedding by simulating how the recursive social diffusion process influences users. 
GraphRec~\cite{fan2019graph} and GraphRec+~\cite{fan2020graph} propose to model three types of aggregations upon social graph, user-item graph and item-item graph. 
Thus, it can comprehensively fuse the social links and item transactions.  ConsisRec~\cite{yang2021consisrec} introduces the social inconsistency problem from context-level and relation-level. It solves this problem by using a sampling-based attention mechanism.  

\begin{figure}
    \centering
    \includegraphics[width=0.8\linewidth]{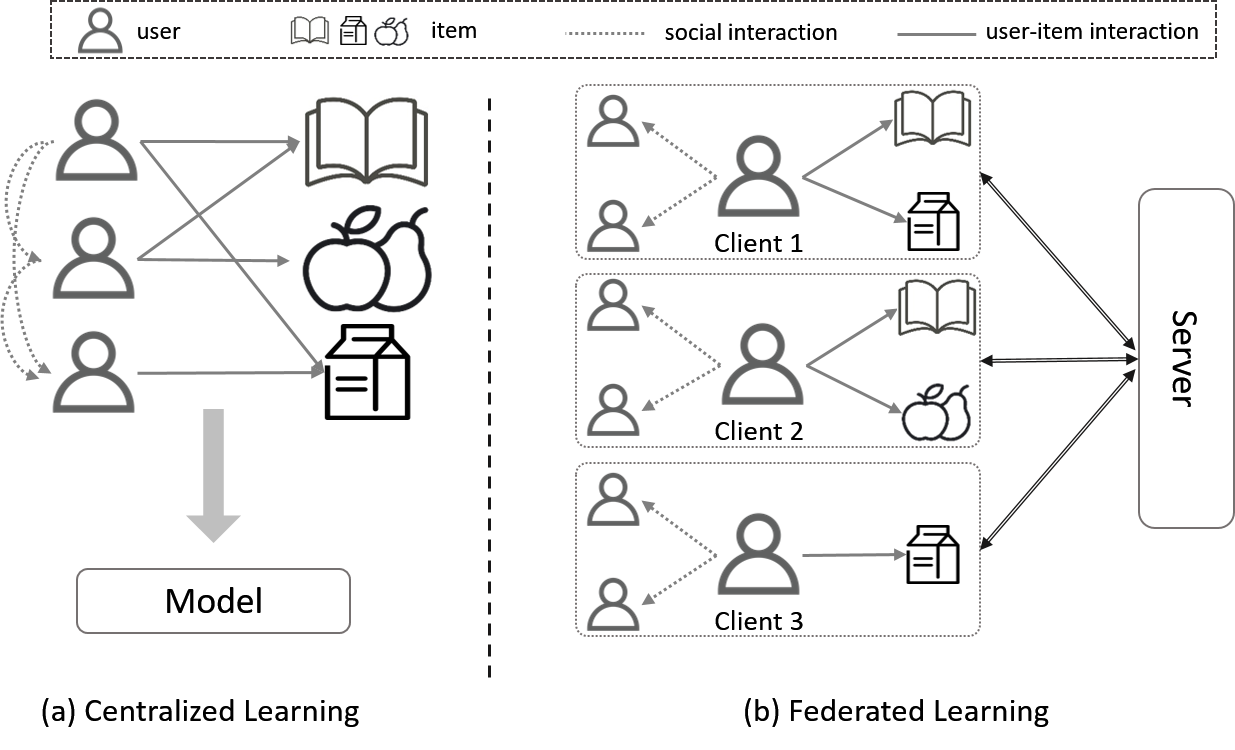}
    \caption{Centralized learning (a) and federated learning (b) for social recommendation. The centralized learning trains the model with all the user privacy data, i.e., both social interactions and user-item interactions, available on the server. In contrast, federated learning locally stores user privacy data, only uploading and requesting non-sensitive data from the server. Dash lines between users denotes social interactions, while solid lines between users and items denotes user-item interactions.}
    \label{fig:intro}
\end{figure}

Though being effective in fusing the social and user-item information, they all require a centralized storage~\cite{wu2021fedgnn,chen2020robust,yang2019federated} of both the social networks and item transaction history of users. 
Existing centralized storage methods pose risks of leaking privacy-sensitive data. 
Additionally, due to the strict privacy protection under General Data Protection Regulation~(GDPR)\footnote{https://gdpr-info.eu/}, centralized data storage may not be the first choice of online platforms in the future. 
Therefore, a new decentralized model training framework for social recommendation is necessary. According to previous researches in federated learning~\cite{mcmahan2017communication,yang2019federated}, the user data can be stored locally in each client while only uploading the necessary gradients for updating the model on a server. 
As for the federated recommender systems~\cite{chen2020robust,chai2020secure,ammad2019federated,wu2021fedgnn}, the sensitive user-item interactions are stored locally and clients only upload gradients to update user/item embeddings. 
% FCF~\cite{ammad2019federated} proposes to train the user and item embeddings locally, while only uploading the gradients of item embeddings to the server. 

However, there is few work discussing how to design a federated learning recommender system to complete the social recommendation task. 
We address the challenges of a federated social recommender system~(FSRS) as follows: 
(1)  \textit{Heterogeneity}. The current federated recommender system stores user-item interactions locally. However, a FSRS requires both user-user and user-item interactions, as shown in Fig.~\ref{fig:intro}. Therefore, we should store and fuse two types of relations simultaneously. 
(2) \textit{Personalization}. Each client has special item interests and social connections, which leads to the non-iid distribution of the local data~\cite{yang2019federated}. The model should be able to characterize the personalized federated learning process~\cite{fallah2020personalized} for those clients, which is rather challenging. 
(3) \textit{Privacy Protection}. Though user privacy data are stored locally, a federated recommender system yet demands collecting necessary gradients~\cite{chai2020secure} from clients for updating embeddings on the server. This uploading process may lead to information leakage of original data~\cite{chai2020secure}. Therefore, we should design a protection module before uploading any information. 

To this end, we devise a novel FSRS framework to address the challenges as mentioned above, which are \textbf{Fe}drated \textbf{So}cial recommendation with \textbf{G}raph neural network~(\modelname). 
Firstly, we propose to use a local graph neural network to learn node embeddings. 
To tackle the heterogeneity of local data, we employ a relation attention mechanism to distinguish user-user and user-item interactions, which characterize the importance of neighbors by assigning different attention weights with respect to their relations.
Secondly, the local graph neural networks on each client are updated only based on the data on devices. 
As such, models on devices possess personalizing training. Last but not least, \modelname employs pseudo-labelling technique for the on-device training of local models. 
This pseudo-labelling can protect the privacy data from leakage when uploading gradients and enhance the robustness of the training process. Extensive experiments on three datasets demonstrate the effectiveness of \modelname. 
Compared with other baselines, \modelname achieves up to 5.26\% in RMSE and 5.46\% in MAE across all three datasets. In ablation studies, \modelname also demonstrates the necessity of each component developed in the federated learning framework.
The contributions are summarized as follows:
\begin{itemize}
    \item \textbf{Novel}: We are the first work proposing a federated learning framework to tackle the social recommendation problem to the best of our knowledge.
    \item \textbf{Substantial}: We address three critical challenges, i.e., heterogeneity, personalization, and privacy protection, by proposing a new model \modelname.
    \item \textbf{Comprehensive}: We conduct extensive experiments on three publicly available datasets to verify the effectiveness of \modelname. Detailed analysis and ablation study further prove the efficacy of our proposed components in \modelname.
\end{itemize}
In the following sections, we first introduce the related work in Sec.~\ref{sec:related_work}. Then, we present some preliminaries in Sec.~\ref{sec:preliminary}, including both the definition and formulation. The detailed descriptions of our proposed \modelname model are in Sec.~\ref{sec:proposed_model}. Experiments are discussed in Sec.~\ref{sec:experiments}. Finally, we conclude this paper and open up possible future work in Sec.~\ref{sec:conclusion}.

\section{Related Work}\label{sec:related_work}
This section presents three relevant areas to this paper: GNN for recommendation, social recommendation, and federated learning for recommendation.

\subsection{Graph Neural Network for Recommendation}
The recent developments of Graph Neural Networks~(GNNs)~\cite{kipf2016semi,hamilton2017inductive,velivckovic2017graph,liu2020alleviating} motivate the community to propose a GNN-based recommender system. 
The intuition of a GNN model is to aggregate neighbors to recursively learn node embeddings~\cite{peng2021streaming}. 
GC-MC~\cite{berg2017graph} first employs the GCN~\cite{kipf2016semi} architecture to complete the user-item rating matrix. 
It uses the GCN as an encoder to train user/item embeddings, which are input to a fully connected neural network to predict the ratings. 
PinSAGE~\cite{pinsage2018ying} proposes to use the GraphSAGE~\cite{hamilton2017inductive} backbone to learn item embeddings over an attributed item graph. 
It first samples fixed-size nodes from multi-hop neighbors and then uses aggregators to aggregate those sampled nodes to learn the embeddings for center nodes. 
NGCF~\cite{wang19neural} is proposed later to explicitly model the collaborative signals upon user-item interaction graph by applying the GNN model. 
DGCF~\cite{liu2020deoscillated} observes the oscillation problem when applying GNN on the bipartite graph and solves it with cross-hop propagation layers. 
BasConv~\cite{liu2020basconv} is a pioneer work that investigates using GNN to complete basket recommendation. 
These works prove the efficacy of using the GNN framework to learn embeddings in a recommender system. 
GNN-based models are advantageous as their aggregation can model high-order structural information crucial for learning user/item embeddings from interactions. 
This paper also adopts the GNN model to embed the local graphs. We employ the graph attention networks~\cite{vaswani2017attention} as a backbone.

\subsection{Social Recommendation}
The social recommendation aims to relieve the data sparsity and cold start problem by inducing information of social links between users~\cite{ma2008sorec,wang2014hgmf,yang2021consisrec}. 
Social recommendation methods can be generally categorized as social matrix factorization based methods and graph neural network based methods. 
% Social matrix factorization incorporates social links into traditional matrix factorization methods.
Existing social matrix factorization approaches either jointly factorize the rating and social relationship matrices or regularize the user/item embeddings with constraints of social connections.
SoRec~\cite{ma2008sorec} co-factorizes the user rating matrix and social link matrix. SocialMF~\cite{jamali2010matrix} adds a regularization term to constrain the difference between the user's taste and his/her trusted friends' average weighted taste. 
SoReg~\cite{ma2011recommender} adds a regularization term to directly minimize the difference in the user latent feature between two trusted users, which can prevent the counteraction of the latent feature of one's trusted friends. 
HGMF~\cite{wang2014hgmf} introduces a hierarchical group matrix factorization technique to learn the user-group feature in a social network for recommendation. 
% Graph neural network based methods are prevalent recently because it is suitable to deal with social networks~\cite{li2020efficient,wu2019dual,wu2018collaborative}. 
Unlike matrix factorization methods, graph neural network methods infer node embeddings directly from graphs and demonstrate the effectiveness from recent social recommendation work~\cite{li2020efficient,wu2019dual,wu2018collaborative}. 
GraphRec~\cite{fan2019graph} and GraphRec+~\cite{fan2020graph} uses graph attention networks to learn user and item embeddings for recommendation. 
\cite{song2019session} utilizes dynamic graph attention networks to capture the dynamic user's interest from the social dimension. 
CUNE~\cite{CUNE} assumes that users hold implicit social links from each other. 
CUNE extracts semantic and reliable social information by graph embedding method.
DiffNet~\cite{wu2019neural} and DiffNet++~\cite{wu2020diffnet++} model the social influence diffusion process to enhance the social recommendation. 
ConsisRec~\cite{yang2021consisrec} examines the inconsistency problems in the social recommendation and introduces a consistent neighbor sampling module in the GNN model. 
The above studies show the effectiveness of incorporating social information into the recommender system.

\subsection{Federated Learning for Recommender System}
Google proposed Federated learning in 2016~\cite{mcmahan2017communication}. 
It calls for data privacy-preserving solutions in machine learning models~\cite{erlingsson2014rappor}, with the raised privacy concerns of existing centralized training based models. 
The fundamental of federated learning is to design a decentralized training framework, which distributes the data to clients rather than storing it in a server~\cite{mcmahan2017communication,konevcny2016federated}. 
% As to the recommendation research area, the exploration of federated recommender systems rises recently~\cite{chai2020secure,wu2021fedgnn,ammad2019federated}. 
% Due to the rising attention of privacy protection, we believe a federated recommender system will become more popular in both academic and industry. 
User transactions are sensitive information and probably cause identity information leakage if used for malicious purposes. 
Several recent works~\cite{chai2020secure,wu2021fedgnn,ammad2019federated} developed federated recommender systems for user information protection while still preserving good enough personalization.
Federated Collaborative Filtering~(FCF)~\cite{ammad2019federated} and FedMF~\cite{chai2020secure} are two pioneering works investigating a novel federated learning framework to learn the user/item embeddings for a recommender system. 
% Their ideas are similar and straightforward, both based on the matrix factorization~\cite{koren2009matrix} of user-item rating matrix. 
Both works develop the federated learning on the top of factorization~\cite{koren2009matrix} of the user-item rating matrix. 
To achieve federated learning, they propose that the user's ratings should be stored locally. 
The user embeddings can be trained locally, and the server only retains the item embeddings. 
This training framework leads to protecting the privacy data, as there is no transfer of users' interactions. 
Ribero et al.~\cite{ribero2020federating} argues that the model updates sent to the server may contain sufficient information to uncover raw data, which leaves privacy concerns. 
They propose to use the differential privacy~\cite{mcsherry2009differentially} to limit the exposure of the data in a federated recommender system.  
FED-MVMF~\cite{flanagan2020federated} extends the Multi-View Matrix Factorization~(MVMF)~\cite{singh2008relational} to a federated learning framework. 
It simultaneously factorizes both feature matrices and interaction matrices. A-FRS~\cite{chen2020robust} proposes a robust federated recommender system against the poisoning attacks of clients. It employs an item similarity model~\cite{kabbur2013fism} in learning the user/item embeddings. 
FedGNN~\cite{wu2021fedgnn} is the most recent work that combines GNN with a federated recommender system, which is also the most relevant work to our paper. 
However, FedGNN fails to solve social recommendations, and the clients' models are not personalized~\cite{fallah2020personalized}. 
We present a comparison of a set of representative social recommender systems and federated learning methods in Table~\ref{tab:model_compare}. 

\begin{table}[htbp]\caption{Comparison of representative models with respect to social information, multi-relation, graph neural network, rating protection, interaction protection and data storage}\label{tab:model_compare}
	\centering
	\begin{tabular}{l|cccc}
	\toprule
	& Social Information & Multi-relation & Graph Neural Network & Data Storage  \\
	\hline
	RSTE~\cite{ma2009learning} & \checkmark & $\times$ & $\times$ & centralized \\
	TrustWalker~\cite{jamali2009trustwalker} & \checkmark & $\times$ & $\times$ & centralized \\
	SoRec~\cite{ma2008sorec} & \checkmark & $\times$ & $\times$ & centralized \\ 
	SoReg~\cite{ma2011recommender} & \checkmark & $\times$ & $\times$ &  centralized\\
	SocialMF~\cite{jamali2010matrix} & \checkmark & $\times$ & $\times$ & centralized\\
	TrustSVD~\cite{guo2015trustsvd} & \checkmark & $\times$ & $\times$ & centralized\\
	CUNE~\cite{CUNE} & \checkmark & \checkmark & $\times$ & centralized\\
	GCMC+SN~\cite{berg2017graph} &\checkmark & $\times$ & \checkmark & centralized\\
	DANSER~\cite{wu2019dual} & \checkmark & \checkmark & \checkmark & centralized \\
	GraphRec~\cite{fan2019graph} & \checkmark&  \checkmark & \checkmark & centralized\\
    ConsisRec~\cite{yang2021consisrec} & \checkmark & \checkmark & \checkmark & centralized\\
    FedMF~\cite{chai2020secure} & $\times$ & $\times$ & $\times$ & local\\
    FedGNN~\cite{wu2021fedgnn} & $\times$ & $\times$ & \checkmark & local\\
    \modelname & \checkmark & \checkmark & \checkmark & local\\
    \bottomrule
	\end{tabular}
\end{table}

% \begin{table}[htbp]\caption{Comparison of different models with respect to social information, multi-relation, graph neural network, rating protection, interaction protection and data storage}\label{tab:model_compare}
% 	\centering
% 	\begin{tabular}{l|cccccccccc}
% 	\toprule
% 	& SoRec~\cite{ma2008sorec} & SoReg~\cite{ma2011recommender} & SocialMF~\cite{jamali2010matrix}  &	CUNE~\cite{CUNE} & GCMC+SN~\cite{berg2017graph} & GraphRec~\cite{fan2019graph} & ConsisRec~\cite{yang2021consisrec} & FedMF~\cite{chai2020secure} &FedGNN~\cite{wu2021fedgnn} & \modelname \\
% 	\hline
% Social Information     & \checkmark  & \checkmark  & \checkmark  & \checkmark  & \checkmark& \checkmark& \checkmark& $\times$ & $\times$ & & \checkmark \\ 
% Multi-relation         & $\times$    & $\times$    & $\times$    & \checkmark  & \checkmark & \checkmark & $\times$ & $\times$ & \checkmark \\
% Graph Neural Network   & $\times$    & $\times$    & $\times$    & $\times$    & \checkmark & \checkmark & \checkmark & $\times$ & \checkmark & \checkmark\\
% Rating Protection      & $\times$    & $\times$    & $\times$    & $\times$    & $\times$ & $\times$ & $\times$ & \checkmark & \checkmark & \checkmark \\
% Interaction Protection & $\times$    & $\times$    & $\times$    & $\times$    & $\times$ & $\times$ & $\times$ $\times$ & \checkmark & \checkmark\\
% Data Storage           & centralized & centralized & centralized & centralized & centralized & centralized & centralized & local & local & local \\
%     \bottomrule
% 	\end{tabular}
% \end{table}

\section{Preliminary}\label{sec:preliminary}
In this section, we present the preliminaries and definitions of essential concepts. The glossary of necessary notations are summarized in Table~\ref{tab:notation}.
\subsection{Definitions}
The target in a social recommendation is to predict the users' ratings to items, when given social interactions and user-item interactions. Denote the $\mathcal{U}=\{u_1, u_2, \dots, u_N\}$ and $\mathcal{T}=\{t_1,t_2,\dots,t_M\}$ as the set of users and item, respectively. $N$ and $M$ are the numbers of users and items, respectively. The social recommendation is to complete the ratings of users to items given both rating matrix $\mathbf{R}\in\mathbb{R}^{N \times M}$ and the social connection matrix $\mathbf{S}\in\{0,1\}^{N\times N}$. We denote the user $n$'s rating value to an item $m$ as $\mathbf{R}_{nm}$. Similarly, the connection between an user $n$ and user $p$ is denoted as $\mathbf{S}_{np}$. In a federated learning scenario, the data of each user is stored locally. Hence, both the rating matrix and social connection matrix are not available. The data of each user are stored in the local client, which is defined as:
\begin{mydef}
\textbf{(Client)}. A client $c$ is defined as a local device storing the rating data and the social data. Each client $c_n$ is associated with a user $n$, whose rating data and social data are $\mathbf{R}_{n\cdot}$ and $\mathbf{S}_{n\cdot}$, respectively.
\end{mydef}

\begin{mydef}
\textbf{(Server).} A server is defined as a central device managing the coordination of multiple clients in training a model. It does not exchange raw data from clients but only requests necessary messages for updating the model.
\end{mydef}

In this paper, we assume clients and server to be honest-but-curious~\cite{kairouz2019advances}. 
In other words, they provide correct information and cannot tamper with the training process.  The Federated Social Recommender System~(FSRS) is to complete the rating matrix given its partially complete rating data and social data, which is defined as follows:
\begin{mydef}
\textbf{(FSRS).} For $n$ clients and a server, given the partially observed rating data and social data as $\mathbf{r}_{n} = [r_{n1},r_{n2},\dots,r_{nk}]$ and  $\mathbf{s}_{n} = [s_{n1},s_{n2},\dots,s_{np}]$ of each client $c_n$, respectively, where $n,p\in\{1,2,\dots,N\}$ and $k\in\{1,2,\dots,M\}$, an FSRS can predict the unobserved rating data of the client $c_n$ without access to the raw data in each client.
\end{mydef}

Note that both the rating and social data are stored locally in the corresponding clients and never be uploaded to the server. Multiple clients collaboratively train an FSRS under the orchestration of the center server~\cite{kairouz2019advances}.  

\subsection{Formulation}
Our FSRS is designed by formulating the data on clients as multiple local graphs, which is illustrated in Fig.~\ref{fig:intro}. The local graph contains the first order neighbors of the client user, including item ratings and social neighbors. We denote the local graph for client $c_n$ as $\mathcal{G}_{n}$, which consists of both user nodes and item nodes. $\mathcal{G}_{n}$ is constructed from partially observed privacy data. Moreover, there are two type of edges in $\mathcal{G}_{n}$, i.e., the user-item edges with rating value as attributes and the user-user edges denoting the social interactions. For each client $c_n$, we denotes its rated items as $\mathcal{T}^{(n)}= \{t_{1}^{(n)},t_{2}^{(n)},\dots,t_{k}^{(n)}\}$ and social neighbors as $ \mathcal{U}^{(n)}=\{u_{1}^{(n)},u_{2}^{(n)},\dots,u_{p}^{(n)}\}$. The FSRS can predict the rating value of an unobserved item $t^{*}\in\mathcal{T}\setminus\mathcal{T}^{(n)}$. In other words, the FSRS can predict the attribute value of the local graph for the edge between the user $u_i$ and a new item $t^{*}$. Thus, the problem can be formulated as follows:
\begin{mydef}
\textbf{(Problem Definition)}: Given the local graphs $\{\mathcal{G}_{n}|_{n=1}^{N}\}$, can we collaboratively train a model to predict the attribute value for an unobserved edges $(u_n, t^{*})$ without access to the raw data of any local graphs? 
\end{mydef}

We specify the social recommendation problem to be a link prediction problem. It indicates that we should learn graph embeddings from local graphs to preserve the structural information. Additionally, it is necessary to tackle the heterogeneity~\cite{liu2020basconv,hu2020heterogeneous,yang2020heterogeneity} of those local graphs. 

\begin{table}
\caption{Glossary of Notations.}  
% \resizebox{\linewidth}{!}{%
\begin{tabular}{r|l}  
\hline\hline
\textbf{Symbol} & \textbf{Definition} \\
\hline
$\mathcal{U};\mathcal{T}$ & user set; item set \\
$\mathbf{R};\mathbf{S}$ & rating matrix; social connection matrix \\
$N;M$ & total number of users; total number of items \\
$c_i$ & client $n$, which is associated with user $n$ \\
$\mathbf{r}_{n};\mathbf{s}_{n}$ & the local observed rating and social data of client $c_n$ \\
$\mathcal{T}^{(n)};\mathcal{U}^{(n)}$ & the local observed rated items; the social connected neighbors of client $c_n$ \\
$\mathbf{E}_{u}(\mathbf{e}_{u});\mathbf{E}_{t}(\mathbf{e}_{t})$ & embedding for users, embedding for items \\
$\mathbf{e}^{*}_{u_n}$ & the local inference embedding of user $n$\\
$\mathbf{a};\mathbf{b};\mathbf{c}$ & attention layer vector for user-user interaction; for item-item interaction; for relation vector \\
$\mathbf{W}_1;\mathbf{W}_2$ & linear mapping matrix for user-user interaction; for item-item interaction \\
$\alpha_{up},\beta_{uk}$ & attention weights for neighbor users; for neighbor items \\
$\mathbf{h}_{u};\mathbf{h}_t$ & hidden embeddings for neighbor users; hidden embeddings for neighbor items \\
$\gamma_{u};\gamma_{t}$ & attention weight for aggregating social relation; for aggregating user-item relation \\
$\mathbf{v}_{u};\mathbf{v}_{t}$ & social relation vector; user-item relation vector \\
$\mathbf{R};\hat{\mathbf{R}}$ & ground truth rating score; predicted rating score \\
$\mathcal{L}_{u};\tilde{\mathcal{L}}_{u}$ & loss for client $u$; protected loss for client $u$ \\
$\mathbf{g}^{(n)};\mathbf{g}_{e}^{(n)};\mathbf{g}_{m}^{(n)}$ & the gradients for client $n$; the embedding gradients; the model gradients \\
$\delta; \lambda$ & the parameters for LDP \\
\hline
\hline
\end{tabular}
% }
\label{tab:notation}
\end{table}

\section{Proposed Framework}\label{sec:proposed_model}
In this section, we illustrate the FSRS framework of our proposed \modelname. It has three crucial modules: embeddings layer, local graph neural networks, and gradient protector. The proposed framework is in Fig.~\ref{fig:framework}.

\subsection{Embeddings}
Node embeddings are crucial components in preserving graph structural information~\cite{wang19neural,liu2020basconv,yang2021consisrec}. The \modelname has embedding layers for user and item nodes. We denote the embeddings for users and items as $\mathbf{E}_{u}\in\mathbb{R}^{d\times N}$ and $\mathbf{E}_{t}\in\mathbb{R}^{d\times N}$, respectively, which are both maintained by the server. $d\in\mathbb{N}_{+}$ is the dimension size for embeddings. Clients request the embeddings tables from the server. Then, they learn a local user/item embeddings and a local GNN model by using their interaction data. Those embeddings will be updated on the server by aggregating the gradients uploaded from clients. 

A client downloads the complete embedding tables and uses the user/item ids in interaction records to infer the corresponding embeddings. To be more specific, for a client $n$, which has rated items as $\mathcal{T}^{(n)}= \{t_{k}^{(n)}|_{k=1}^{K}\}$ and social neighbors as $ \mathcal{U}^{(n)}=\{u_{p}^{(n)}|_{p=1}^{P}\}$, its rated item embeddings are $\{\mathbf{e}^{(n)}_{t_k}|_{k=1}^{K}\}$ and social neighbor embeddings are $\{\mathbf{e}^{(n)}_{u_p}|_{p=1}^{P}\}$, where $\mathbf{e}^{(n)}_{t_k},\mathbf{e}^{(n)}_{u_p}\in\mathbb{R}^{d}$ and $K,P$ denote the total number of item neighbors and user neighbors, respectively. Those embeddings are input the local GNN model to learn client user embedding and predict item scores. 

\subsection{Local Graph Neural Network}
The local GNN module is the major component in \modelname to learn node embeddings and make a prediction. It consists of heterogeneous graph attention layers, relational graph aggregation layers, and rating prediction layers. 

\subsubsection{Relational Graph Attention}
\begin{figure}
    \centering
    \includegraphics[width=0.9\textwidth]{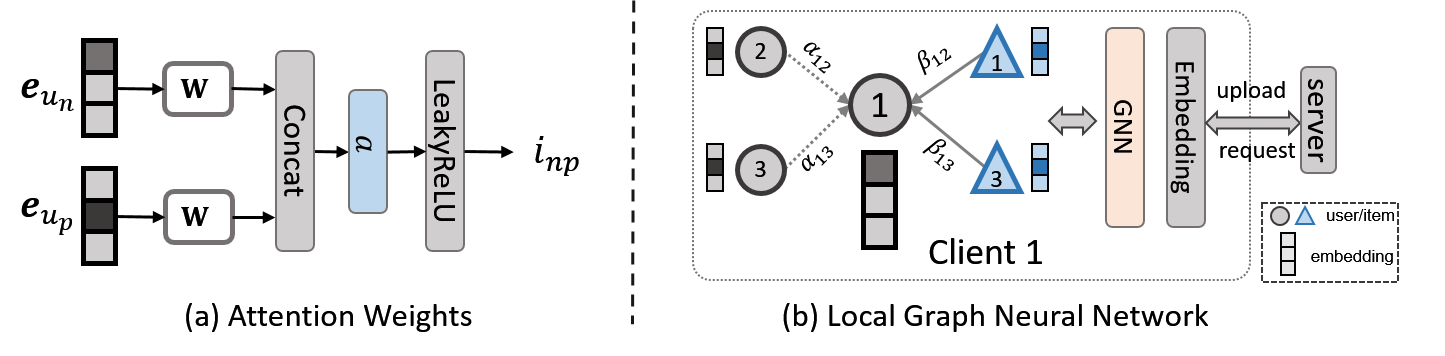}
    \caption{(a): The calculation of the attention weights between two embeddings, which consists of linear mapping matrix $\mathbf{W}$, concatenation of two embeddings, attention layer $a$, non-linear activation, and softmax function. (b): The local GNN learns the embedding of $u_1$ by aggregating the neighboring embeddings. }
    \label{fig:local_GNN}
\end{figure}
In general, we have no constraints on the local graph neural networks. There can be arbitrary GNN models, such as GCN~\cite{kipf2016semi}, GAT~\cite{velivckovic2017graph}, and GraphSAGE~\cite{hamilton2017inductive}, etc. This paper focuses on proposing a new FSRS framework. We directly adopt the GAT layer for learning node embedding and leave other GNN layers for future investigation. The GAT layer is designed by employing the self-attention mechanism~\cite{vaswani2017attention}. To learn the node embedding of $u_n$, we aggregate neighbor embeddings of $u_n$. However, since neighbors contribute unequally to the center node $u_n$, we should first learn a weight for each neighbor by employing an attention layer. Specifically, for a social pair $(u_n,u_p)$, their attention scores are formulated as:
\begin{equation}\label{eq:linear_mapping}
    o_{np} = \text{Attention}(\mathbf{W}_{1}\mathbf{e}_{u_n},\mathbf{W}_{1}\mathbf{e}_{u_p}),
\end{equation}
where $o_{np}$ is a scalar denoting the attention weight, $\mathbf{W}_{1}\in \mathbb{R}^{d\times d}$ is a linear mapping matrix, and Attention is the attention layer. We define the attention layer as a single-layer feed-forward neural network. It is parameterized with a weight vector $a\in\mathbb{R}^{2d}$ and employs a LeakyReLU activation~\cite{vaswani2017attention}, which is formulated as:
\begin{equation}\label{eq:attention_layer}
    o_{np} = \text{LeakyReLU}\left(\mathbf{a}^{\top}\left[\mathbf{W}_{1}\mathbf{e}_{u_n} \| \mathbf{W}_{1}\mathbf{e}_{u_p}\right]\right),
\end{equation}
where $\mathbf{a}^{\top}$ denotes the transpose of the attention layer parameter and $\|$ denotes the concatenation operation of two vectors. An illustration of the attention weight is in Fig.~\ref{fig:local_GNN}(a). The attention weight should be calculated for all the neighbors of the center node $u$, which forms a probability distribution by employing the softmax function:
\begin{equation}
\alpha_{np}=\operatorname{softmax}_{p}\left(o_{np}\right)=\frac{\exp \left(o_{np}\right)}{\sum_{i=1}^{P} \exp \left(o_{ni}\right)},
\end{equation}
where $\alpha_{np}$ is the final attention weights, exp denotes the exponential function. Note that $\alpha$ is calculated as the attention weights for user neighbors. We should learn attention weights for user-item pair $(u_n,i_k)$ in a similar way, which employs another linear mapping as in Eq.~(\ref{eq:linear_mapping}) and another attention parameters as in Eq.~(\ref{eq:attention_layer}), as following:
\begin{equation}\label{eq:attention_layer}
    v_{nk} = \text{LeakyReLU}\left(\mathbf{b}^{\top}\left[\mathbf{W}_{2}\mathbf{e}_{u_n} \| \mathbf{W}_{2}\mathbf{e}_{i_k}\right]\right),
\end{equation}
where $\mathbf{W}_2\in\mathbb{R}^{d\times d}$ is the mapping matrix and $\mathbf{b}\in\mathbb{R}^{2d}$ is the weights for the user-item interaction attention layer. By employing the softmax to all the item neighbors, we derive the attention weights for item neighbors as:
\begin{equation}
\beta_{nk}=\operatorname{softmax}_{k}\left(v_{nk}\right)=\frac{\exp \left(v_{ni}\right)}{\sum_{i=1}^{K} \exp \left(v_{ni}\right)},
\end{equation}
where $\beta_{nk}$ denotes the attention weights for items normalized over all neighboring items, next, we present the relational aggregation for both user neighbors and item neighbors. 

\subsubsection{Relational Graph Aggregation}
\begin{figure}
    \centering
    \includegraphics[width=0.95\textwidth]{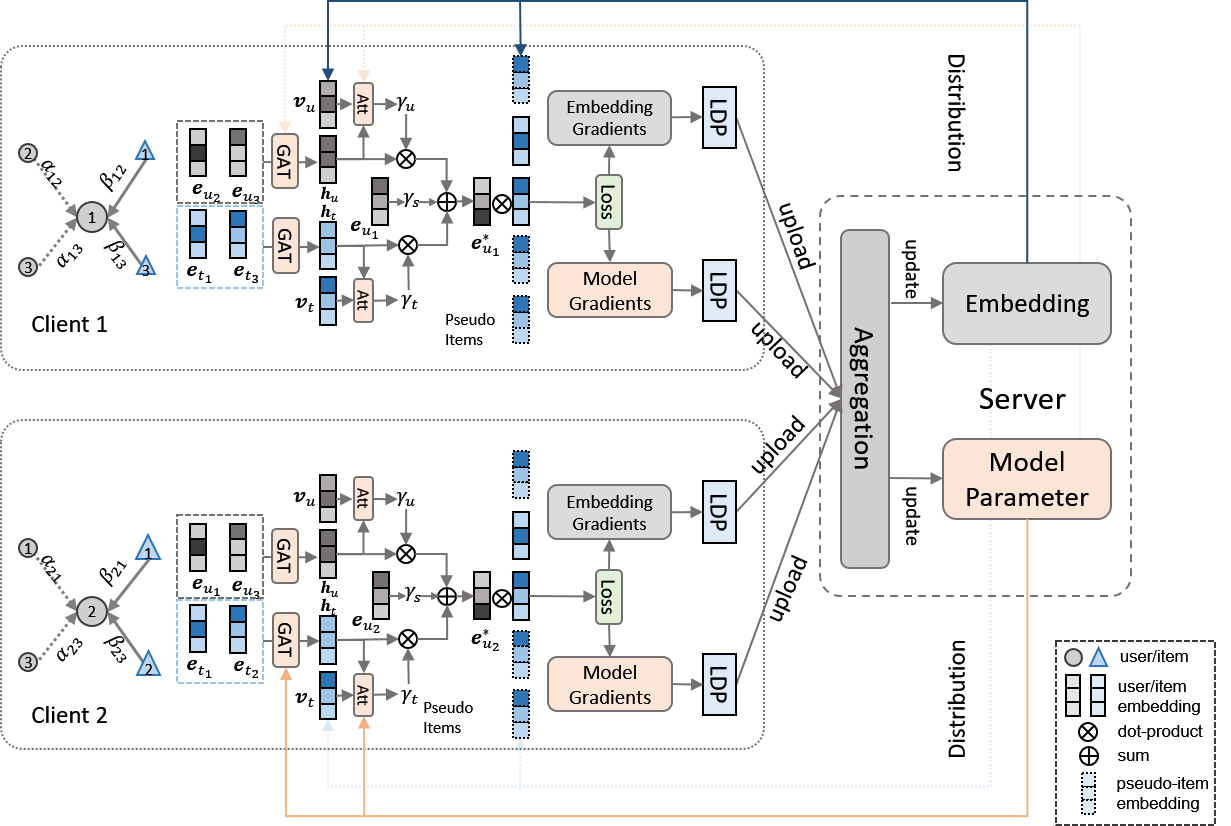}
    \caption{The framework of \modelname. For simplicity and without loss of generality, we present a two-client scenario. In each client, we use the local GAT layer to infer node embeddings and adopt the attention layer to aggregate social neighbors and item neighbors. Then, we sample a set of pseudo items bundled with local data to calculate the loss and gradients. Both the embedding gradients and model gradients are uploaded to the server for aggregation after LDP operation. }
    \label{fig:framework}
\end{figure}

Inferring the center user embeddings requires aggregating both neighbor user nodes and neighbor item nodes with their associated attention weights, which are formulated as follows:
\begin{equation}
    \mathbf{h}_{u}^{(n)} = \sum_{p=1}^{P}\alpha_{np}\mathbf{W}_{h}\mathbf{e}_{u_p}, \quad \mathbf{h}_{t}^{(n)} = \sum_{k=1}^{K}\beta_{nk}\mathbf{W}_{h}\mathbf{e}_{t_k},
\end{equation}
where $\mathbf{W}_{h}\in\mathbb{R}^{d\times d}$ is the linear mapping weight matrix, and $\mathbf{h}_{u}^{(n)},\mathbf{h}_{t}^{(n)}\in\mathbb{R}^{d}$ denotes the hidden embeddings for aggregating user neighbors and item neighbors, respectively. The general aggregation process is illustrated in Figure~\ref{fig:local_GNN}(b). Intuitively, we should aggregate hidden embeddings and the center node embedding to infer the embedding of $u_n$. However, social relation, user-item interactions and center node embedding are not equally contributed to the learning process~\cite{yang2021consisrec}. We should also handle the heterogeneity during the aggregation step. Therefore, we propose to use three relation vectors to preserve their semantics, i.e., $\mathbf{v}_{u}, \mathbf{v}_{t}, \mathbf{v}_{s}\in \mathbb{R}^{d}$ preserving the social semantics, user-item semantics and center node itself semantics respectively.  To be more specific, we concatenate hidden embeddings with their relation vectors and employing the self-attention mechanism to learn the weights for aggregation:
\begin{equation}
    \gamma_u = \frac{\exp \left(\mathbf{c}^{\top}\left[\mathbf{h}_{u}^{(n)}\| \mathbf{v}_{u} \right]\right)}{\exp \left(\mathbf{c}^{\top}\left[\mathbf{h}_{u}^{(n)}\| \mathbf{v}_{u} \right]\right) + \exp \left(\mathbf{c}^{\top}\left[\mathbf{h}_{t}^{(n)}\| \mathbf{v}_{t} \right]\right) + \exp \left(\mathbf{c}^{\top}\left[\mathbf{h}_{s}^{(n)}\| \mathbf{v}_{s} \right]\right)},
\end{equation}
\begin{equation}
    \gamma_t = \frac{\exp \left(\mathbf{c}^{\top}\left[\mathbf{h}_{t}^{(n)}\| \mathbf{v}_{t} \right]\right)}{\exp \left(\mathbf{c}^{\top}\left[\mathbf{h}_{u}^{(n)}\| \mathbf{v}_{u} \right]\right) + \exp \left(\mathbf{c}^{\top}\left[\mathbf{h}_{t}^{(n)}\| \mathbf{v}_{t} \right]\right) + \exp \left(\mathbf{c}^{\top}\left[\mathbf{h}_{s}^{(n)}\| \mathbf{v}_{s} \right]\right)},
\end{equation}
\begin{equation}
    \gamma_{s} = \frac{\exp \left(\mathbf{c}^{\top}\left[\mathbf{h}_{s}^{(n)}\| \mathbf{v}_{s} \right]\right)}{\exp \left(\mathbf{c}^{\top}\left[\mathbf{h}_{u}^{(n)}\| \mathbf{v}_{u} \right]\right) + \exp \left(\mathbf{c}^{\top}\left[\mathbf{h}_{t}^{(n)}\| \mathbf{v}_{t} \right]\right) + \exp \left(\mathbf{c}^{\top}\left[\mathbf{h}_{s}^{(n)}\| \mathbf{v}_{s} \right]\right)},
\end{equation}
where $\gamma_u$, $\gamma_t$ and $\gamma_{s}$ are the attention weights for hidden user neighbors embedding, hidden item neighbor embedding and center node itself embedding, respectively. $\mathbf{c}\in\mathbb{R}^{2d}$ is the weight vector for the attention layer. Given that, we infer the node embeddings of $u_n$ as:
\begin{equation}\label{eq:embedding_inference}
    \mathbf{e}^{*}_{u_n} = \gamma_{s}\mathbf{e}_{u_n} + \gamma_{u}\mathbf{h}_{u}^{(n)} + \gamma_{t}\mathbf{h}_{t}^{(n)},
\end{equation}
where $\mathbf{e}_{u_n}^{*}$ is the local user embedding for prediction. The aggregation is illustrated in Fig.~\ref{fig:framework}.  As such, local clients preserve their user embeddings, which tackles the personalization problem. Next, we will use the learned embeddings and downloaded item embeddings to make a prediction. 

\subsubsection{Prediction}
To predict the local item ratings, we adopt the dot-product between the inferred user embedding and item embeddings. For a user $u$ and an item $t$ with embedding $\mathbf{e}^{*}_{u}$ and $\mathbf{e}_{t}$, respectively, the rating $\mathbf{R}_{ut}$ is:
\begin{equation}\label{eq:rating_prediction}
    \hat{\mathbf{R}}_{ut} = \mathbf{e}_{u}^{*} \cdot \mathbf{e}_{t},
\end{equation}
where $\cdot$ denotes the dot-product operation. We use the local user-item rating values to optimize the prediction by employing the Root Mean Squared Error~(RMSE) between the predicted score $\hat{\mathbf{R}}_{ut}$ and the ground-truth rating score $\mathbf{R}_{ut}$:
\begin{equation}\label{eq:loss}
    \mathcal{L}_{u} = \sqrt{\frac{\sum_{t\in\mathcal{T}^{(u)}}(\mathbf{R}_{ut}-\hat{\mathbf{R}}_{ut})^2}{|\mathcal{T}^{(u)}|}},
\end{equation}
where $\mathcal{T}^{(u)}$ denotes the rated items of user $u$ and $\mathcal{L}_{u}$ is the local loss for user $u$. Recall that each user is associated with a client. This loss function will be used to calculate the gradient for clients. Then, the gradients are collected from multiple clients to the server for further updates. However, directly uploading gradients leads the user-item interaction data to be vulnerable~\cite{chai2020secure,wu2021fedgnn}. It urges us to design a privacy protection mechanism regarding the gradients, which will be introduced next.

\subsection{Privacy Protection}
This section introduces two techniques to protect the local user-item interaction data when uploading the gradients: dynamic Local Differential Privacy~(LDP) and pseudo-item labelling. 
\subsubsection{Local Differential Privacy}
According to FedMF~\cite{chai2020secure}, the user's rating information can be inferred if given the gradients of a user uploaded in two continuous steps. Though our case is more complicated than in FedMF, it is still problematic if directly optimizing the local data and uploading the gradients. Moreover, for embedding gradients, only items with ratings in a local client have non-zero gradients to upload to the server.  FedMF~\cite{chai2020secure} proposes using encryption for the gradients so that the server cannot inverse the encoding process. However, it requires generating the public key and secret key and introducing additional computation for encryption. Additionally, to solve the zero-gradient for non-rated items, FedGNN~\cite{wu2021fedgnn} proposes to sample pseudo-interacted items and add Gaussian noise with the same mean and variance as the ground-truth items to their gradients. Another technique used broadly is the Local Differential Privacy~(LDP) module~\cite{erlingsson2014rappor,ribero2020federating,qi2020privacy}. It adopts clipping the local gradients based on their L$\infty$-norm with a threshold $\delta$ and applies a LDP module with zero-mean Laplacian noise to the unified gradients to achieve privacy protection. 

To be more specific, we instantiate the item embedding gradients, user embedding gradients and model gradients from client $n$ as $\mathbf{g}_t^{(n)}$, $\mathbf{g}_u^{(n)}$ and $\mathbf{g}_{m}^{(n)}$, respectively. Combining them gives the gradients as $\mathbf{g}^{(n)}=\{\mathbf{g}_t^{(n)},\mathbf{g}_u^{(n)},\mathbf{g}_{m}^{(n)}\}=\frac{\partial\mathcal{L}_{u}}{\partial\mathbf{\Theta}}$, where $\mathbf{\Theta}$ denotes all trainable parameters. Then, the LDP is formulated as:
\begin{equation}\label{eq:LDP_gradient}
    \tilde{\mathbf{g}}^{(n)} = \text{clip}(\mathbf{g}^{(n)}, \delta) + \text{Laplacian}(0,\lambda),
\end{equation}
where $\tilde{\mathbf{g}}^{(n)}$ is the randomized gradients, clip$(x,\delta)$ denotes limiting $x$ with the threshold $\delta$, and Laplacian$(0,\lambda)$ is the Laplacian noise with $0$ mean and $\lambda$ strength. However, a constant noise strength is inappropriate when dealing with gradients at different magnitudes. The gradient magnitude of different parameters varies during training. Hence, we propose to add dynamic noise based on the gradient, which is formulated as follows:
\begin{equation}\label{eq:LDP_gradient_dynamic}
    \tilde{\mathbf{g}}^{(n)} = \text{clip}(\mathbf{g}^{(n)}, \delta) + \text{Laplacian}(0,\lambda\cdot\text{mean}(\mathbf{g}^{(n)})),
\end{equation}

\subsubsection{Pseudo-Item Labelling}
Based on existing work, we propose a new privacy protection module, which is advantageous as it can protect the training gradients and enhance the model with more robustness. In a local client, before calculating the training loss, we first sample $q$ items not in the neighbor items, which are the pseudo-items in Fig.~\ref{fig:framework}, denoting as $\tilde{\mathcal{T}}^{(u)} = \{\Tilde{t}^{(u)}_1,\tilde{t}^{(u)}_2,\dots,\tilde{t}^{(u)}_q\}$. Then, we use the local model to predict the ratings for these pseudo items. The predicted ratings are rounded to be the pseudo ratings. Hence, the loss in Eq.~(\ref{eq:loss}) is changed to:
\begin{equation}\label{eq:loss_new}
    \tilde{\mathcal{L}}_{u} = \sqrt{\frac{\sum_{t\in\mathcal{T}^{(u)}\cup\tilde{\mathcal{T}}^{(u)}}(\mathbf{R}_{ut}-\hat{\mathbf{R}}_{ut})^2}{|\mathcal{T}^{(u)}|}}.
\end{equation}
Note that compared with Eq.(\ref{eq:loss}), Eq.(\ref{eq:loss_new}) is calculated from both the true interacted items and pseudo items. The ground-truth ratings for pseudo items are the rounded predicted score. The difference between the predicted score and the rounded one contributes to the gradients for those pseudo items.  Here, we assume that $\mathbf{R}_{ut}\in\mathbb{N}$, while $\hat{\mathbf{R}}_{ut}\in\mathbb{R}$. The gradients derived from Eq.~(\ref{eq:loss_new}) contain both ground-truth rating information and the pseudo item rating information, which prevents the data leakage problem. Additionally, the pseudo labels of items provide additional rating information, which can alleviate the cold-start issue of the data. Intuitively, this technique works as a data augmentation method~\cite{liu2021augmenting,xia2020composed,liu2020kg}. We sample those pseudo items and view the difference between rounded ratings with a predicted rating as the randomness, which enhances the robustness of the local model.

\begin{algorithm}[b]
\caption{\textbf{\modelname} (\textbf{Fe}derated \textbf{So}cial Recommendation with \textbf{G}raph Neural Network)}
\label{alg:fesog}
\SetKwData{False}{False}\SetKwData{This}{this}\SetKwData{Up}{up}
\SetKwFunction{Union}{Union}
\SetKwFunction{server}{ServerRun}
\SetKwFunction{ClientUpdate}{ClientUpdate}
\SetKwInOut{Input}{Input}\SetKwInOut{Output}{Output}
\SetKwProg{Fn}{Function}{:}{\KwRet}

\Input{Embedding Size, learning rate:$d$, $\eta$ \\
Total number of clients, items: $N,M,T$ \\
The number of pseudo items: $p$ \\
LDP parameter: $\delta, \lambda$ \\
Clients local graph: $\{\mathcal{G}_{n} | _{n=1}^{N}\}$

}

\Output{Model parameters and embeddings $\mathbf{\Theta}$; Local client embeddings $\{\mathbf{e}_{u_{n}}^{*}|_{n=1}^{N}\}$}
\BlankLine
 Initializing $\mathbf{\Theta}$; 
 
% (\tcp*[f]{federated training})
\While{not converge\label{alg:server_loop_start}}{
sampling a fractions of clients $\mathcal{N}$; 
\label{alg:clients_sampling}
\For{$n\in\mathcal{N}$}{
    $\mathbf{g}^{(n)},|\mathbf{R}_{n}|$ = \ClientUpdate{$n$, $\mathbf{\Theta}$}; \tcp*[f]{collecting gradients from clients}\\
    }
$\bar{\mathbf{g}}^{(n)}\leftarrow$ Eq.~(\ref{eq:avg_gradients}); \tcp*[f]{averaging gradients from clients} \\
$\mathbf{\Theta}=\mathbf{\Theta} - \eta\cdot\bar{\mathbf{g}}^{(n)}$; \tcp*[f]{updating parameters}
}\label{alg:server_loop_end}

\Fn{\ClientUpdate{$n$, $\mathbf{\Theta}$}}{
downloading $\mathbf{\Theta}$ from server;\label{alg:downloading} \\
$\mathbf{e}_{u_n}^{*} \leftarrow$ Eq.~(\ref{eq:embedding_inference});\label{alg:user_inference} \tcp*[f]{local user embedding inference} \\
sampling $p$ pseudo items;\label{alg:pesudo-item sampling} \\
calculating the ratings of those pseudo items using Eq.~(\ref{eq:rating_prediction}); \tcp*[f]{pseudo-labelling}\\
$\tilde{\mathcal{L}}_{n} \leftarrow $ Eq.~(\ref{eq:loss_new}); \\
$\mathbf{g}^{(n)} = \frac{\partial\tilde{\mathcal{L}}_{u}}{\partial\mathbf{\Theta}}$;  \tcp*[f]{computing the gradients} \\
$\tilde{\mathbf{g}}^{(n)}\leftarrow$ Eq.~(\ref{eq:LDP_gradient}); \tcp*[f]{LDP for gradients}\label{alg:LDP} \\
\KwRet $\tilde{\mathbf{g}}^{(n)}$, $ p+|\mathcal{T}^{(u)}|$;\label{alg:func_ret} \tcp*[f]{return gradients and the number of interactions}\\ 
}
\end{algorithm}

\subsection{Optimization}
In this section, we first present the optimization process of the \modelname framework before we present the pseudo-code of the algorithm. 
\subsubsection{Gradient Collection for Optimization}
The server in \modelname collects the gradients uploaded from clients to update both the model parameters and embeddings, which collaboratively optimize the model. Recall that the gradients from client $n$ is $\tilde{\mathbf{g}}^{(n)}$ and the parameters is $\mathbf{\Theta}$, which includes $\mathbf{\Theta}_m$, $\mathbf{\Theta}_t$ and $\mathbf{\Theta}_u$ indicating model parameters, item embedding and user embedding, separately. In each round, the server builds a connection with a batch (e.g., 128) of clients, denoted as $\mathcal{N}$. It first sends the current model parameters $\mathbf{\Theta}_m$ and embeddings $\mathbf{\Theta}_e$ to those clients. Then, it aggregates local gradients from those clients as follows:
\begin{equation}\label{eq:avg_gradients}
\bar{\mathbf{g}}_m=\frac{\sum_{n \in \mathcal{N}}\left|\mathcal{R}_{n}\right| \cdot \tilde{\mathbf{g}}_m^{(n)}}{\sum_{n \in \mathcal{N}}\left|\mathcal{R}_{n}\right|},
\quad
\bar{\mathbf{g}}_t=\frac{\sum_{n \in \mathcal{N}}\left|\mathcal{R}_{n}\right| \cdot \tilde{\mathbf{g}}_t^{(n)}}{\sum_{n \in \mathcal{N}}\left|\mathcal{R}_{n}^t\right|},
\quad
\bar{\mathbf{g}}_u=\frac{\sum_{n \in \mathcal{N}}\left|\mathcal{R}_{n}\right| \cdot \tilde{\mathbf{g}}_u^{(n)}}{\sum_{n \in \mathcal{N}}\left|\mathcal{R}_{n}^u\right|},
\end{equation}
where $\mathcal{R}_{n}$ is the total number interaction for calculating the gradients, including both the real interactions and pseudo interactions. 
$\mathcal{R}_{n}^t$ and $\mathcal{R}_{n}^u$ indicate interactions involving item $t$ and user $u$, separately. 
Intuitively, $\bar{\mathbf{g}}_m$,  $\bar{\mathbf{g}}_t$, $\bar{\mathbf{g}}_u$ are weighted average of the gradients from clients. After aggregation, the server updates the parameter $\mathbf{\Theta}$ with gradient descent as:
\begin{equation}\label{eq:gradient_update}
\mathbf{\Theta}^{*}=\mathbf{\Theta}-\eta \cdot \bar{\mathbf{g}},
\end{equation}
where $\eta$ is the learning rate. This learning process is operated multiple rounds until convergence. 
\subsubsection{Algorithm}
The pseudo-code of the algorithm of \modelname is presented in Algorithm~\ref{alg:fesog}. The input are consist of the training hyper-parameters such as the embedding size $d$ and learning $\eta$. Additionally, the client data should also be given, i.e. the clients local graphs $\{\mathcal{G}_{n} | _{n=1}^{N}\}$. Though the target is to predict item ratings, we output the parameters $\mathbf{\Theta}$ and the local inferred embeddings $\{\mathbf{e}_{u_{n}}^{*}|_{n=1}^{N}\}$, which is sufficient for clients to predict ratings. In the algorithm, the line~\ref{alg:server_loop_start} to the line~\ref{alg:server_loop_end} is the loop operated on the server, which sends parameters to clients and collects their gradients for updating. The function \textsf{ClientUpdate}() is the operation on local devices. It downloads the parameters to infer the local user embeddings (line~\ref{alg:downloading}). Then, the pseudo items are sampled~(line~\ref{alg:pesudo-item sampling}). Pseudo-labelling and LDP are combined ~(line~\ref{alg:pesudo-item sampling} to line~\ref{alg:LDP}) to protect the gradients from privacy leakage. 
This function returns the gradients and the number of interactions~(line~\ref{alg:func_ret}) for the server to collect. 

% \subsection{Model Analysis}
% In this section, we compare existing work to address the strengths of \modelname. 

\section{Experiments}\label{sec:experiments}
In this section, we conduct experiments to evaluate the effectiveness of \modelname. We will answer the following Research Questions~(RQs):
\begin{itemize}
    \item \textbf{RQ1}: Does \modelname outperform existing methods in social recommendation?  
    \item \textbf{RQ2}: What is the impact of the hyper-parameters in \modelname? 
    \item \textbf{RQ3}: Are those components in \modelname necessary? 
\end{itemize}

\subsection{Experimental Setup}
\subsubsection{Datasets}
In this paper, we adopt three commonly used social recommendation datasets to conduct experimental analyses, which are Ciao, Epinions~\cite{tang2012etrust,tang2012mtrust,tang2013exploiting} and Filmtrust~\cite{guo2013novel}. Ciao and Epinions\footnote{\url{https://www.cse.msu.edu/~tangjili/datasetcode/truststudy.htm}} \cite{tang2012etrust,tang2012mtrust,tang2013exploiting} are crawled from shopping website. Both datasets contain user rating scores on items and trust links between users as social relations. Each user can give an integer score in $\{1,2,3,4,5\}$ to rate an item, where $1$ indicates least like while $5$ represents most. Filmtrust\footnote{\url{https://guoguibing.github.io/librec/datasets.html}}~\cite{guo2013novel} is built from online film rating website and the trust relationship between users. The rating scale ranges from 1 to 8. Social relations are also the trust links between users in these datasets. Data statistics are shown in Table \ref{datasets}. In our FSRS scenario, each user is treated as a local client, and the user's interactions are local privacy data on the device. The global graph information is transferred from user embeddings.

\begin{table}[htbp]\caption{Statistics of datasets}\label{datasets}
	\centering
	\begin{tabular}{l|c|c|c}
		\toprule
		Dataset & Ciao & Epinions & Filmtrust \\
		\hline
		Users & 7,317 & 18,069 & 874 \\
% 		\hline
		Items & 104,975 & 261,246 & 1,957\\
		\hline
		\# of ratings & 283,320 & 762,938 & 18,662\\
% 		\hline
		Rating density & 0.0369$\%$ & 0.0162$\%$ & 1.0911$\%$\\
		\hline
% 		\hline
		\# of social connections & 111,781 & 355,530 & 1,853\\
% 		\hline
		Social connection density & 0.2088$\%$ & 0.1089$\%$	& 0.2426$\%$\\
		\bottomrule
	\end{tabular}
\end{table}

\subsubsection{Baselines}
We adopt three types of baselines for comparison: traditional Matrix Factorization~(MF) based methods for social recommendation, recent GNN-based methods for social recommendation, and federated learning frameworks.  MF-based and GNN-based methods are based on centralized learning, which is unable to protect user privacy. Federated learning methods are not able to handle the fusion of local social information and rating information.  
% The baseline methods are listed as follows:

\textbf{MF-based methods}
\begin{itemize}
    \item SoRec~\cite{ma2008sorec}: It co-factorizes user-item rating matrix and user-user social matrix.
    \item SoReg~\cite{ma2011recommender}: It develops a social regularization with social links to regularize on matrix factorization.
    \item SocialMF~\cite{jamali2010matrix}: Compared with SoReg, social matrix factorization also considers social trust propagation.
    \item CUNE~\cite{CUNE}: Collaborative user network embedding assumes users hold implicit social links from each other, and it tries to extract semantic and reliable social information by graph embedding method.
\end{itemize}

\textbf{GNN-based methods}
\begin{itemize}
    \item GCMC+SN~\cite{berg2017graph}: GCMC is a graph neural network based method. User nodes are initialized as vectors learned by node2vec~\cite{grover2016node2vec} from the social graph to obtain social information. The dense representation learned upon the social graph can include more information than the random initialized feature.
    \item GraphRec~\cite{fan2019graph}: Graph recommendation uses graph neural network to learn user embedding and item embedding from their neighbors and uses several fully connected layers as the rating predictor.
    \item ConsisRec~\cite{yang2021consisrec}: It is the state-of-the-art method in social recommendation. ConsisRec modifies graph neural network to mitigate the inconsistency problems in social recommendation.
  \end{itemize}
  
\textbf{Federated learning methods}
\begin{itemize}
    \item FedMF~\cite{chai2020secure}: It separates the matrix factorization computation to different users and uses an encryption method to avoid information leakage.
    \item FedGNN~\cite{wu2021fedgnn}: Federated graph neural network is the state-of-the-art federated recommendation method. It adopts local differential privacy methods to protect user's interaction with items.
\end{itemize}

% \begin{table}[htbp]\caption{Statistics of datasets}\label{datasets}
% 	\centering
% 	\begin{tabular}{l|cccccccccc}
% 	& SoRec & SoReg & SocialMF & GCMC+SN & GraphRec & CUNE & & ConsisRec & FedMF & FedGNN \\
% 	Social Information &&&&&&&& \\
% 	Multi-relation &&&&&&&& \\
% 	Graph Embedding &&&&&&&& \\
% 	Data Storage &&&&&&&& \\
% 	\end{tabular}
% \end{table}

\subsubsection{Evaluation Metrics}
To evaluate the performance and compare, we adopt Mean Absolute Error (MAE) and Root Mean Square Error (RMSE) to measure the model performance because they are the most commonly used metrics in social recommendation. Smaller values of both two metrics indicate better performance in the test data. The two metrics are calculated as follows:
\begin{equation}
	\text{MAE}=\frac{\sum_{n=1}^N\sum_{t\in\mathcal{T}^{(n)}}|\mathbf{R}_{nt}-\hat{\mathbf{R}}_{nt}|}{\sum_{n=1}^N|\mathcal{T}^{(n)}|},
\end{equation}
% \begin{equation}
% 	\text{RMSE}=\frac{\sum_{n=1}^N\sum_{t\in\mathcal{T}^{(n)}}(\mathbf{R}_{nt}-\hat{\mathbf{R}}_{nt})^{2}}{\sum_{n=1}^N|\mathcal{T}^{(n)}|}
% \end{equation}
\begin{equation}
	\text{RMSE}=\sqrt{\frac{\sum_{n=1}^N\sum_{t\in\mathcal{T}^{(n)}}(\mathbf{R}_{nt}-\hat{\mathbf{R}}_{nt})^{2}}{\sum_{n=1}^N|\mathcal{T}^{(n)}|}},
	\end{equation}
where $\mathbf{R}_{nt}$ and $\hat{\mathbf{R}}_{nt}$ are the true rating value and predicted rating value of user $n$ for item $t$, respectively. $N$ is the total number of users for testing. $\mathcal{T}^{(n)}$ denotes the rated items for user $n$. Again, the evaluation is conducted on devices locally since the server has no access to the local privacy data. The lower MAE and RMSR both indicate better performance. 

\subsubsection{Experimental Setups}
The ratings in each dataset are randomly split into training set ($60\%$), validation set ($20\%$), and test set ($20\%$). Hyper-parameters are tuned based on the validation performance. Then, we report the final performance on the test dataset. In all experiments, we initialize the parameters with standard Gaussian distribution. For the LDP technique used in \modelname, the gradient clipping threshold is set to $0.3$, and the strength of Laplacian noise is set to 0.1. Other hyperparameters are tuned based on grid searching. The number of pseudo interacted items $p$ is search in $\{10,50,100,500,1000\}$. Embedding size $d$ is tuned from $\{4,8,16,32,64\}$. User batch size in each training round is searched in $\{16,32,64,128,256\}$. Learning rate $\eta$ is searched in $\{0.1,0.05,0.01\}$. Training is stopped if RMSE on the validation set does not improve for $5$ successive validations.

\subsection{Overall Comparison (RQ1)}
% We compare FeSoG with baseline models from different category. SoRec, SoReg and SocialMF are matrix factorization based methods. These methods are simple and efficient. GCMC, Graph, CUNE and ConsisRec are GNN-based methods, GNN based models have a high performance in social recommendation. The recent state-of-the-art social recommendation methods are all based on GNN model. FedMF and FedGNN are two federated learning architecture for recommendation, which aims to protect user's private data from the central server. 
In this section, we conduct the overall comparison of different models. The experimental results are shown in Table \ref{Comparison Experiment}, which are categorized into three groups. We have the following observations:
\begin{itemize}
    \item \modelname significantly outperforms the State-Of-The-Art~(SOTA) federated recommender systems in all datasets. Compared with FedGNN, \modelname achieves on average $2.99\%$ and $4.03\%$ relative improvements on RMSE and MAE, respectively. Several advantages of \modelname support its superiority: (1) social information helps the recommendation, which the improvement can demonstrate over FedMF; (2) the relational graph attention and aggregation can effectively integrate both user-item interactions and social information; (3) the local pseudo-item sampling technique enhances the performance.
    \item The GNN-based models perform better than those MF-based models. ConsisRec is the SOTA GNN model that employs relation attention and consistent neighbor aggregation, which leads to its best performance. Compared with SocialMF, which is the best MF-based method, ConsisRec achieves in average $3.47\%$ and $5.27\%$ relative improvements in RMSE and MAE, respectively. GNN-based models are better as they can directly model structure information and simultaneously aggregate social and user-item interactions. Between the two federated learning baselines, FedGNN also significantly outperforms FedMF, which again supports the claims that GNN-based models are better than MF-based methods. \modelname is also based on GNN aggregation. Its local GNN aggregation employs relation attention, which leads to its better performance against FedGNN. 
    \item Federated learning impairs the performance compared with centralized learning. Even a simple GCMC+SN model is better than the FedGNN model. There are two reasons: On the one hand, to achieve privacy protection, the federated learning framework has no access to the local data, limiting its capacity to model the global structures. According to~\cite{liu2020alleviating,fan2019graph}, the core of graph embedding is to aggregate high-order neighbors and select informative contexts. On the other hand, the local gradients are protected by adding random noise. Though it theoretically will not hurt the performance in expectation, it still prevents the server from receiving qualitative gradients from clients.  We should find a trade-off between performance and privacy protection. This observation also brings opportunities for federated learning research. 
\end{itemize}

\begin{table}[htb]
\caption{Experiment results compared with baseline methods. The best federated learning results are in bold, and the best results for non-federated learning methods are underlined. Improvement indicates the percent that \modelname improves against the second-best federated learning result.}
\label{Comparison Experiment}
\begin{tabular}{l|cccccc}
\toprule
\multirow{2}{*}{Method} & \multicolumn{2}{c}{Ciao} & \multicolumn{2}{c}{Epinions}  & \multicolumn{2}{c}{Filmtrust} \\
\cmidrule(r){2-3} \cmidrule(r){4-5} \cmidrule(r){6-7}
&  RMSE      &  MAE   
&  RMSE      &  MAE 
&  RMSE      &  MAE   \\
\midrule
SoRec  & 1.2024  & 0.8693  & 1.3389  & 1.0618 & 1.8094 & 1.4529 \\
SoReg  & 1.0066  & 0.7595  & 1.0751  & 0.8309 & 1.7950 & 1.4413 \\
SocialMF  & 1.0013  & 0.7535  & 1.0706  & 0.8264 & 1.8077 & 1.4557 \\
\hline
GCMC+SN & 1.0301 & 0.7970 & 1.1070 & 0.8480 & 1.8025& 1.4325\\
GraphRec  & 1.0040  & 0.7591  & 1.0799  & 0.8219 & \underline{1.6775} & 1.3194 \\
CUNE  & 1.0002  & 0.7591  & 1.0681 & 0.8284 & 1.7675 & 1.4178\\
ConsisRec & \underline{0.9722} & \underline{0.7394} & \underline{1.0495} & \underline{0.8046} &1.7148 & \underline{1.3093} \\
\hline
FedMF & 2.4216 & 2.0792 & 2.0685 & 1.5254 & 2.795 & 2.1713\\
FedGNN & 2.02 & 1.58 & 1.8346 & 1.4238 & 2.13 & 1.65\\
\modelname & \textbf{1.9136} & \textbf{1.4937} & \textbf{1.7969} & \textbf{1.3847} & \textbf{2.0942} & \textbf{1.5855}\\
\midrule
improvement & \text{5.26\%} & \text{5.46\%} & \text{2.05\%} & \text{2.74\%} & \text{1.68\%} & \text{3.9\%}\\

\bottomrule
\end{tabular}
\end{table}

\subsection{Sensitivity Analysis (RQ2)}
In this section, we emphasize on analyzing the impacts of those hyper-parameters involving in \modelname and some other baselines. We include user batch size $|\mathcal{N}|$ as in the line~\ref{alg:clients_sampling} of Algorithm~\ref{alg:fesog}, embedding size $d$, learning rate $\eta$, the number of pseudo items $p$, local differential privacy parameter $\delta$ and $\lambda$.
\subsubsection{User Batch Size}

\begin{figure}[htb]
    % \begin{center}
 \subfigure[Ciao]{
\includegraphics[width=.32\textwidth]{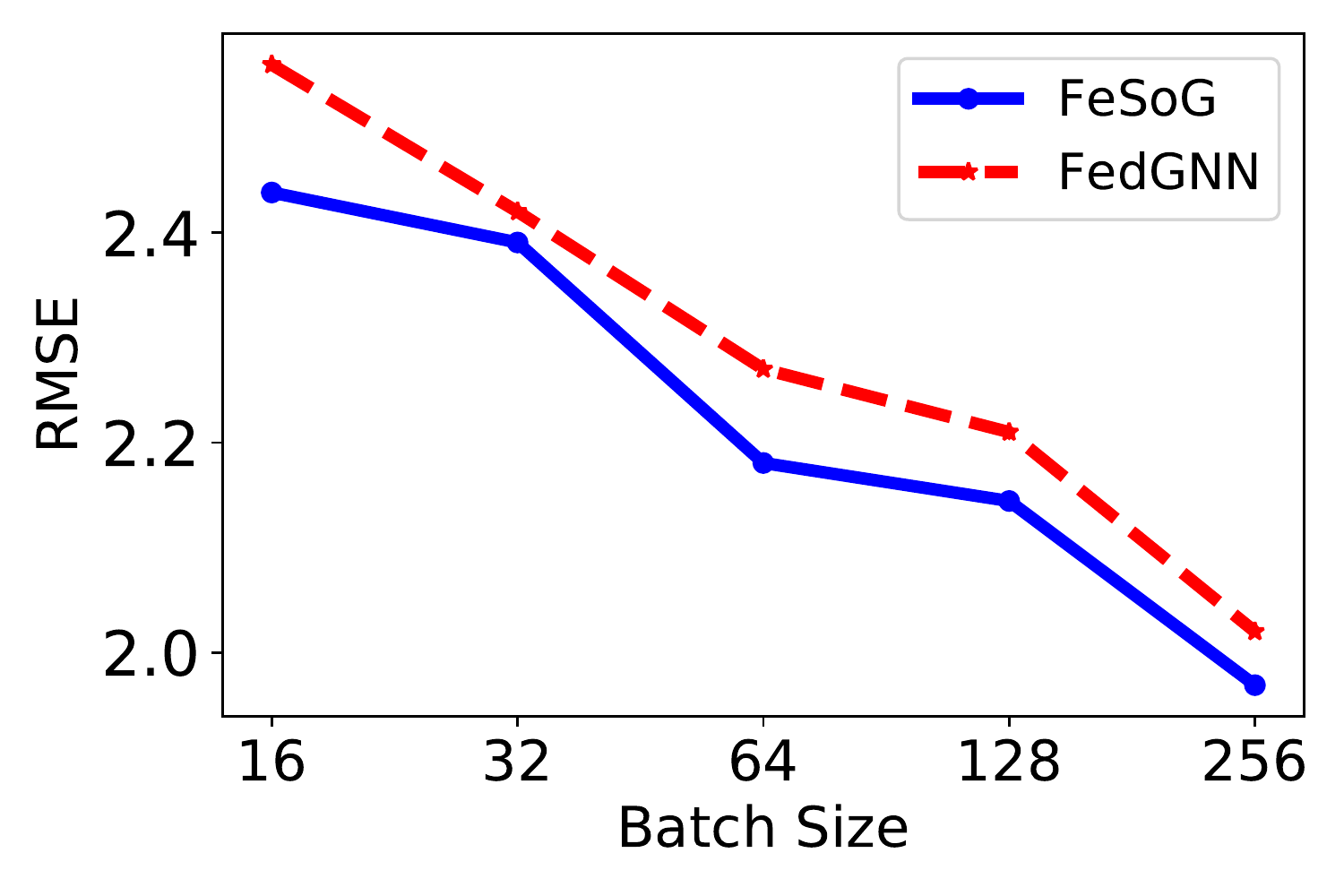}}
 \subfigure[Epinions]{
\includegraphics[width=.32\textwidth]{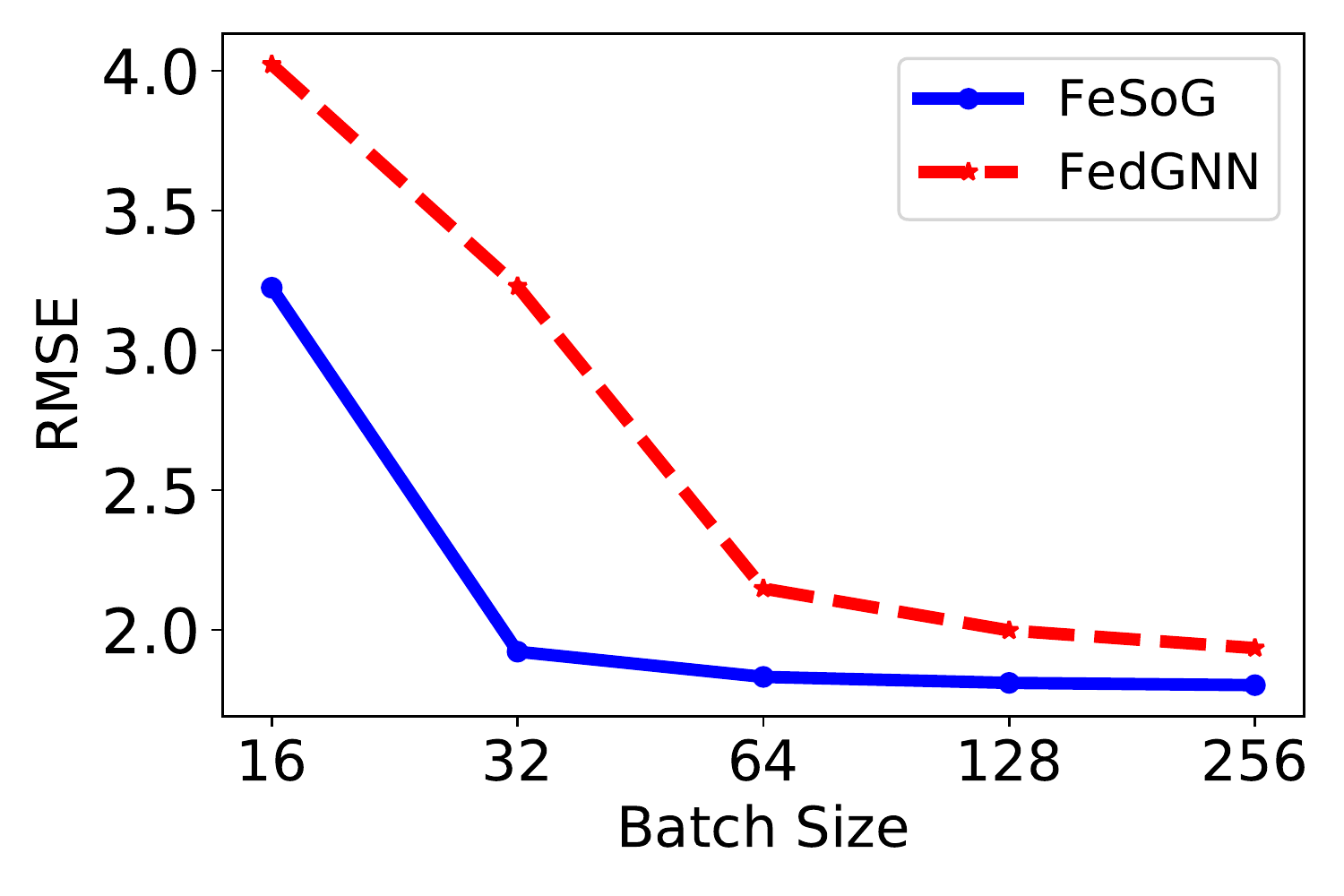}}
 \subfigure[Filmtrust]{
\includegraphics[width=.32\textwidth]{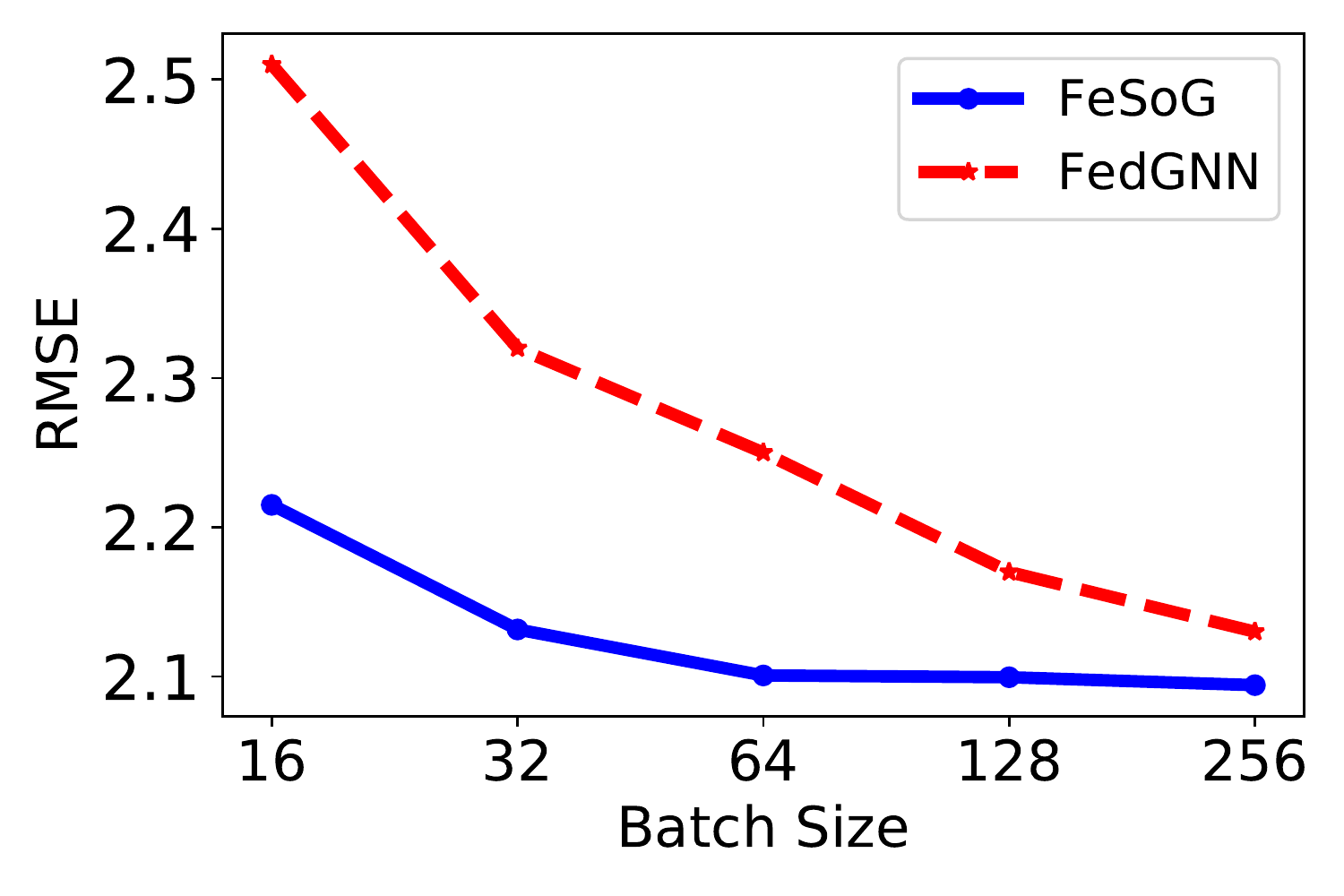}}
% \end{center}
\caption{RMSE performance with respect to user batch size on three datasets. }
\label{fig:batch_size_rmse}
\end{figure}
\begin{figure}[htb]
    % \begin{center}
 \subfigure[Ciao]{
\includegraphics[width=.32\textwidth]{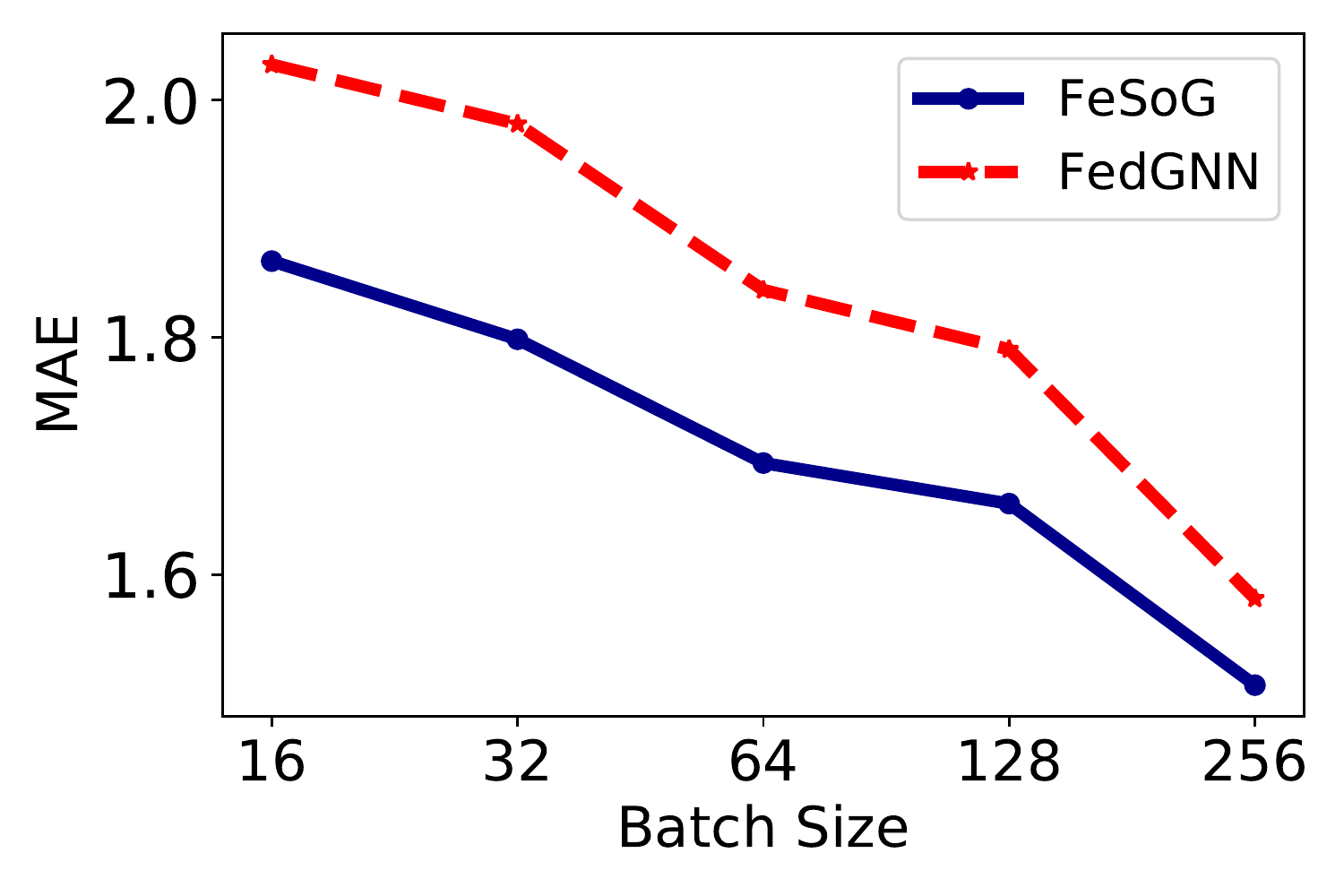}}
 \subfigure[Epinions]{
\includegraphics[width=.32\textwidth]{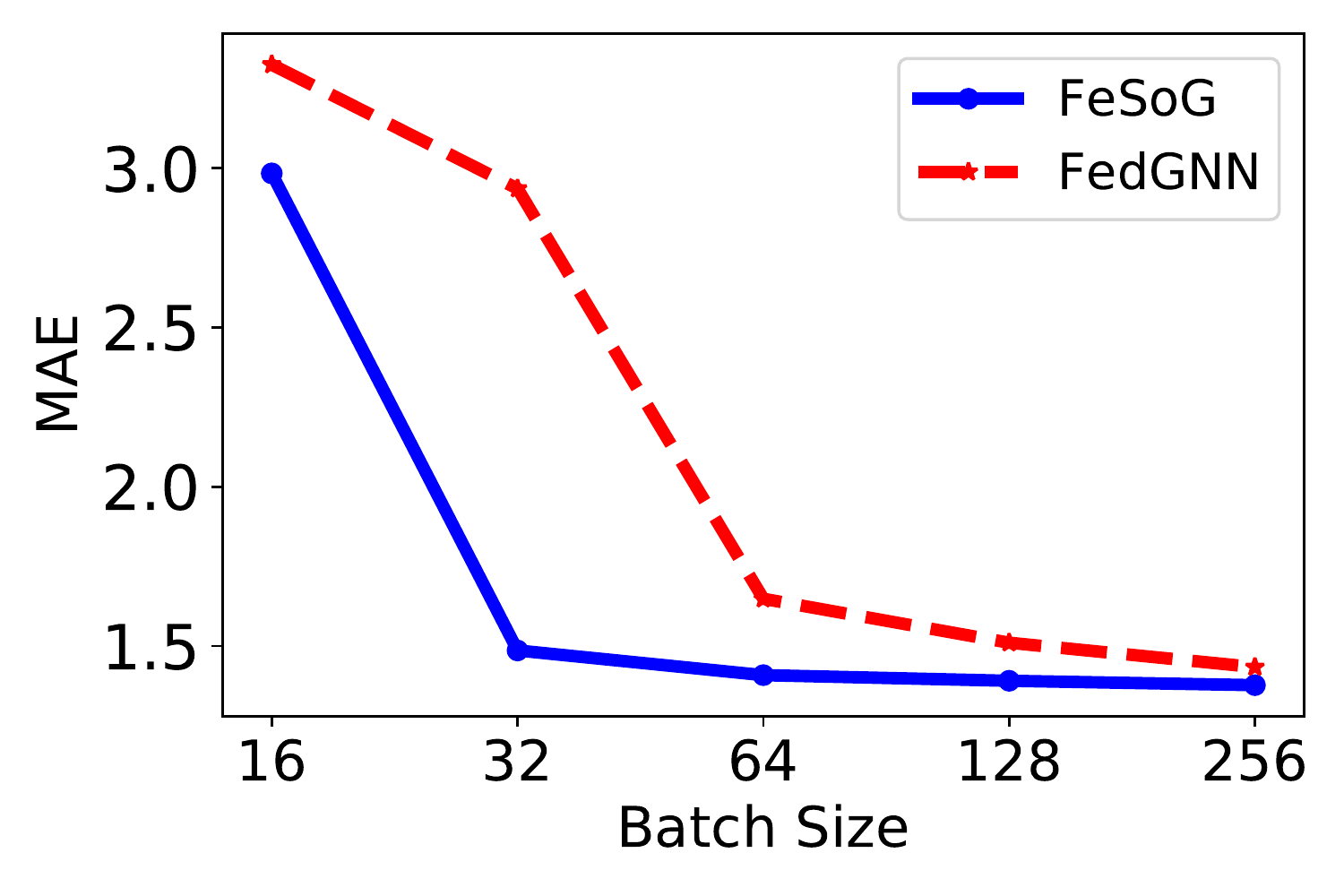}}
 \subfigure[Filmtrust]{
\includegraphics[width=.32\textwidth]{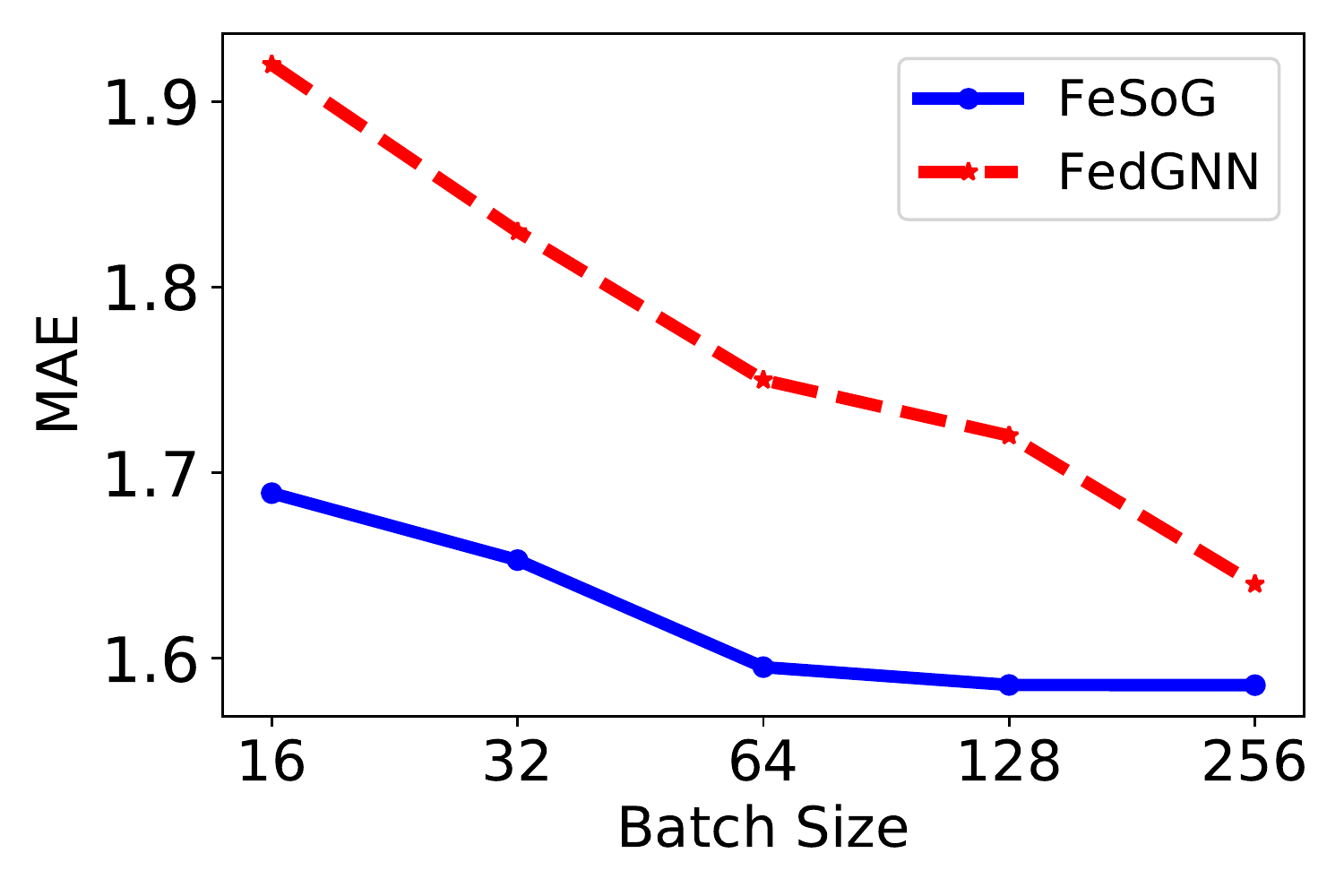}}
% \end{center}
\caption{MAE performance with respect to user batch size on three datasets. }
\label{fig:batch_size_mae}
\end{figure}
In this section, we analyze the impact of user batch size. Intuitively, choosing a small user batch size increases the communication rounds for the server to train a model. However, it is unclear how it affects the prediction performance. We report the performance of FedGNN and FeSoG with respect to RMSE and MAE across three datasets, which is illustrated in Fig.~\ref{fig:batch_size_rmse} and Fig.~\ref{fig:batch_size_mae}, respectively. We have the following observations:
\begin{itemize}
    \item \modelname performs better than FedGNN. On all three datasets, the RMSE of \modelname is consistently lower than FedGNN, which results from its powerful embedding ability of relational local GNN. 
    \item The performance of \modelname becomes better with the increase of user batch size across datasets. With a larger user batch size, the server can obtain a more accurate global information estimation, which leads to a better performance. However, in practice, aggregating more users at each training step would lead to more computational cost and more time to converge.
\end{itemize}

\subsubsection{Number of Pseudo Items}\label{sec:pseduo_item}
\begin{figure}[htb]
 \subfigure[Ciao]{
\includegraphics[width=.32\textwidth]{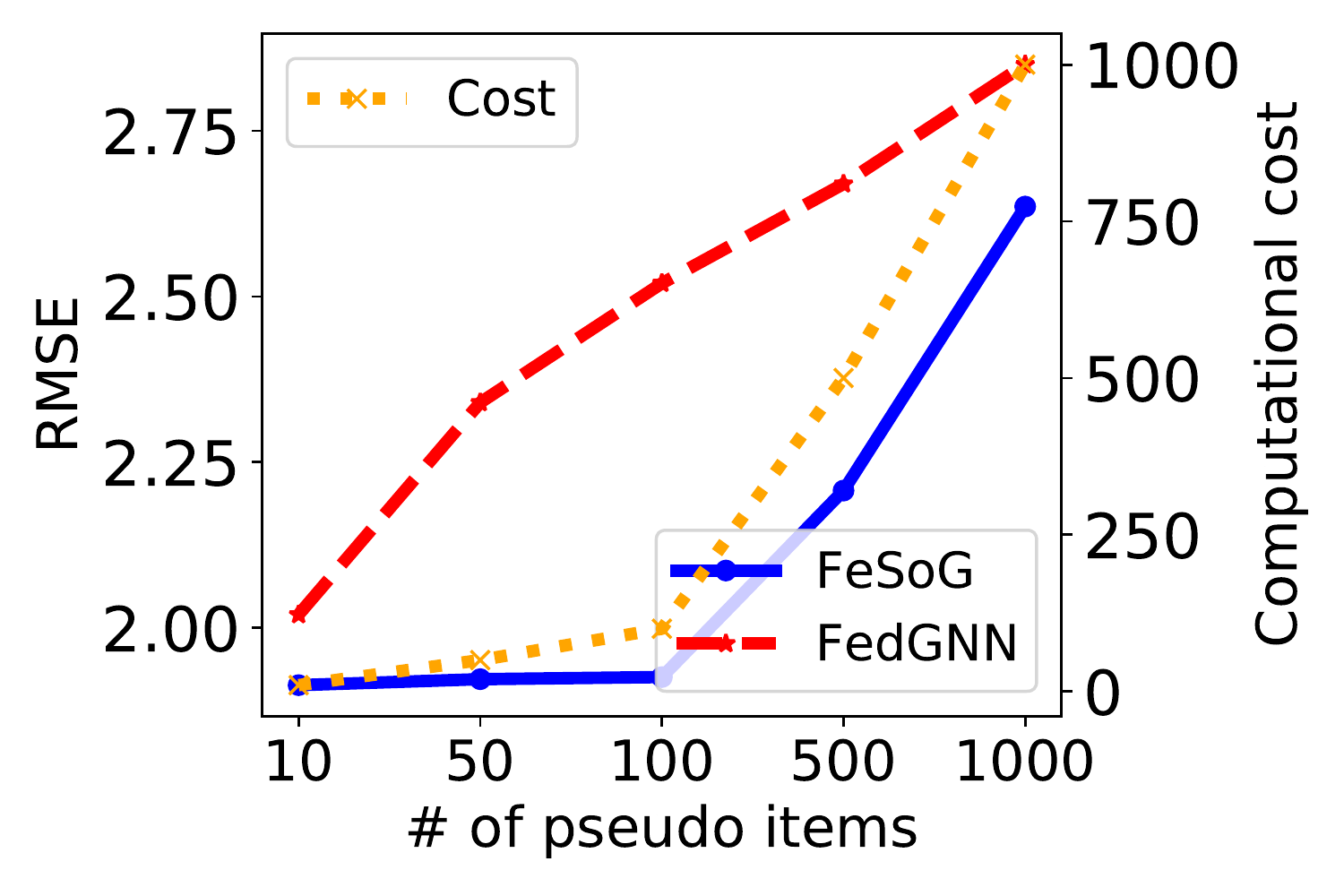}}
 \subfigure[Epinions]{
\includegraphics[width=.32\textwidth]{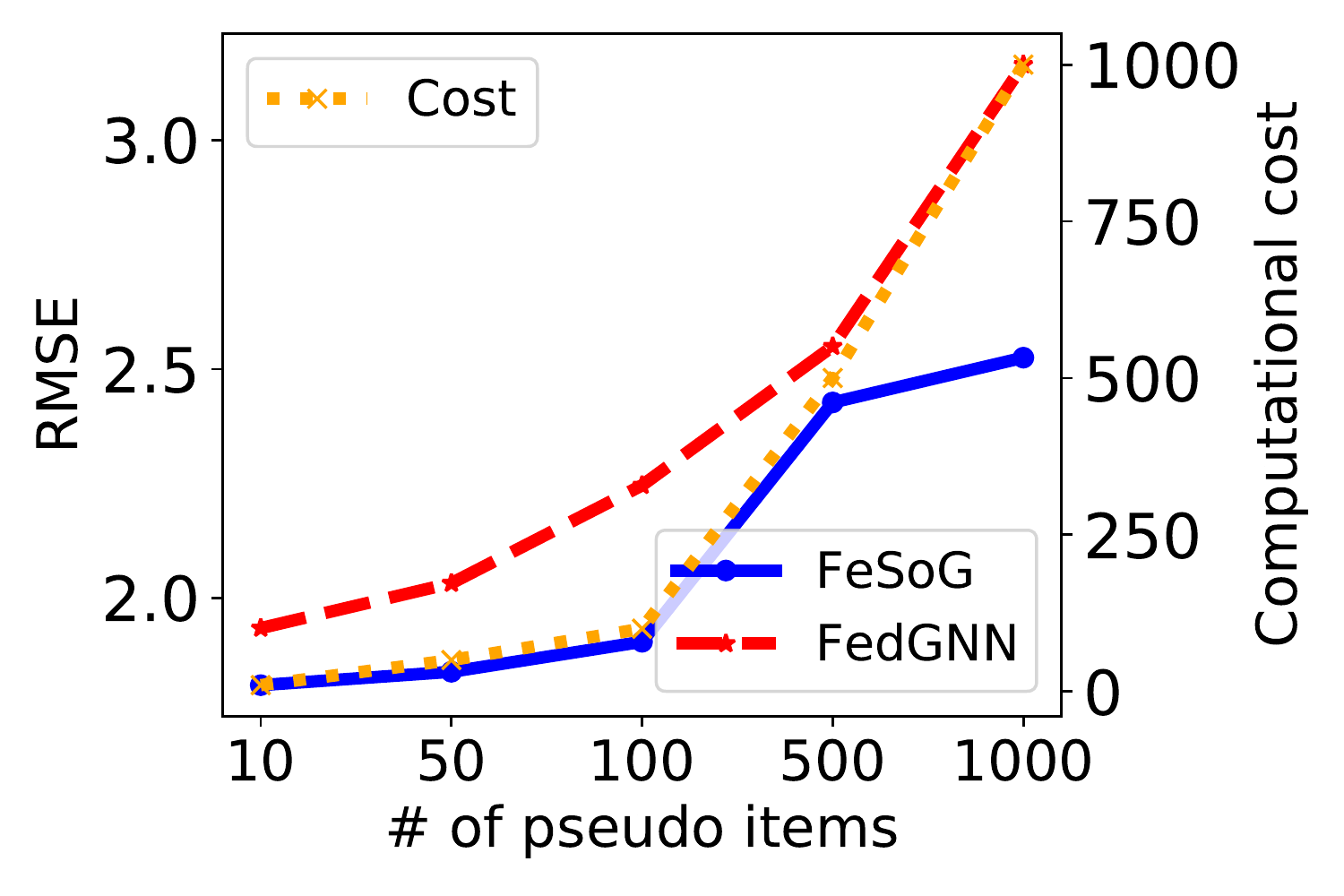}}
 \subfigure[Filmtrust]{
\includegraphics[width=.32\textwidth]{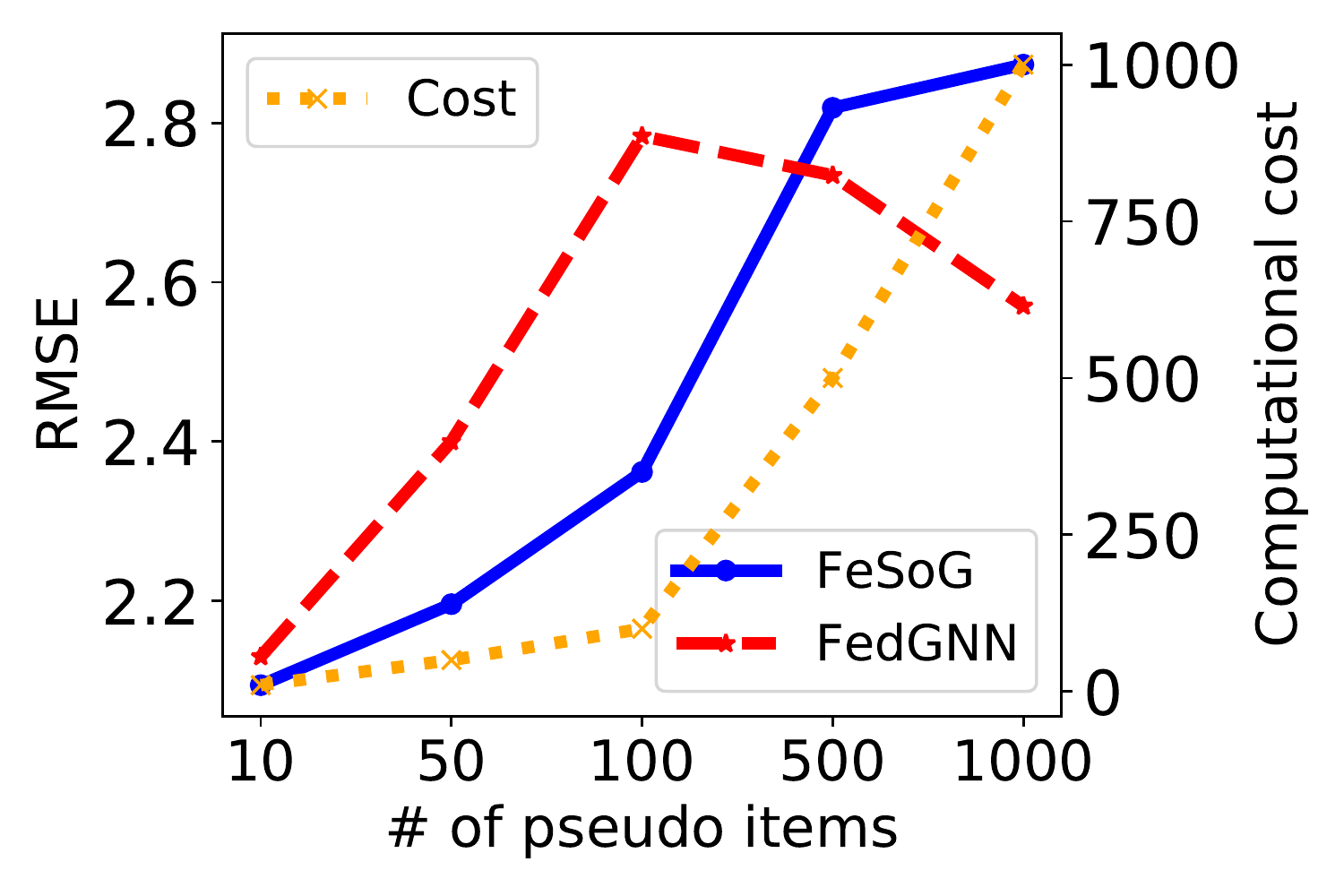}}
\caption{RMSE performance with respect to different pseudo item numbers on three datasets.}
\label{fig:pseudo_item_rmse}
\end{figure}
\begin{figure}[htb]
 \subfigure[Ciao]{
\includegraphics[width=.32\textwidth]{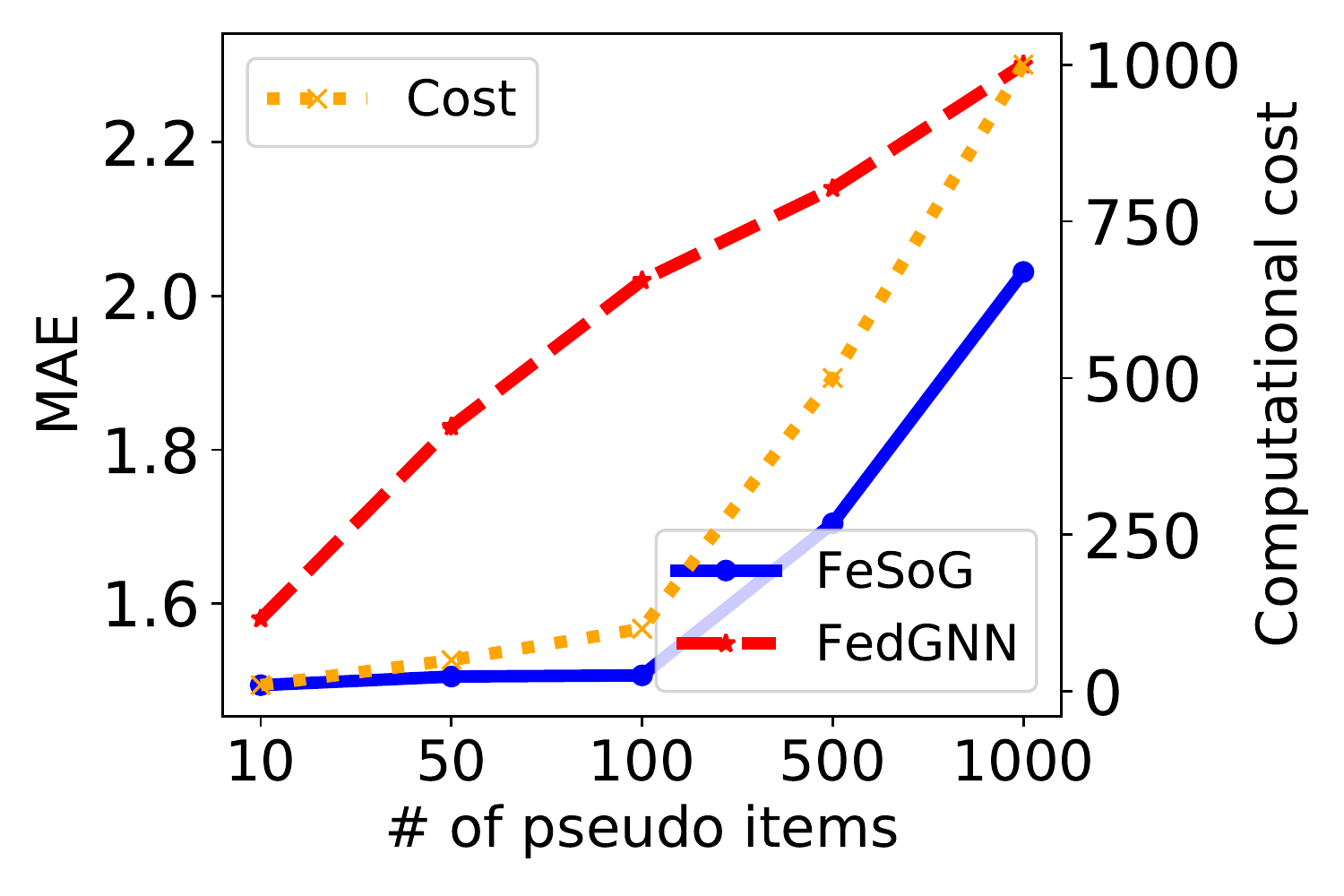}}
 \subfigure[Epinions]{
\includegraphics[width=.32\textwidth]{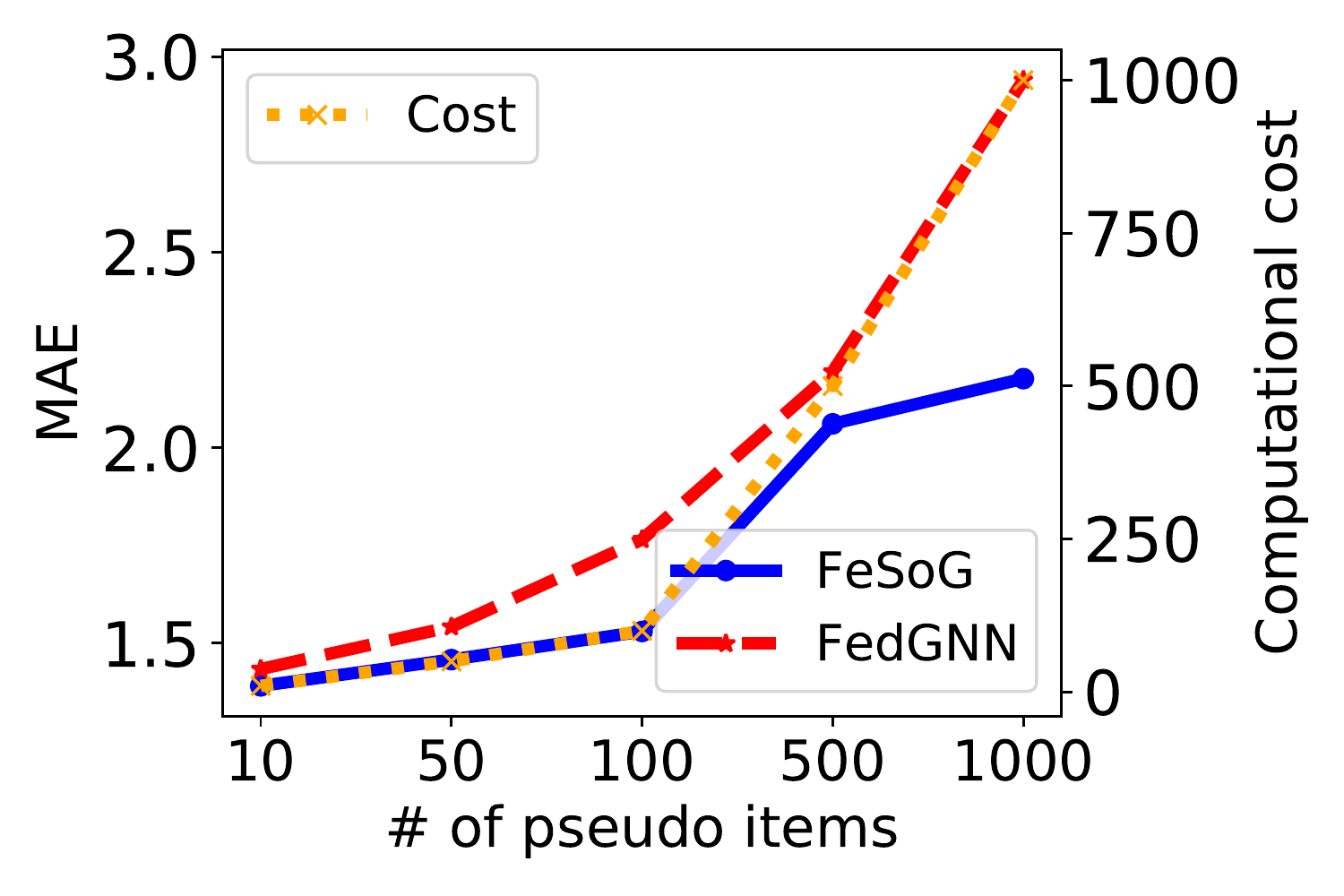}}
 \subfigure[Filmtrust]{
\includegraphics[width=.32\textwidth]{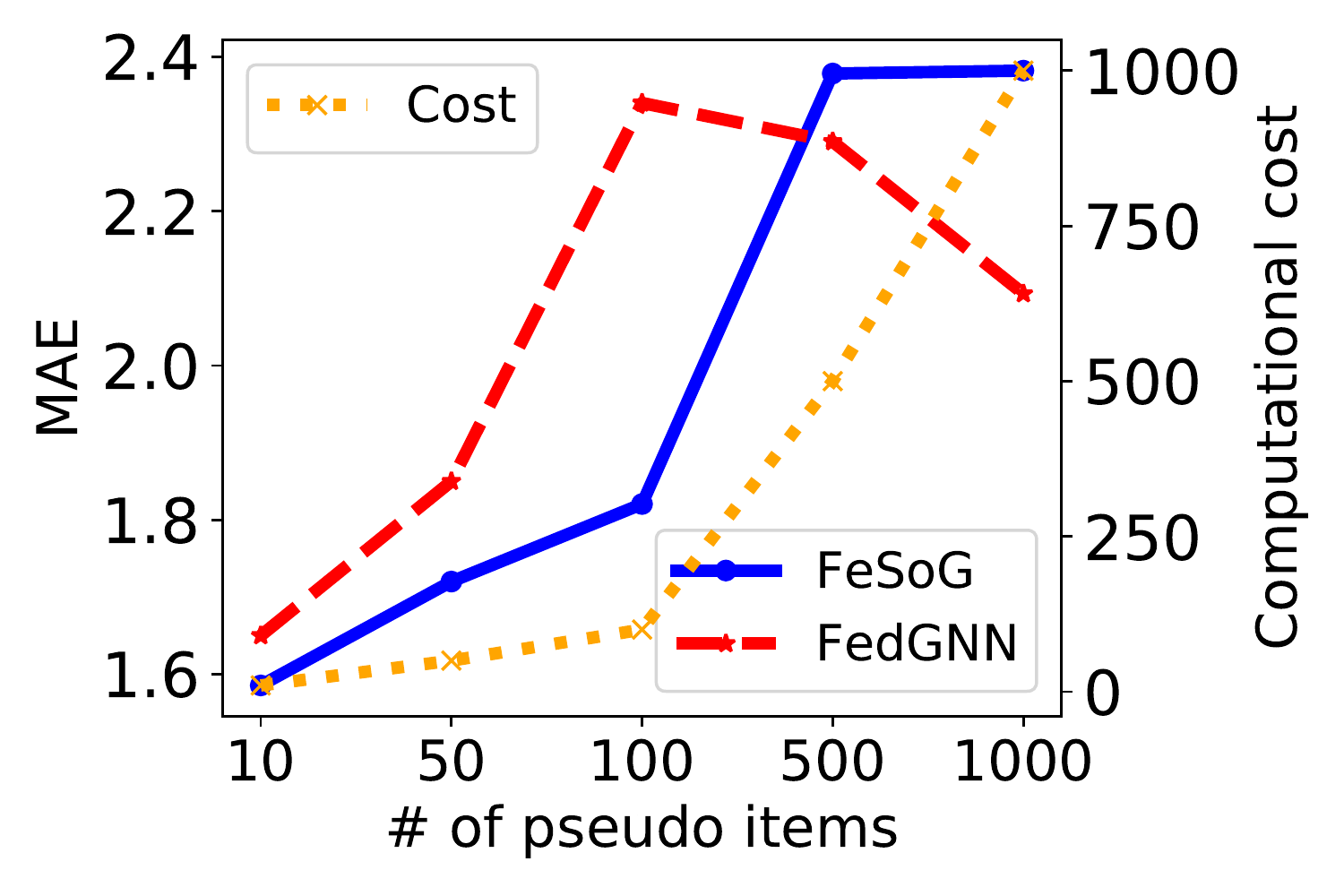}}
\caption{MAE performance with respect to different pseudo item numbers on three datasets.}
\label{fig:pseudo_item_mae}
\end{figure}
Pseudo items are sampled to protect the gradients from privacy leakage. Sampling more pseudo items requires more computational cost as more ratings should be predicted. However, it is unclear how many samples should be selected to achieve satisfying results. Hence, we conduct experiments on three datasets to study the impacts of the number of sampled pseudo items.  We also compare FedGNN, which samples a set of negative items and assigns them with random gradients. The influence of the number of pseudo items on three datasets with respect to RMSE and MAE is reported in Fig.~\ref{fig:pseudo_item_rmse} and Fig.~\ref{fig:pseudo_item_mae}. Besides the performance value, we also present the computational cost with respect to the number of sampled pseudo items. We have the following observations:
\begin{itemize}
    \item \modelname yields better performance compared with FedGNN. On Ciao and Epinions datasets, the performance of \modelname and FedGNN gets worse with the increase of pseudo items. However, the value of \modelname is much lower than FedGNN. It suggests that our pseudo item sampling and the pseudo labeling techniques are robust. Therefore, we can conclude that the pseudo item sampling in \modelname can protect privacy data and enhance the training process. 
    \item If increasing the number of pseudo items, the error value all increases for both FedGNN and \modelname on Ciao and Epinions datasets, which results from more noise caused by pseudo items. However, on the Filmtrust dataset, the error value of FedGNN first increases and then drops. This is because FedGNN generates gradients of pseudo items from the same Gaussian distribution, and at the same time Filmtrust dataset only has $1957$ items. So when sampling more than $500$ pseudo items, the gradient of pseudo items from different users would counteract with each other to reduce noise impact. \modelname generates gradients of pseudo items by user-specific labelling, which does not have this characteristic.
    
    \item We should find a trade-off between pseudo-item sampling and model performance. As illustrated in Fig.~\ref{fig:pseudo_item_rmse} and Fig.~\ref{fig:pseudo_item_mae}, if increasing the number of pseudo items, the extra computational cost increases linearly and the server would be harder to infer the true interacted items. At the same time, prediction accuracy would become worse. So we should find a suitable pseudo item number to balance privacy protection and model performance. For example, $100$ pseudo items for Ciao dataset would be an appropriate setting because the model performance does not deteriorate much when the number of pseudo items is lower than $100$. 
\end{itemize}

\subsubsection{Embedding Size}
\begin{figure}[!hbt]
 \subfigure[Ciao]{
\includegraphics[width=.32\textwidth]{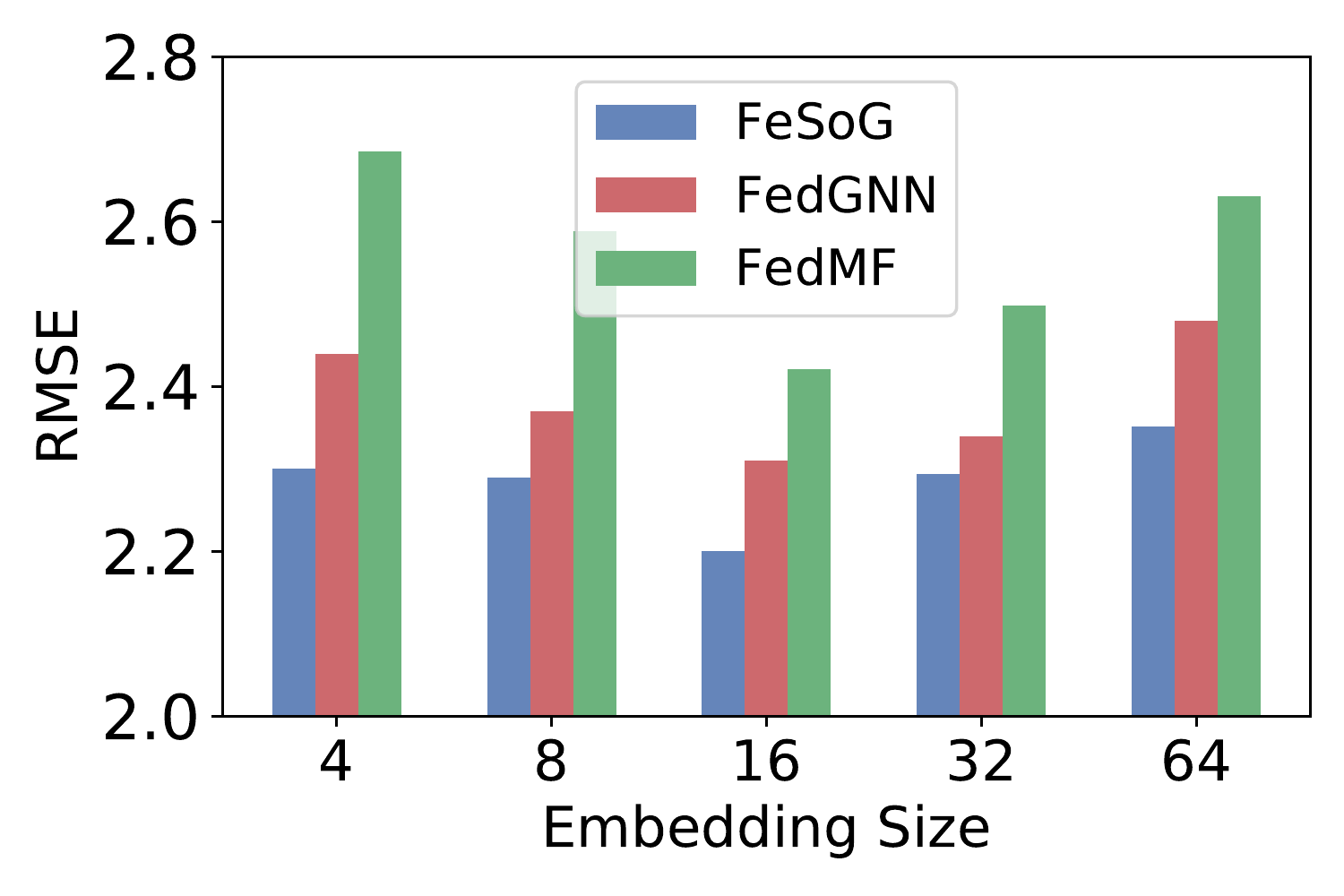}}
 \subfigure[Epinions]{
\includegraphics[width=.32\textwidth]{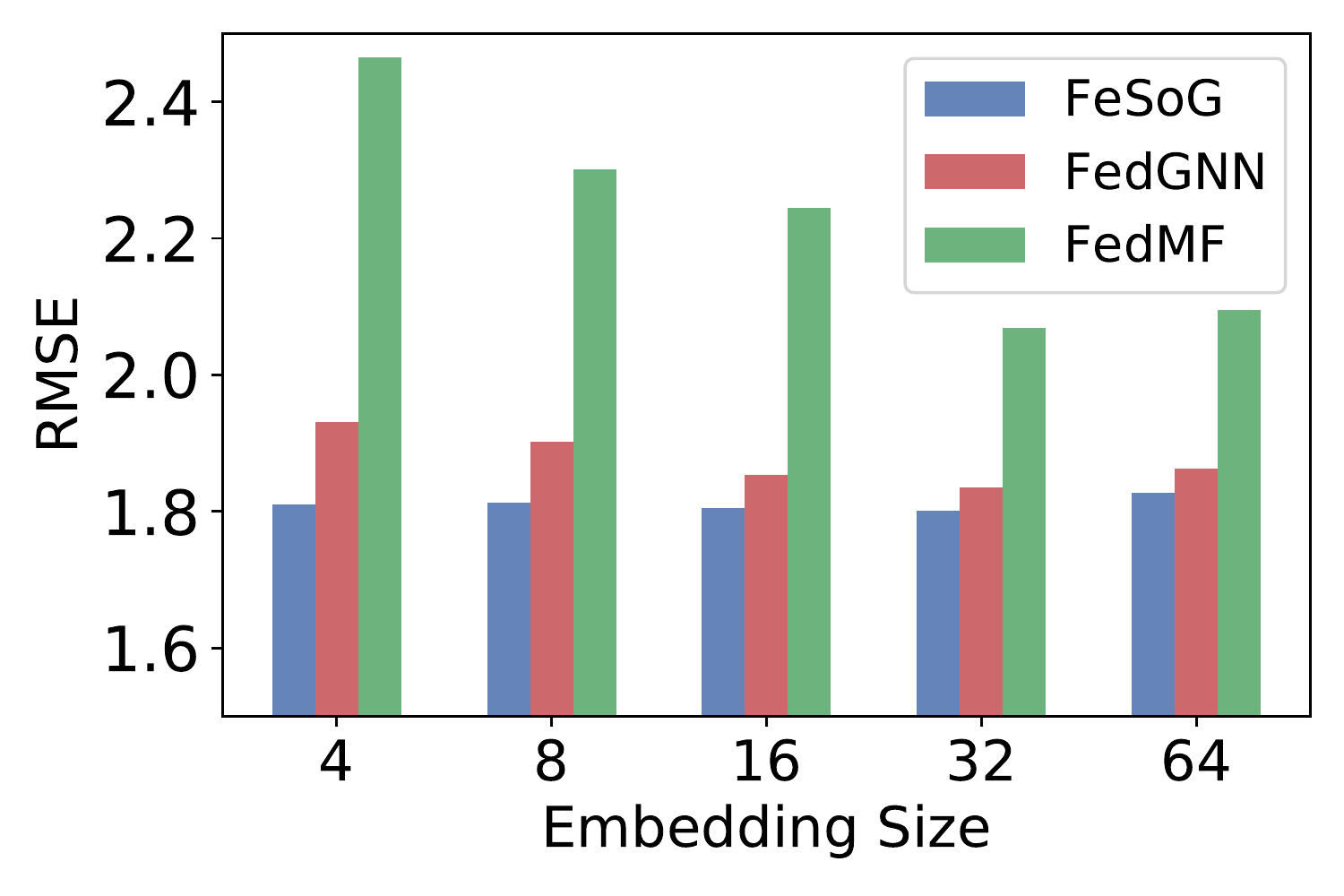}}
 \subfigure[Filmtrust]{
\includegraphics[width=.32\textwidth]{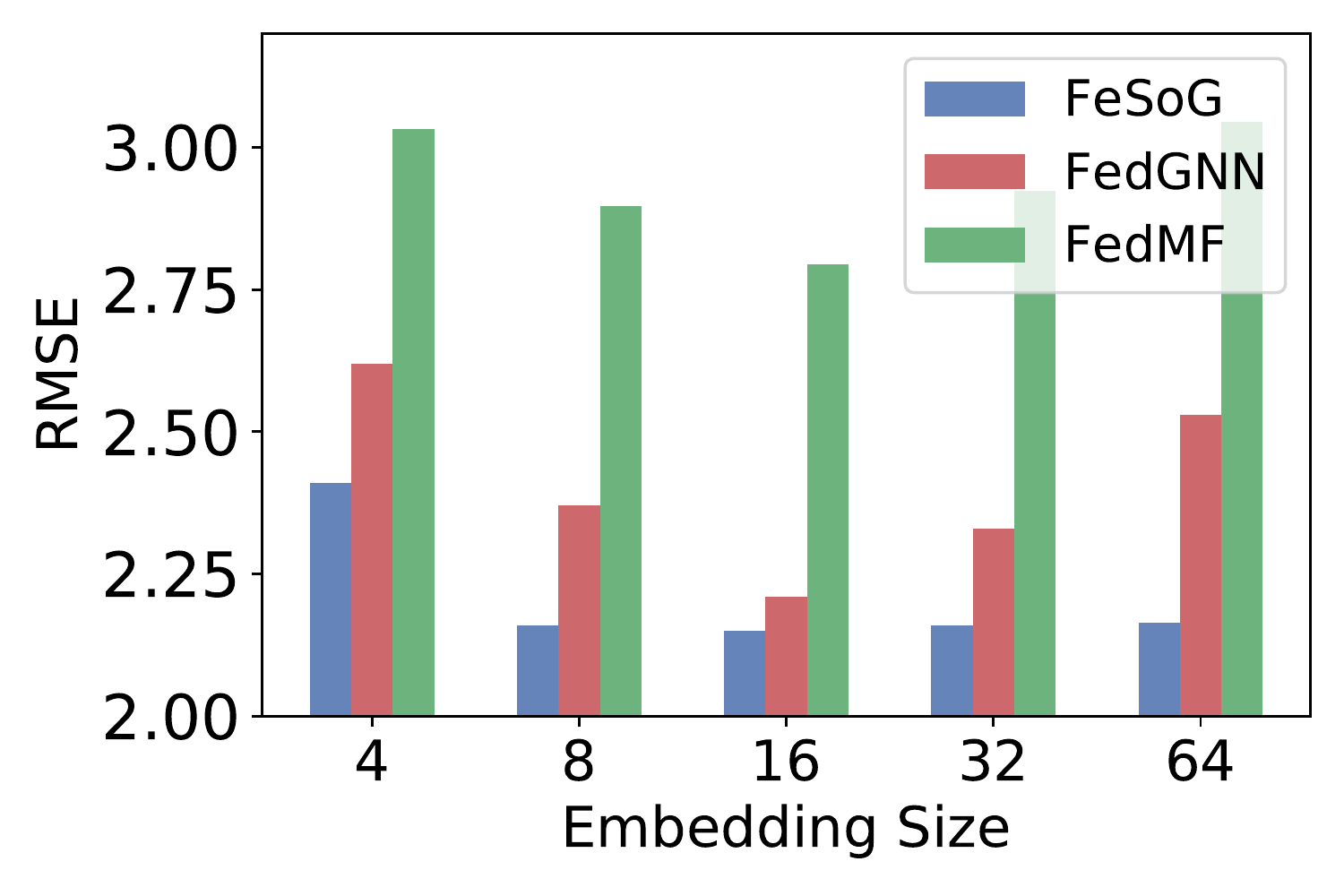}}
\caption{RMSE performance with respect to different embedding sizes $d$ on three datasets.}
\label{fig:emb_size_rmse}
\end{figure}
\begin{figure}[!hbt]
 \subfigure[Ciao]{
\includegraphics[width=.32\textwidth]{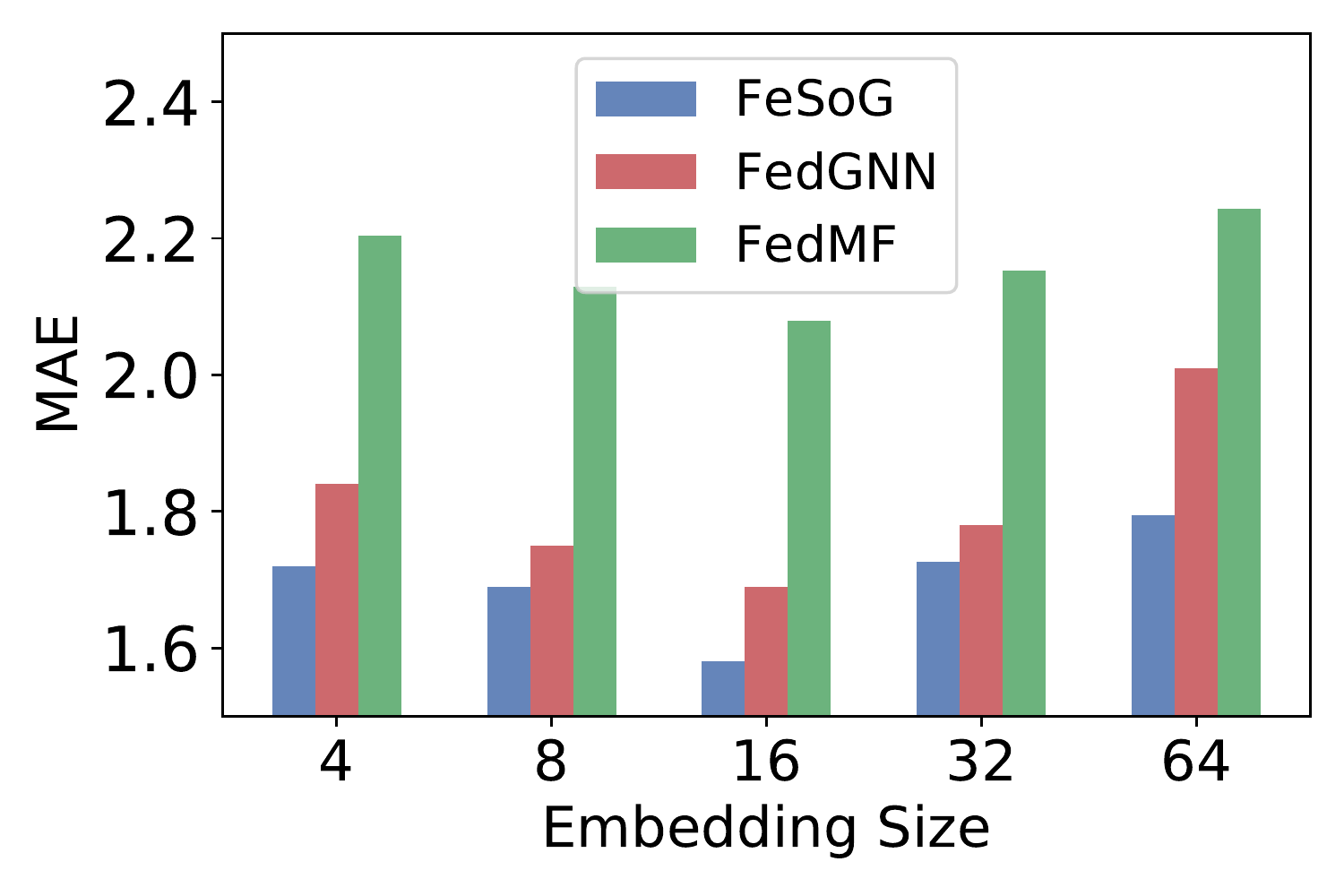}}
 \subfigure[Epinions]{
\includegraphics[width=.32\textwidth]{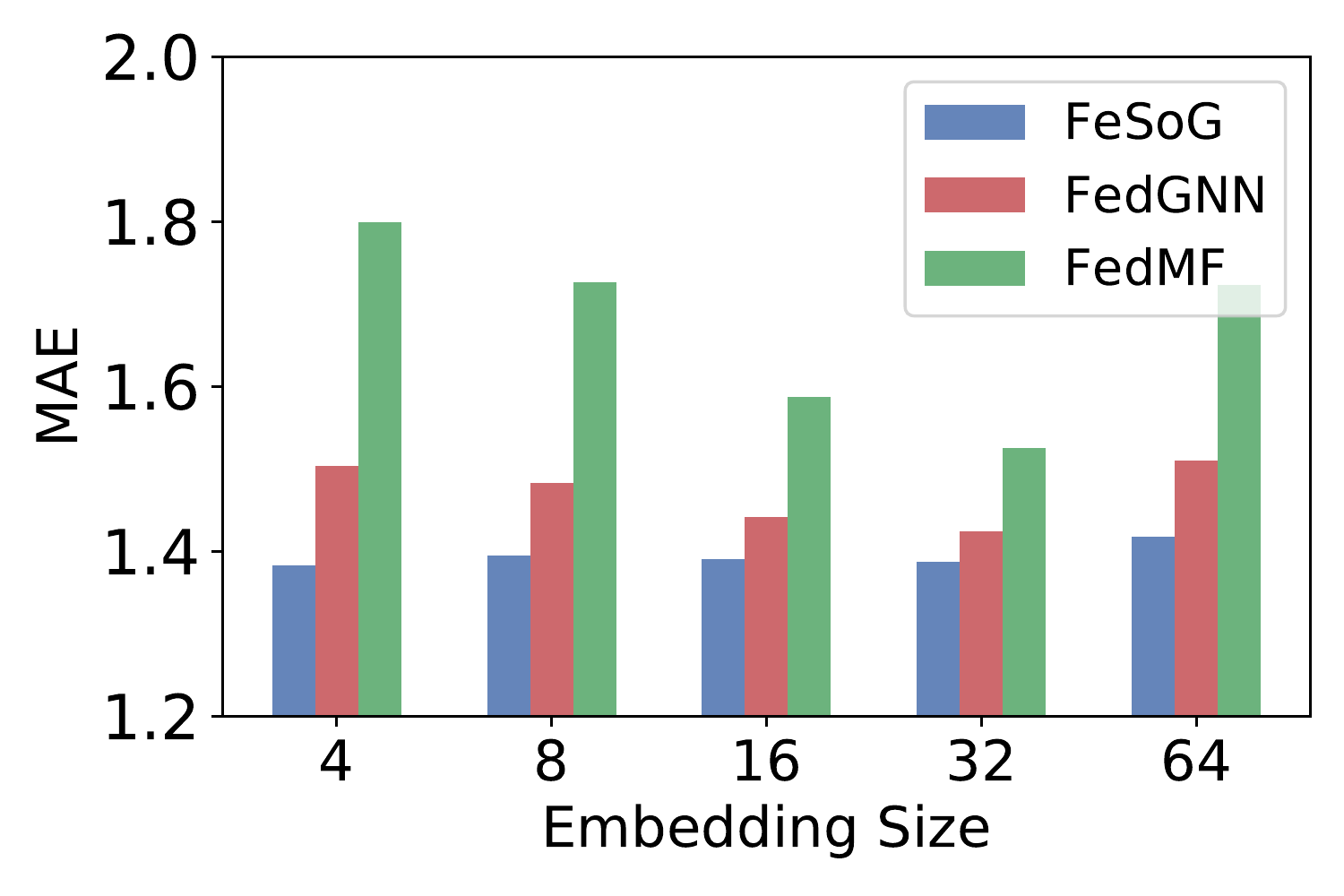}}
 \subfigure[Filmtrust]{
\includegraphics[width=.32\textwidth]{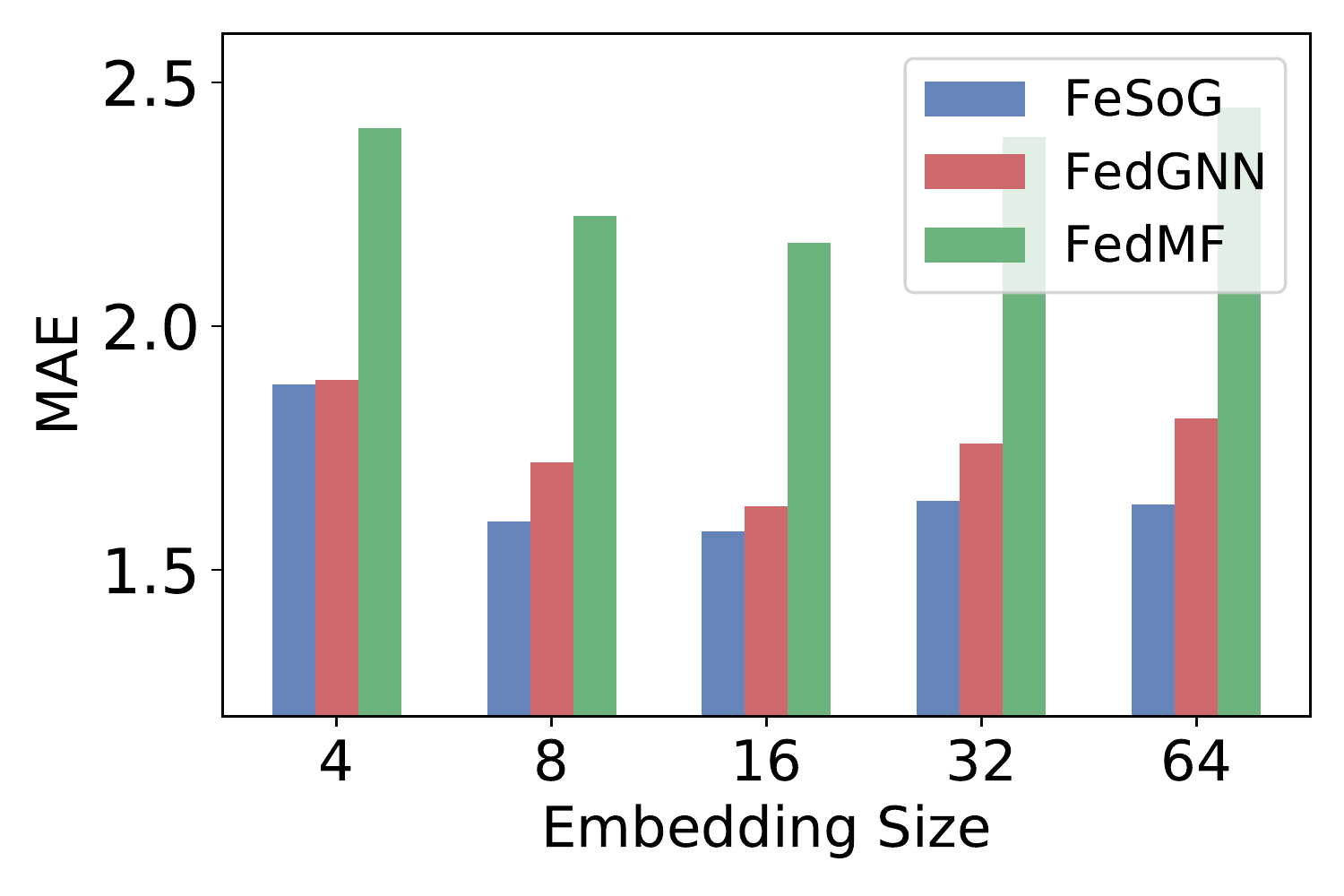}}
\caption{MAE performance with respect to different embedding sizes $d$ on three datasets.}
\label{fig:emb_size_mae}
\end{figure}
This section studies the performance with respect to the embedding size. The values of RMSE and MAE on three datasets are reported in Fig.~\ref{fig:emb_size_rmse} and Fig.~\ref{fig:emb_size_mae}, respectively. For comparison, we also select two representative baselines, which are FedGNN and FedMF. We have the following observations:
\begin{itemize}
    \item The proposed \modelname consistently outperforms other federated recommender system methods across all embedding sizes. This observation suggests that \modelname can learn informative structures from local graphs.  Compared with FedMF, FedGNN demonstrates a much better performance, which justifies the necessity to employ GNN for graph embedding. 
    \item \modelname has lower fluctuations as embedding sizes change compared with the other two baselines. It indicates that \modelname is more robust than FedGNN on different embedding sizes, demonstrating the effectiveness of using pseudo-item protection to enhance the training. 
    \item Furthermore, suitable embedding sizes are crucial for different datasets to achieve satisfying performance. All federated learning methods follow the same trend and obtain the best performance in either $d=16$ or $d=32$. With a smaller embedding size~(e.g., $d=4$), the model has insufficient representation ability. However, with large embedding sizes~(e.g., $d=64$), it may cause the overfitting issue due to limited data.
\end{itemize}

\subsubsection{Learning Rate}
\begin{figure}[htb]
 \subfigure[Ciao]{
\includegraphics[width=.32\textwidth]{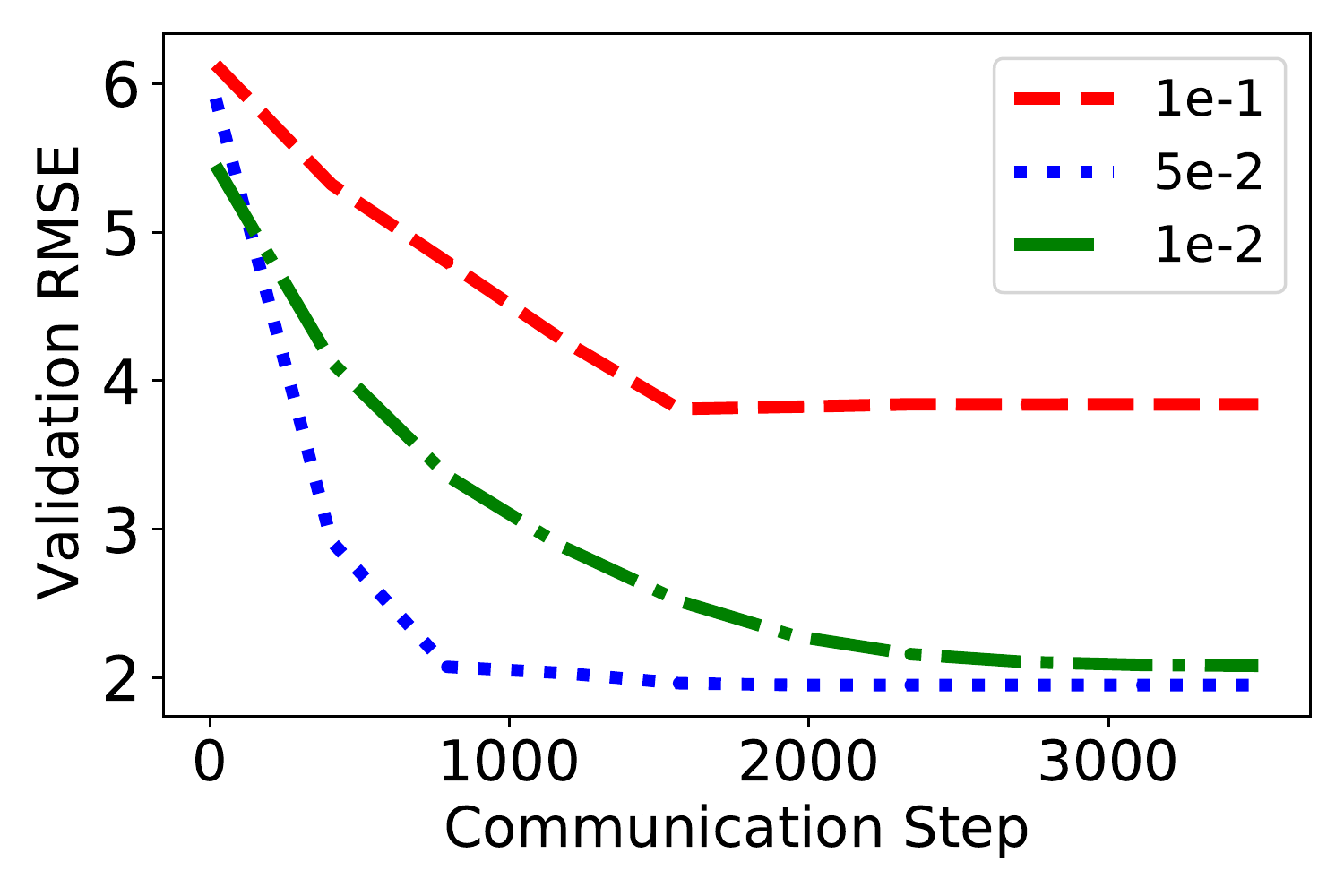}}
 \subfigure[Epinions]{
\includegraphics[width=.32\textwidth]{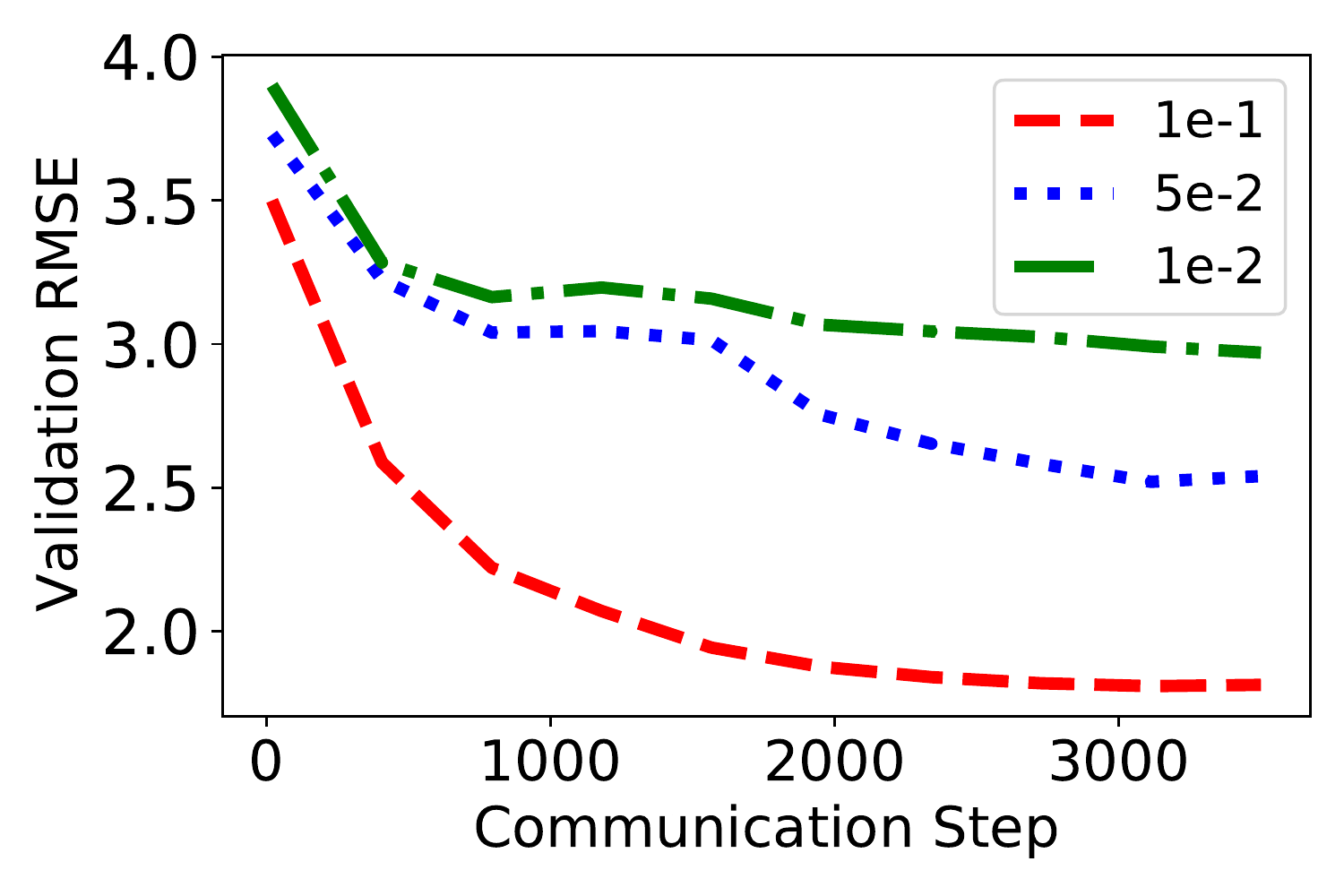}}
 \subfigure[Filmtrust]{
\includegraphics[width=.32\textwidth]{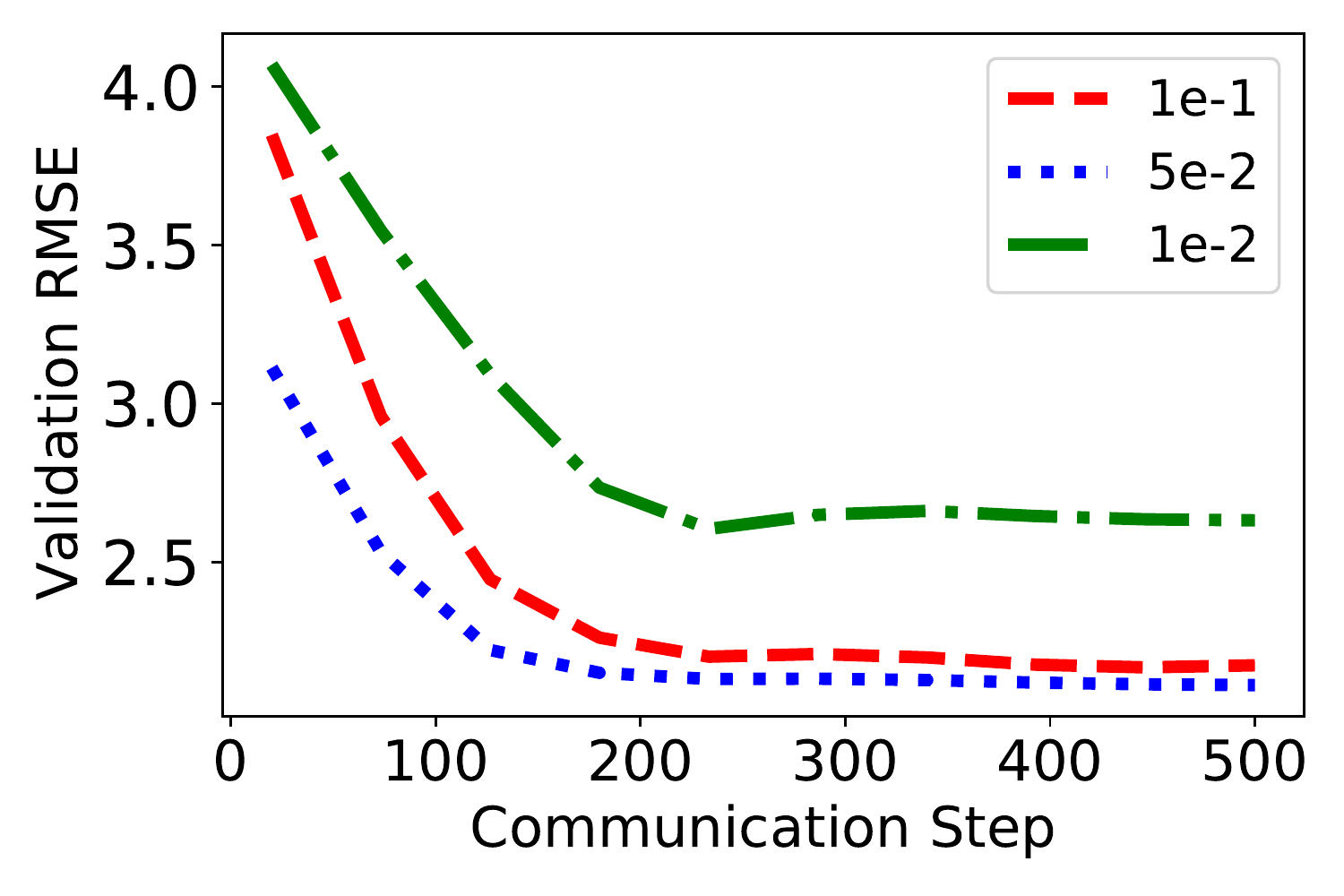}}
\caption{Loss curves for different learning rate $\eta$ on three datasets.}
\label{fig:lr_loss}
\end{figure}

Learning rate affects the number of communication rounds for the convergence of a federated learning framework. Intuitively, fewer communication steps can decrease the number of communication times between the server and clients, which implies less risks of information leakage. We investigate the impact of learning rate with respect to the training loss of \modelname and report the results in Fig.~\ref{fig:lr_loss}. The learning rather is selected from $\{0.1,0.05,0.01\}$. We have the following observations:
\begin{itemize}
    \item The training of \modelname is smooth. The loss on three datasets all converges smoothly when increasing the number of communications steps. It suggests that the federated learning framework can effectively transfer informative gradients for the \modelname to converge.  
    \item Different datasets prefer different learning rates. For example, on the Epinions dataset, the best learning rate $\eta=0.1$. While both Ciao and Filmtrust datasets converge the fastest when the learning rate $\eta=0.05$. Learning rate has a critical influence on different datasets.
    % \item All three datasets show that \modelname converges after 1,000 communication steps with a small learning rate. It demonstrates the robust training of having pseudo-item protection in \modelname. It also demonstrates that \modelname can converge to a locally optimal point with lower communication cost and information leakage risks. 
\end{itemize}

\subsubsection{Local Differential Privacy parameters $\delta$ and $\lambda$}

\begin{figure}[htb]
 \subfigure[Ciao]{
\includegraphics[width=.32\textwidth]{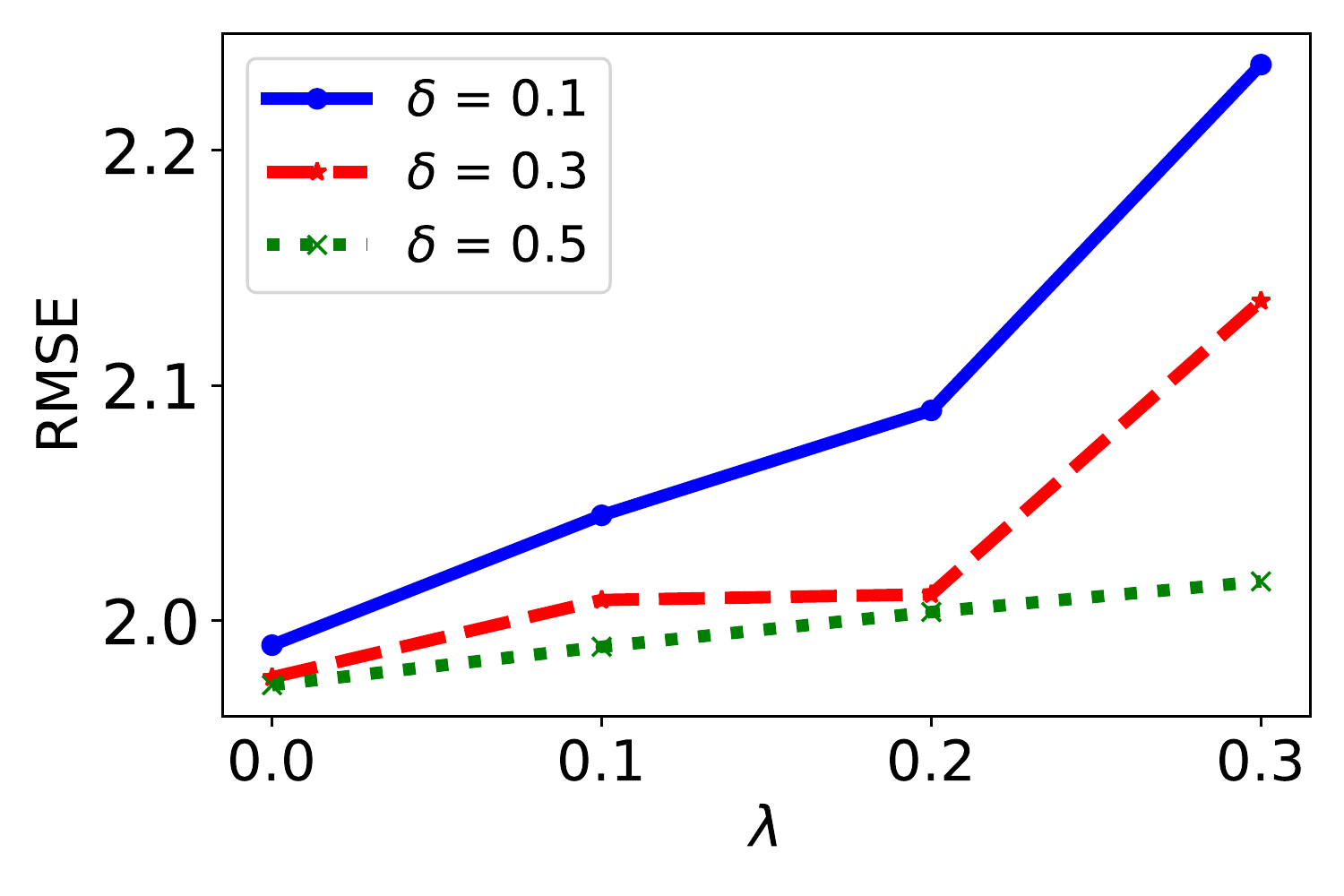}}
 \subfigure[Epinions]{
\includegraphics[width=.32\textwidth]{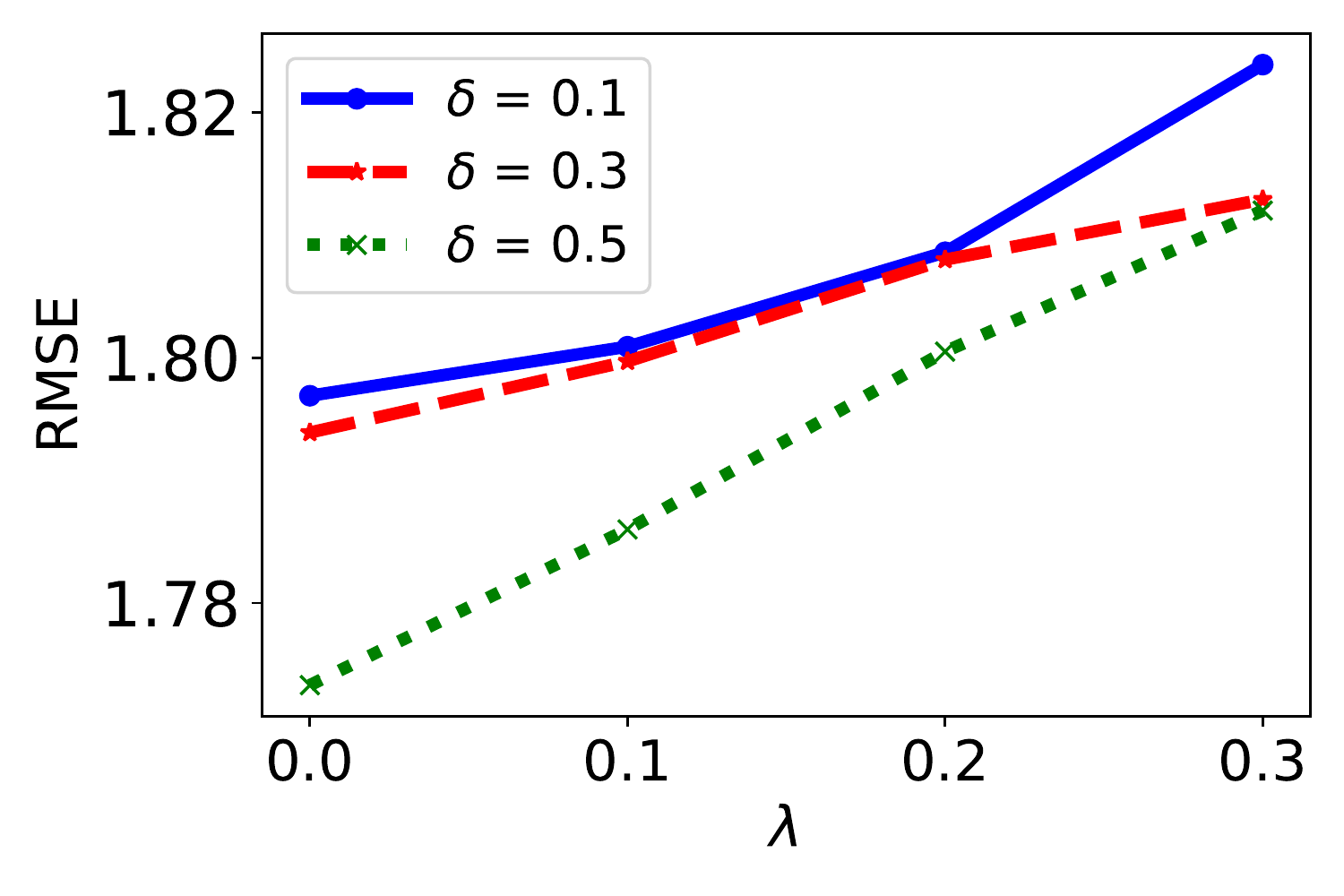}}
 \subfigure[Filmtrust]{
\includegraphics[width=.32\textwidth]{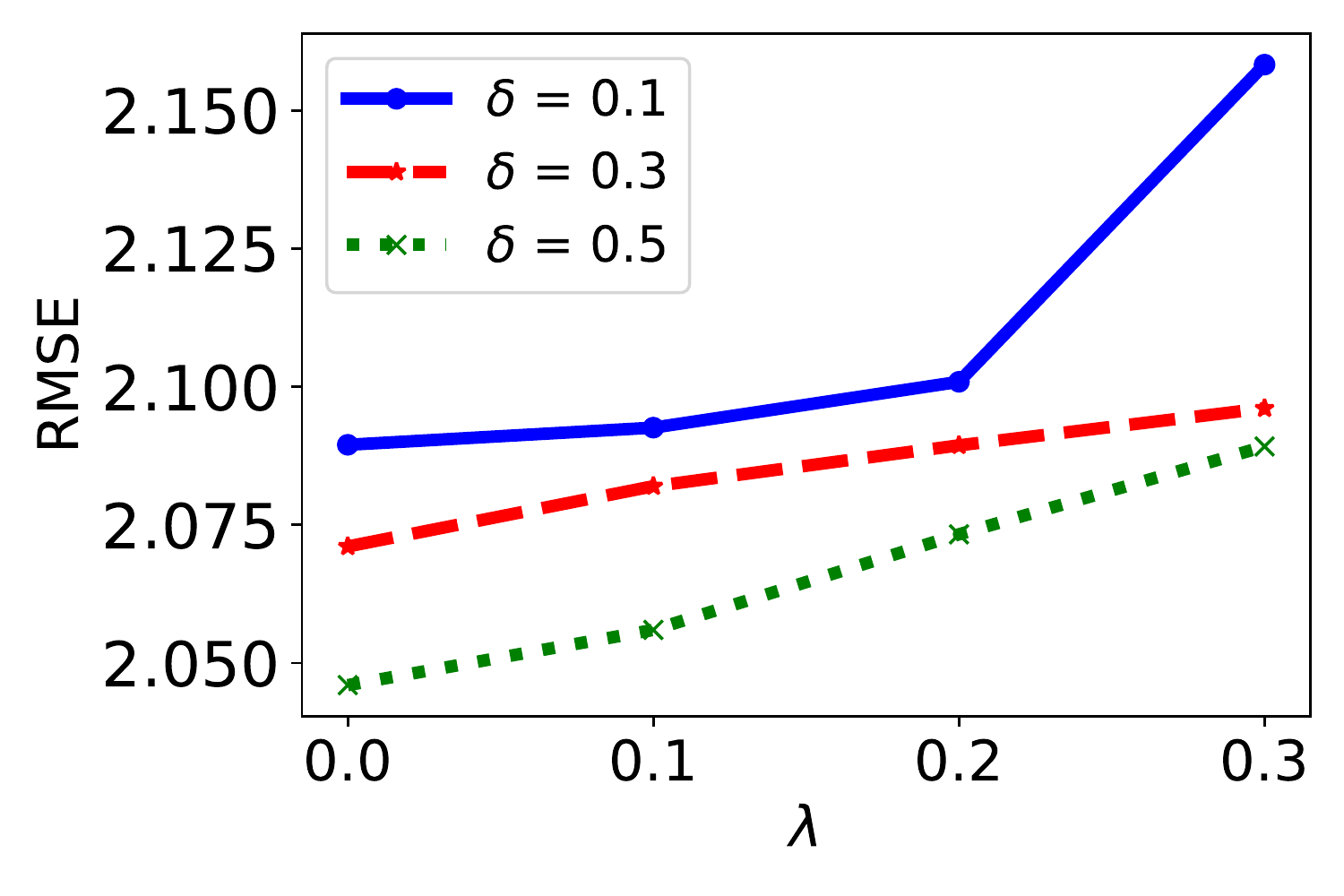}}
\caption{RMSE performance with respect to different $\delta$ and $\lambda$ on three datasets.}
\label{fig:ldp_para_rmse}
\end{figure}

\begin{figure}[htb]
 \subfigure[Ciao]{
\includegraphics[width=.32\textwidth]{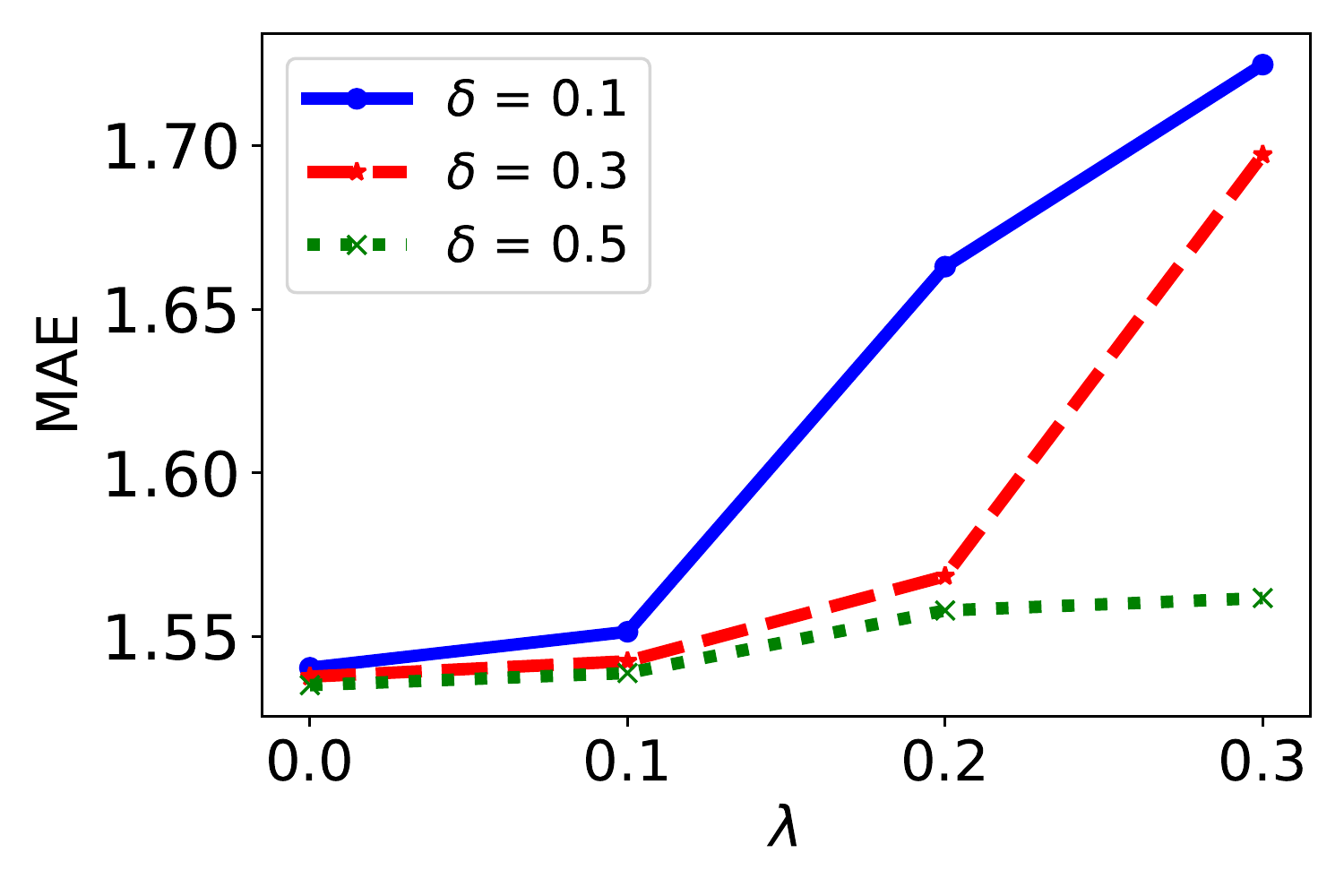}}
 \subfigure[Epinions]{
\includegraphics[width=.32\textwidth]{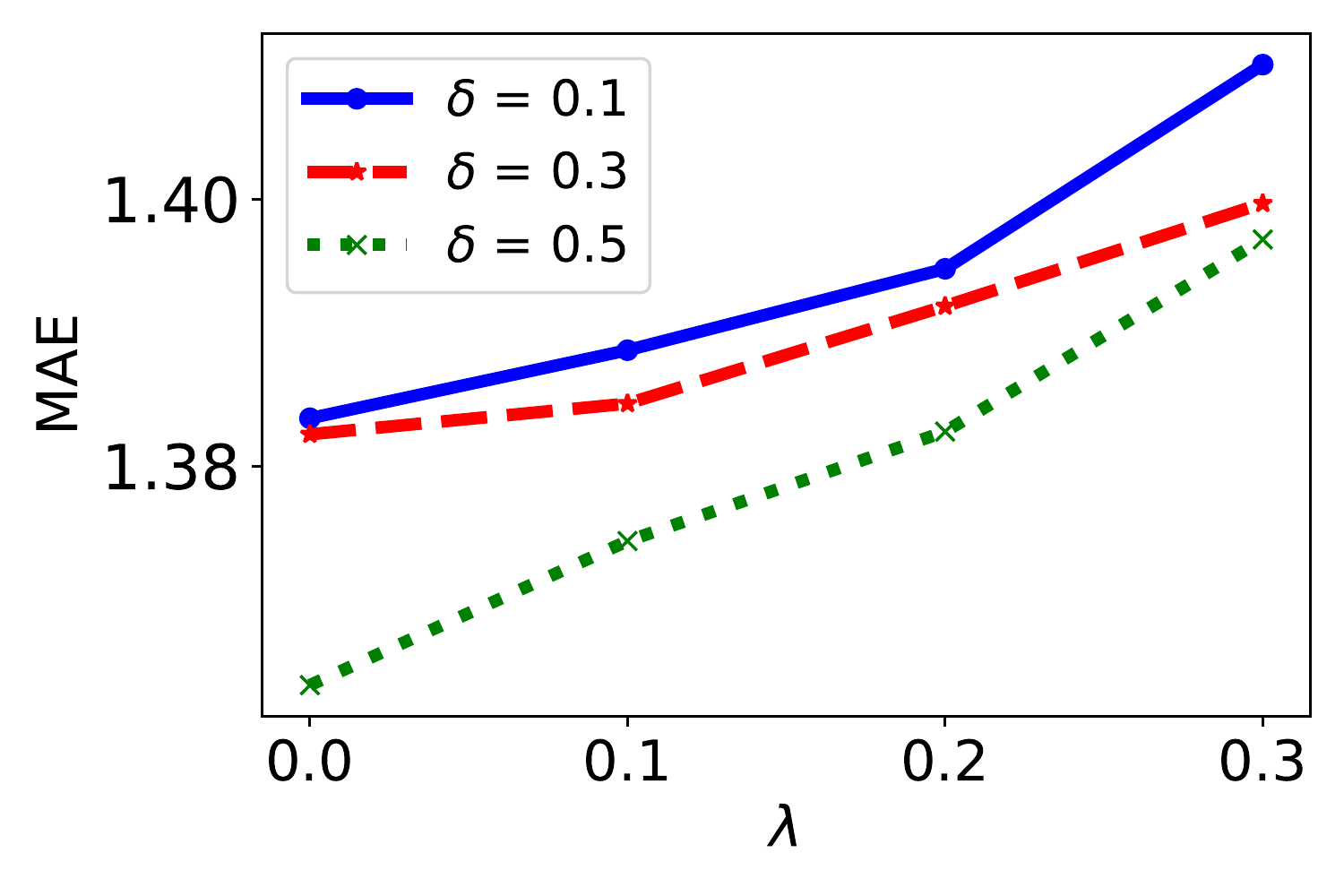}}
 \subfigure[Filmtrust]{
\includegraphics[width=.32\textwidth]{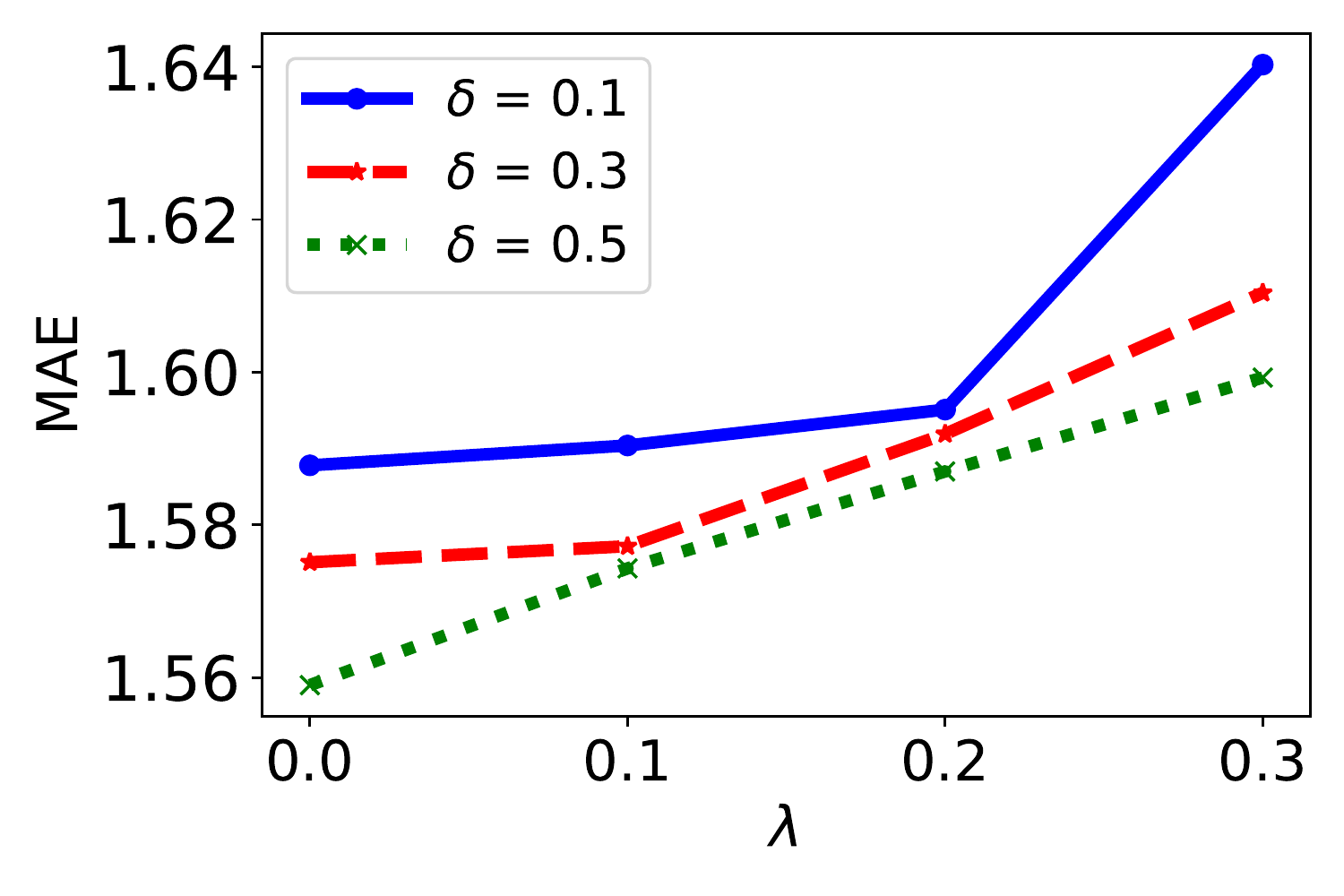}}
\caption{MAE performance with respect to different $\delta$ and $\lambda$ on three datasets.}
\label{fig:ldp_para_mae}
\end{figure}

This section studies the correlations between model performance and the LDP module. It contains two parameters: gradient clip threshold $\delta$ and Laplace noise strength $\lambda$, as shown in Eq.~(\ref{eq:LDP_gradient_dynamic}). Intuitively, this module alters gradients by injecting noise to gradients to protect the user's privacy. The injected gradients contribute to the training because the server also aggregates them for updating. Therefore, we conduct experiments to study the model performance with respect to the variations of these two parameters, which is shown in Fig.~\ref{fig:ldp_para_rmse} and Fig.~\ref{fig:ldp_para_mae} for RMSE and MAE, respectively. We have the following observations:

\begin{itemize}
    \item With a fixed $\lambda$, \modelname performs better when increasing $\delta$. The reason is that a large $\delta$ tends to clip less gradients. Therefore, the aggregated gradient information would be more accurate to reflect the true gradients. 
    
    \item With fixed $\delta$, \modelname performs worse when increasing $\lambda$. It is because a larger $\lambda$ injects more substantial Laplace noise to the model gradient. Hence, the gradient learned from data would be overwhelmed by generated noises. Thus, a smaller $\lambda$ is preferred.
    
    \item There is a trade-off in selecting optimal values. Although larger $\delta$ or smaller $\lambda$ lead to better performance, it would increase the risk of privacy leakage. If $\delta$ is infinitely large and $\lambda$ is $0$, the server can revert the interactions by checking uploaded gradients from clients. Therefore, we should choose an optimal pair to achieve acceptable performance with a small enough privacy budget.

\end{itemize}

% \begin{figure}[htb]
%     \centering
%     \includegraphics[width=.5\textwidth]{figure_experiment/privacy/delta.pdf}
%     \caption{Privacy budget $\epsilon$'s relation with $\delta$ and $\lambda$}
%     \label{Privacy budget}
% \end{figure}

\subsection{Ablation Study (RQ3)}

\begin{table}[htb]
\caption{Ablation study on FeSoG. The relative difference represents the performance difference between the corresponding variant and \modelname. A positive difference indicates worse performance, while negative ones indicate better performance.  }
\label{tab:Ablation_Study}
\begin{tabular}{l|cccccc}
\toprule
\multirow{2}{*}{Variant} & \multicolumn{2}{c}{Ciao} & \multicolumn{2}{c}{Epinions}  & \multicolumn{2}{c}{Filmtrust} \\
\cmidrule(r){2-3} \cmidrule(r){4-5} \cmidrule(r){6-7}
&  RMSE      &  MAE   
&  RMSE      &  MAE 
&  RMSE      &  MAE   \\
\midrule
\textbf{sharing GAT layer} & 2.0262 & 1.5863 & 1.8046 &  1.3876 & 2.148 & 1.5939\\
{\small relative difference} & {\small $5.8\%$} & {\small $6.2\%$} & {\small $0.43\%$} &  {\small $0.21\%$} & {\small $2.57\%$} & {\small $0.53\%$}\\
\hline
\textbf{w/o relational vector} &2.1545 &  1.7023 & 1.8044 & 1.3903 & 2.1379 & 1.6441\\
{\small relative difference} & {\small 12.58\%} & {\small 13.96\%} & {\small 0.42\%} & {\small 0.40\%} & {\small 2.09\%} & {\small 3.7\%} \\
\hline
\textbf{w/o pseudo items} & \textbf{1.9026} & \textbf{1.4849} & \textbf{1.7053} & \textbf{1.3021} & \textbf{2.0825} & \textbf{1.5798}\\
{\small relative difference} & {\small -0.57\%} & {\small -0.58\%} & {\small -5.1\%} & {\small -5.97\%} & {\small -0.56\%} & {\small -0.36\%} \\
\hline
\modelname & \underline{1.9136} & \underline{1.4937} & \underline{1.7969} & \underline{1.3847} & \underline{2.0942} & \underline{1.5855}\\
\bottomrule
\end{tabular}
\end{table}
In this section, we conduct an ablation study to analyze those components in \modelname to validate their effectiveness. We create three other variants of \modelname:
\begin{itemize} %[leftmargin=*]
    \item \textbf{sharing GAT layer:} \modelname applies different GAT layers to learn the attention weights for social neighbors and item neighbors. This variant shares the GAT layer for all neighbors. Hence, we only have one GAT layer to learn weight. Note that this variant also employs the relational vector during aggregation. 
    \item \textbf{w/o relational vectors:} this variant ignores the social relation and user-item relation vectors during aggregation. As such, it directly aggregates all the neighbors with their associated attention weights. Since the attention weights are learned separately for user neighbors and item neighbors, we should normalize those weights for all neighbors. 
    \item \textbf{w/o pseudo items:} this variant is a special case of Sec.~\ref{sec:pseduo_item} where the number of the pseudo item is $0$. It cannot protect the user privacy data from a protection perspective because. Without pseudo items, we only use the true interacted items that have non-zero gradients. However, we also include this variant for a comprehensive study.
\end{itemize}
The performance comparison of these methods on three datasets is reported in Table~\ref{tab:Ablation_Study}. We also present the corresponding difference between \modelname and the variants in a group. We have the following observations:
\begin{itemize}
    \item We should employ different GAT layers for users neighbors and item neighbors. Compared with \modelname, sharing GAT layer has worse performance. On the Filmtrust dataset, it has $2.57\%$ and $0.53\%$ relative difference on RMSE and MAE, respectively. On the Epinions dataset, it has $0.43\%$ and $0.21\%$ relative difference on RMSE and MAE, respectively. The worst performance of this variant is on Ciao, which has $5.8\%$ and $6.2\%$ relative difference on RMSE and MAE, respectively. 
    \item We should apply the relational vectors for aggregation. Compared with \modelname, without relational vectors has worse performance. On the Epinions dataset, it has $0.42\%$ and $0.40\%$ relative difference on RMSE and MAE, respectively. On the Filmtrust dataset, it has $2.09\%$ and $3.7\%$ relative difference on RMSE and MAE, respectively. The worst performance of this variant is on Ciao, which has $12.58\%$ and $13.96\%$ relative difference on RMSE and MAE, respectively. 
    \item Sampling pseudo items always worsen performance. However, compared with \modelname, without pseudo items is unable to protect the privacy because the server can easily infer the true interacted items by checking which item gradient is not $0$. 
\end{itemize}

\section{Conclusion and Future Work}\label{sec:conclusion}
In this paper, we propose a new federated learning framework, \modelname, for social recommendation. It decentralizes the data storage compared with existing social recommender systems. Moreover, it comprehensively fuses the local user privacy data in clients and uses a server to train an FSRS collaboratively. We address three challenges in designing this model: the heterogeneity of the data, the personalization requirements of the local modeling, and privacy protection for communication. The components in \modelname jointly tackle these challenges: the relational attention and aggregation of the local graph neural network distinguish social and item neighbors; the local user embedding inference preserves the personalizing information for clients; the pseudo-item labeling, as well as the dynamic LDP technique, protect the gradients from privacy data leakage. 
To verify the effectiveness of \modelname, we conduct extensive experiments. The overall comparing experiments demonstrate that \modelname significantly outperforms SOTA federated learning framework in solving social recommendation problems. Detailed sensitivity analysis regarding the hyper-parameters further justifies the efficacy of \modelname infusing the social information and user-item interaction and preserving the user privacy data locally. Moreover, the ablation study by dropping the components in \modelname demonstrates the necessity of our designing. 

Though being effective in solving the social recommendation problem, there are still several future directions. Firstly, we randomly sample pseudo items and predict their pseudo labels to protect gradients. However, as the items may have their relations, we may investigate employing an adaptive sampling of items rather than randomly. For example, we may train a local reinforcement learning model to explore the non-interacted item, which can decrease the noise. Secondly, we train the relational graph neural network only by leveraging local data. It is also possible to extend the local graph to be a high-order graph. However, this requires data transferring among clients. To protect privacy, we may design a peer-to-peer communication of clients preserving the decentralized storage characteristics of a federated recommender system. Finally, we can study the efficiency of communication. Since it requires numerous communication rounds to train a federated learning framework, it will be satisfying to decrease the time for communication or increase the bandwidth for communication with large-scale clients.

\section*{Acknowledgment}
Hao Peng is supported by the National Key R\&D Program of China through grant 2021YFB1714800, NSFC through grants 62002007 and U20B2053, S\&T Program of Hebei through grant 21340301D, Fundamental Research Funds for the Central Universities. 
Philip S. Yu is partially supported by NSF under grants III-1763325, III-1909323,  III-2106758, and SaTC-1930941.

%%
%% The next two lines define the bibliography style to be used, and
%% the bibliography file.
\bibliographystyle{ACM-Reference-Format}
\bibliography{reference}

%%% -*-BibTeX-*-
%%% Do NOT edit. File created by BibTeX with style
%%% ACM-Reference-Format-Journals [18-Jan-2012].

\begin{thebibliography}{66}

%%% ====================================================================
%%% NOTE TO THE USER: you can override these defaults by providing
%%% customized versions of any of these macros before the \bibliography
%%% command.  Each of them MUST provide its own final punctuation,
%%% except for \shownote{}, \showDOI{}, and \showURL{}.  The latter two
%%% do not use final punctuation, in order to avoid confusing it with
%%% the Web address.
%%%
%%% To suppress output of a particular field, define its macro to expand
%%% to an empty string, or better, \unskip, like this:
%%%
%%% \newcommand{\showDOI}[1]{\unskip}   % LaTeX syntax
%%%
%%% \def \showDOI #1{\unskip}           % plain TeX syntax
%%%
%%% ====================================================================

\ifx \showCODEN    \undefined \def \showCODEN     #1{\unskip}     \fi
\ifx \showDOI      \undefined \def \showDOI       #1{#1}\fi
\ifx \showISBNx    \undefined \def \showISBNx     #1{\unskip}     \fi
\ifx \showISBNxiii \undefined \def \showISBNxiii  #1{\unskip}     \fi
\ifx \showISSN     \undefined \def \showISSN      #1{\unskip}     \fi
\ifx \showLCCN     \undefined \def \showLCCN      #1{\unskip}     \fi
\ifx \shownote     \undefined \def \shownote      #1{#1}          \fi
\ifx \showarticletitle \undefined \def \showarticletitle #1{#1}   \fi
\ifx \showURL      \undefined \def \showURL       {\relax}        \fi
% The following commands are used for tagged output and should be
% invisible to TeX
\providecommand\bibfield[2]{#2}
\providecommand\bibinfo[2]{#2}
\providecommand\natexlab[1]{#1}
\providecommand\showeprint[2][]{arXiv:#2}

\bibitem[\protect\citeauthoryear{Ammad-Ud-Din, Ivannikova, Khan, Oyomno, Fu,
  Tan, and Flanagan}{Ammad-Ud-Din et~al\mbox{.}}{2019}]%
        {ammad2019federated}
\bibfield{author}{\bibinfo{person}{Muhammad Ammad-Ud-Din},
  \bibinfo{person}{Elena Ivannikova}, \bibinfo{person}{Suleiman~A Khan},
  \bibinfo{person}{Were Oyomno}, \bibinfo{person}{Qiang Fu},
  \bibinfo{person}{Kuan~Eeik Tan}, {and} \bibinfo{person}{Adrian Flanagan}.}
  \bibinfo{year}{2019}\natexlab{}.
\newblock \showarticletitle{Federated collaborative filtering for
  privacy-preserving personalized recommendation system}.
\newblock \bibinfo{journal}{\emph{arXiv preprint arXiv:1901.09888}}
  (\bibinfo{year}{2019}).
\newblock


\bibitem[\protect\citeauthoryear{Berg, Kipf, and Welling}{Berg
  et~al\mbox{.}}{2017}]%
        {berg2017graph}
\bibfield{author}{\bibinfo{person}{Rianne van~den Berg},
  \bibinfo{person}{Thomas~N Kipf}, {and} \bibinfo{person}{Max Welling}.}
  \bibinfo{year}{2017}\natexlab{}.
\newblock \showarticletitle{Graph convolutional matrix completion}.
\newblock \bibinfo{journal}{\emph{arXiv preprint arXiv:1706.02263}}
  (\bibinfo{year}{2017}).
\newblock


\bibitem[\protect\citeauthoryear{Cao, Peng, Wu, Dou, Li, and Yu}{Cao
  et~al\mbox{.}}{2021}]%
        {cao2021knowledge}
\bibfield{author}{\bibinfo{person}{Yuwei Cao}, \bibinfo{person}{Hao Peng},
  \bibinfo{person}{Jia Wu}, \bibinfo{person}{Yingtong Dou},
  \bibinfo{person}{Jianxin Li}, {and} \bibinfo{person}{Philip~S Yu}.}
  \bibinfo{year}{2021}\natexlab{}.
\newblock \showarticletitle{Knowledge-Preserving Incremental Social Event
  Detection via Heterogeneous GNNs}. In \bibinfo{booktitle}{\emph{Proceedings
  of the Web Conference 2021}}. \bibinfo{pages}{3383--3395}.
\newblock


\bibitem[\protect\citeauthoryear{Chai, Wang, Chen, and Yang}{Chai
  et~al\mbox{.}}{2020}]%
        {chai2020secure}
\bibfield{author}{\bibinfo{person}{Di Chai}, \bibinfo{person}{Leye Wang},
  \bibinfo{person}{Kai Chen}, {and} \bibinfo{person}{Qiang Yang}.}
  \bibinfo{year}{2020}\natexlab{}.
\newblock \showarticletitle{Secure federated matrix factorization}.
\newblock \bibinfo{journal}{\emph{IEEE Intelligent Systems}}
  (\bibinfo{year}{2020}).
\newblock


\bibitem[\protect\citeauthoryear{Chen, Zhang, Tung, Kankanhalli, and Chen}{Chen
  et~al\mbox{.}}{2020}]%
        {chen2020robust}
\bibfield{author}{\bibinfo{person}{Chen Chen}, \bibinfo{person}{Jingfeng
  Zhang}, \bibinfo{person}{Anthony~KH Tung}, \bibinfo{person}{Mohan
  Kankanhalli}, {and} \bibinfo{person}{Gang Chen}.}
  \bibinfo{year}{2020}\natexlab{}.
\newblock \showarticletitle{Robust federated recommendation system}.
\newblock \bibinfo{journal}{\emph{arXiv preprint arXiv:2006.08259}}
  (\bibinfo{year}{2020}).
\newblock


\bibitem[\protect\citeauthoryear{Dou, Liu, Sun, Deng, Peng, and Yu}{Dou
  et~al\mbox{.}}{2020}]%
        {dou2020enhancing}
\bibfield{author}{\bibinfo{person}{Yingtong Dou}, \bibinfo{person}{Zhiwei Liu},
  \bibinfo{person}{Li Sun}, \bibinfo{person}{Yutong Deng}, \bibinfo{person}{Hao
  Peng}, {and} \bibinfo{person}{Philip~S Yu}.} \bibinfo{year}{2020}\natexlab{}.
\newblock \showarticletitle{Enhancing graph neural network-based fraud
  detectors against camouflaged fraudsters}. In
  \bibinfo{booktitle}{\emph{Proceedings of the 29th ACM International
  Conference on Information \& Knowledge Management}}.
  \bibinfo{pages}{315--324}.
\newblock


\bibitem[\protect\citeauthoryear{Erlingsson, Pihur, and Korolova}{Erlingsson
  et~al\mbox{.}}{2014}]%
        {erlingsson2014rappor}
\bibfield{author}{\bibinfo{person}{{\'U}lfar Erlingsson},
  \bibinfo{person}{Vasyl Pihur}, {and} \bibinfo{person}{Aleksandra Korolova}.}
  \bibinfo{year}{2014}\natexlab{}.
\newblock \showarticletitle{Rappor: Randomized aggregatable privacy-preserving
  ordinal response}. In \bibinfo{booktitle}{\emph{Proceedings of the 2014 ACM
  SIGSAC conference on computer and communications security}}.
  \bibinfo{pages}{1054--1067}.
\newblock


\bibitem[\protect\citeauthoryear{Fallah, Mokhtari, and Ozdaglar}{Fallah
  et~al\mbox{.}}{2020}]%
        {fallah2020personalized}
\bibfield{author}{\bibinfo{person}{Alireza Fallah}, \bibinfo{person}{Aryan
  Mokhtari}, {and} \bibinfo{person}{Asuman Ozdaglar}.}
  \bibinfo{year}{2020}\natexlab{}.
\newblock \showarticletitle{Personalized federated learning: A meta-learning
  approach}.
\newblock \bibinfo{journal}{\emph{arXiv preprint arXiv:2002.07948}}
  (\bibinfo{year}{2020}).
\newblock


\bibitem[\protect\citeauthoryear{Fan, Ma, Li, He, Zhao, Tang, and Yin}{Fan
  et~al\mbox{.}}{2019}]%
        {fan2019graph}
\bibfield{author}{\bibinfo{person}{Wenqi Fan}, \bibinfo{person}{Yao Ma},
  \bibinfo{person}{Qing Li}, \bibinfo{person}{Yuan He}, \bibinfo{person}{Eric
  Zhao}, \bibinfo{person}{Jiliang Tang}, {and} \bibinfo{person}{Dawei Yin}.}
  \bibinfo{year}{2019}\natexlab{}.
\newblock \showarticletitle{Graph neural networks for social recommendation}.
  In \bibinfo{booktitle}{\emph{The World Wide Web Conference, {WWW} 2019, San
  Francisco, CA, USA, May 13-17, 2019}}. \bibinfo{publisher}{{ACM}},
  \bibinfo{pages}{417--426}.
\newblock


\bibitem[\protect\citeauthoryear{Fan, Ma, Li, Wang, Cai, Tang, and Yin}{Fan
  et~al\mbox{.}}{2020}]%
        {fan2020graph}
\bibfield{author}{\bibinfo{person}{Wenqi Fan}, \bibinfo{person}{Yao Ma},
  \bibinfo{person}{Qing Li}, \bibinfo{person}{Jianping Wang},
  \bibinfo{person}{Guoyong Cai}, \bibinfo{person}{Jiliang Tang}, {and}
  \bibinfo{person}{Dawei Yin}.} \bibinfo{year}{2020}\natexlab{}.
\newblock \showarticletitle{A Graph Neural Network Framework for Social
  Recommendations}.
\newblock \bibinfo{journal}{\emph{IEEE Transactions on Knowledge and Data
  Engineering}} (\bibinfo{year}{2020}).
\newblock


\bibitem[\protect\citeauthoryear{Flanagan, Oyomno, Grigorievskiy, Tan, Khan,
  and Ammad-Ud-Din}{Flanagan et~al\mbox{.}}{2020}]%
        {flanagan2020federated}
\bibfield{author}{\bibinfo{person}{Adrian Flanagan}, \bibinfo{person}{Were
  Oyomno}, \bibinfo{person}{Alexander Grigorievskiy},
  \bibinfo{person}{Kuan~Eeik Tan}, \bibinfo{person}{Suleiman~A Khan}, {and}
  \bibinfo{person}{Muhammad Ammad-Ud-Din}.} \bibinfo{year}{2020}\natexlab{}.
\newblock \showarticletitle{Federated multi-view matrix factorization for
  personalized recommendations}.
\newblock \bibinfo{journal}{\emph{arXiv preprint arXiv:2004.04256}}
  (\bibinfo{year}{2020}).
\newblock


\bibitem[\protect\citeauthoryear{Grover and Leskovec}{Grover and
  Leskovec}{2016}]%
        {grover2016node2vec}
\bibfield{author}{\bibinfo{person}{Aditya Grover} {and} \bibinfo{person}{Jure
  Leskovec}.} \bibinfo{year}{2016}\natexlab{}.
\newblock \showarticletitle{node2vec: Scalable feature learning for networks}.
  In \bibinfo{booktitle}{\emph{Proceedings of the 22nd ACM SIGKDD international
  conference on Knowledge discovery and data mining}}.
  \bibinfo{pages}{855--864}.
\newblock


\bibitem[\protect\citeauthoryear{Guo, Zhang, and Yorke-Smith}{Guo
  et~al\mbox{.}}{2013}]%
        {guo2013novel}
\bibfield{author}{\bibinfo{person}{G. Guo}, \bibinfo{person}{J. Zhang}, {and}
  \bibinfo{person}{N. Yorke-Smith}.} \bibinfo{year}{2013}\natexlab{}.
\newblock \showarticletitle{A Novel Bayesian Similarity Measure for Recommender
  Systems}. In \bibinfo{booktitle}{\emph{Proceedings of the 23rd International
  Joint Conference on Artificial Intelligence (IJCAI)}}.
  \bibinfo{pages}{2619--2625}.
\newblock


\bibitem[\protect\citeauthoryear{Guo, Zhang, and Yorke-Smith}{Guo
  et~al\mbox{.}}{2015}]%
        {guo2015trustsvd}
\bibfield{author}{\bibinfo{person}{Guibing Guo}, \bibinfo{person}{Jie Zhang},
  {and} \bibinfo{person}{Neil Yorke-Smith}.} \bibinfo{year}{2015}\natexlab{}.
\newblock \showarticletitle{Trustsvd: Collaborative filtering with both the
  explicit and implicit influence of user trust and of item ratings}. In
  \bibinfo{booktitle}{\emph{Proceedings of the AAAI Conference on Artificial
  Intelligence}}, Vol.~\bibinfo{volume}{29}.
\newblock


\bibitem[\protect\citeauthoryear{Hamilton, Ying, and Leskovec}{Hamilton
  et~al\mbox{.}}{2017}]%
        {hamilton2017inductive}
\bibfield{author}{\bibinfo{person}{William~L Hamilton}, \bibinfo{person}{Rex
  Ying}, {and} \bibinfo{person}{Jure Leskovec}.}
  \bibinfo{year}{2017}\natexlab{}.
\newblock \showarticletitle{Inductive representation learning on large graphs}.
  In \bibinfo{booktitle}{\emph{Advances in Neural Information Processing
  Systems 30: Annual Conference on Neural Information Processing Systems 2017,
  December 4-9, 2017, Long Beach, CA, {USA}}}. \bibinfo{pages}{1024--1034}.
\newblock


\bibitem[\protect\citeauthoryear{Hu, Dong, Wang, and Sun}{Hu
  et~al\mbox{.}}{2020}]%
        {hu2020heterogeneous}
\bibfield{author}{\bibinfo{person}{Ziniu Hu}, \bibinfo{person}{Yuxiao Dong},
  \bibinfo{person}{Kuansan Wang}, {and} \bibinfo{person}{Yizhou Sun}.}
  \bibinfo{year}{2020}\natexlab{}.
\newblock \showarticletitle{Heterogeneous graph transformer}. In
  \bibinfo{booktitle}{\emph{Proceedings of The Web Conference 2020}}.
  \bibinfo{pages}{2704--2710}.
\newblock


\bibitem[\protect\citeauthoryear{Jamali and Ester}{Jamali and Ester}{2009}]%
        {jamali2009trustwalker}
\bibfield{author}{\bibinfo{person}{Mohsen Jamali} {and} \bibinfo{person}{Martin
  Ester}.} \bibinfo{year}{2009}\natexlab{}.
\newblock \showarticletitle{Trustwalker: a random walk model for combining
  trust-based and item-based recommendation}. In
  \bibinfo{booktitle}{\emph{Proceedings of the 15th ACM SIGKDD international
  conference on Knowledge discovery and data mining}}.
  \bibinfo{pages}{397--406}.
\newblock


\bibitem[\protect\citeauthoryear{Jamali and Ester}{Jamali and Ester}{2010}]%
        {jamali2010matrix}
\bibfield{author}{\bibinfo{person}{Mohsen Jamali} {and} \bibinfo{person}{Martin
  Ester}.} \bibinfo{year}{2010}\natexlab{}.
\newblock \showarticletitle{A matrix factorization technique with trust
  propagation for recommendation in social networks}. In
  \bibinfo{booktitle}{\emph{Proceedings of the fourth ACM conference on
  Recommender systems}}. \bibinfo{pages}{135--142}.
\newblock


\bibitem[\protect\citeauthoryear{Kabbur, Ning, and Karypis}{Kabbur
  et~al\mbox{.}}{2013}]%
        {kabbur2013fism}
\bibfield{author}{\bibinfo{person}{Santosh Kabbur}, \bibinfo{person}{Xia Ning},
  {and} \bibinfo{person}{George Karypis}.} \bibinfo{year}{2013}\natexlab{}.
\newblock \showarticletitle{Fism: factored item similarity models for top-n
  recommender systems}. In \bibinfo{booktitle}{\emph{Proceedings of the 19th
  ACM SIGKDD international conference on Knowledge discovery and data mining}}.
  \bibinfo{pages}{659--667}.
\newblock


\bibitem[\protect\citeauthoryear{Kairouz, McMahan, Avent, Bellet, Bennis,
  Bhagoji, Bonawitz, Charles, Cormode, Cummings, et~al\mbox{.}}{Kairouz
  et~al\mbox{.}}{2019}]%
        {kairouz2019advances}
\bibfield{author}{\bibinfo{person}{Peter Kairouz}, \bibinfo{person}{H~Brendan
  McMahan}, \bibinfo{person}{Brendan Avent}, \bibinfo{person}{Aur{\'e}lien
  Bellet}, \bibinfo{person}{Mehdi Bennis}, \bibinfo{person}{Arjun~Nitin
  Bhagoji}, \bibinfo{person}{Keith Bonawitz}, \bibinfo{person}{Zachary
  Charles}, \bibinfo{person}{Graham Cormode}, \bibinfo{person}{Rachel
  Cummings}, {et~al\mbox{.}}} \bibinfo{year}{2019}\natexlab{}.
\newblock \showarticletitle{Advances and open problems in federated learning}.
\newblock \bibinfo{journal}{\emph{arXiv preprint arXiv:1912.04977}}
  (\bibinfo{year}{2019}).
\newblock


\bibitem[\protect\citeauthoryear{Kipf and Welling}{Kipf and Welling}{2017}]%
        {kipf2016semi}
\bibfield{author}{\bibinfo{person}{Thomas~N Kipf} {and} \bibinfo{person}{Max
  Welling}.} \bibinfo{year}{2017}\natexlab{}.
\newblock \showarticletitle{Semi-supervised classification with graph
  convolutional networks}. In \bibinfo{booktitle}{\emph{5th International
  Conference on Learning Representations, {ICLR} 2017, Toulon, France, April
  24-26, 2017, Conference Track Proceedings}}.
  \bibinfo{publisher}{OpenReview.net}.
\newblock


\bibitem[\protect\citeauthoryear{Kone{\v{c}}n{\`y}, McMahan, Yu, Richt{\'a}rik,
  Suresh, and Bacon}{Kone{\v{c}}n{\`y} et~al\mbox{.}}{2016}]%
        {konevcny2016federated}
\bibfield{author}{\bibinfo{person}{Jakub Kone{\v{c}}n{\`y}},
  \bibinfo{person}{H~Brendan McMahan}, \bibinfo{person}{Felix~X Yu},
  \bibinfo{person}{Peter Richt{\'a}rik}, \bibinfo{person}{Ananda~Theertha
  Suresh}, {and} \bibinfo{person}{Dave Bacon}.}
  \bibinfo{year}{2016}\natexlab{}.
\newblock \showarticletitle{Federated learning: Strategies for improving
  communication efficiency}.
\newblock \bibinfo{journal}{\emph{arXiv preprint arXiv:1610.05492}}
  (\bibinfo{year}{2016}).
\newblock


\bibitem[\protect\citeauthoryear{Koren, Bell, and Volinsky}{Koren
  et~al\mbox{.}}{2009}]%
        {koren2009matrix}
\bibfield{author}{\bibinfo{person}{Yehuda Koren}, \bibinfo{person}{Robert
  Bell}, {and} \bibinfo{person}{Chris Volinsky}.}
  \bibinfo{year}{2009}\natexlab{}.
\newblock \showarticletitle{Matrix factorization techniques for recommender
  systems}.
\newblock \bibinfo{journal}{\emph{Computer}} \bibinfo{volume}{42},
  \bibinfo{number}{8} (\bibinfo{year}{2009}), \bibinfo{pages}{30--37}.
\newblock


\bibitem[\protect\citeauthoryear{Li, Tei, and Fukazawa}{Li
  et~al\mbox{.}}{2020}]%
        {li2020efficient}
\bibfield{author}{\bibinfo{person}{Munan Li}, \bibinfo{person}{Kenji Tei},
  {and} \bibinfo{person}{Yoshiaki Fukazawa}.} \bibinfo{year}{2020}\natexlab{}.
\newblock \showarticletitle{An Efficient Adaptive Attention Neural Network for
  Social Recommendation}.
\newblock \bibinfo{journal}{\emph{IEEE Access}}  \bibinfo{volume}{8}
  (\bibinfo{year}{2020}), \bibinfo{pages}{63595--63606}.
\newblock


\bibitem[\protect\citeauthoryear{Liu, Zhang, and Gulla}{Liu
  et~al\mbox{.}}{2019}]%
        {liu2019real}
\bibfield{author}{\bibinfo{person}{Peng Liu}, \bibinfo{person}{Lemei Zhang},
  {and} \bibinfo{person}{Jon~Atle Gulla}.} \bibinfo{year}{2019}\natexlab{}.
\newblock \showarticletitle{Real-time social recommendation based on graph
  embedding and temporal context}.
\newblock \bibinfo{journal}{\emph{International Journal of Human-Computer
  Studies}}  \bibinfo{volume}{121} (\bibinfo{year}{2019}),
  \bibinfo{pages}{58--72}.
\newblock


\bibitem[\protect\citeauthoryear{Liu, Wan, He, Peng, and Yu}{Liu
  et~al\mbox{.}}{2020d}]%
        {liu2020kg}
\bibfield{author}{\bibinfo{person}{Ye Liu}, \bibinfo{person}{Yao Wan},
  \bibinfo{person}{Lifang He}, \bibinfo{person}{Hao Peng}, {and}
  \bibinfo{person}{Philip~S Yu}.} \bibinfo{year}{2020}\natexlab{d}.
\newblock \showarticletitle{KG-BART: Knowledge Graph-Augmented BART for
  Generative Commonsense Reasoning}.
\newblock \bibinfo{journal}{\emph{arXiv preprint arXiv:2009.12677}}
  (\bibinfo{year}{2020}).
\newblock


\bibitem[\protect\citeauthoryear{Liu, Dou, Yu, Deng, and Peng}{Liu
  et~al\mbox{.}}{2020a}]%
        {liu2020alleviating}
\bibfield{author}{\bibinfo{person}{Zhiwei Liu}, \bibinfo{person}{Yingtong Dou},
  \bibinfo{person}{Philip~S Yu}, \bibinfo{person}{Yutong Deng}, {and}
  \bibinfo{person}{Hao Peng}.} \bibinfo{year}{2020}\natexlab{a}.
\newblock \showarticletitle{Alleviating the Inconsistency Problem of Applying
  Graph Neural Network to Fraud Detection}. In
  \bibinfo{booktitle}{\emph{Proceedings of the 43rd International {ACM} {SIGIR}
  conference on research and development in Information Retrieval, {SIGIR}
  2020, Virtual Event, China, July 25-30, 2020}}. \bibinfo{publisher}{{ACM}},
  \bibinfo{pages}{1569--1572}.
\newblock


\bibitem[\protect\citeauthoryear{Liu, Fan, Wang, and Yu}{Liu
  et~al\mbox{.}}{2021}]%
        {liu2021augmenting}
\bibfield{author}{\bibinfo{person}{Zhiwei Liu}, \bibinfo{person}{Ziwei Fan},
  \bibinfo{person}{Yu Wang}, {and} \bibinfo{person}{Philip~S. Yu}.}
  \bibinfo{year}{2021}\natexlab{}.
\newblock \showarticletitle{Augmenting Sequential Recommendation with
  Pseudo-PriorItems via Reversely Pre-training Transformer}.
\newblock \bibinfo{journal}{\emph{Proceedings of the 44th international ACM
  SIGIR conference on Research and development in information retrieval}}.
\newblock


\bibitem[\protect\citeauthoryear{Liu, Meng, Zhang, and Yu}{Liu
  et~al\mbox{.}}{2020b}]%
        {liu2020deoscillated}
\bibfield{author}{\bibinfo{person}{Zhiwei Liu}, \bibinfo{person}{Lin Meng},
  \bibinfo{person}{Jiawei Zhang}, {and} \bibinfo{person}{Philip~S Yu}.}
  \bibinfo{year}{2020}\natexlab{b}.
\newblock \showarticletitle{Deoscillated Graph Collaborative Filtering}.
\newblock \bibinfo{journal}{\emph{arXiv preprint arXiv:2011.02100}}
  (\bibinfo{year}{2020}).
\newblock


\bibitem[\protect\citeauthoryear{Liu, Wan, Guo, Achan, and Yu}{Liu
  et~al\mbox{.}}{2020c}]%
        {liu2020basconv}
\bibfield{author}{\bibinfo{person}{Zhiwei Liu}, \bibinfo{person}{Mengting Wan},
  \bibinfo{person}{Stephen Guo}, \bibinfo{person}{Kannan Achan}, {and}
  \bibinfo{person}{Philip~S Yu}.} \bibinfo{year}{2020}\natexlab{c}.
\newblock \showarticletitle{Basconv: aggregating heterogeneous interactions for
  basket recommendation with graph convolutional neural network}. In
  \bibinfo{booktitle}{\emph{Proceedings of the 2020 SIAM International
  Conference on Data Mining}}. SIAM, \bibinfo{pages}{64--72}.
\newblock


\bibitem[\protect\citeauthoryear{Ma, King, and Lyu}{Ma et~al\mbox{.}}{2009}]%
        {ma2009learning}
\bibfield{author}{\bibinfo{person}{Hao Ma}, \bibinfo{person}{Irwin King}, {and}
  \bibinfo{person}{Michael~R Lyu}.} \bibinfo{year}{2009}\natexlab{}.
\newblock \showarticletitle{Learning to recommend with social trust ensemble}.
  In \bibinfo{booktitle}{\emph{Proceedings of the 32nd international ACM SIGIR
  conference on Research and development in information retrieval}}.
  \bibinfo{pages}{203--210}.
\newblock


\bibitem[\protect\citeauthoryear{Ma, Yang, Lyu, and King}{Ma
  et~al\mbox{.}}{2008}]%
        {ma2008sorec}
\bibfield{author}{\bibinfo{person}{Hao Ma}, \bibinfo{person}{Haixuan Yang},
  \bibinfo{person}{Michael~R Lyu}, {and} \bibinfo{person}{Irwin King}.}
  \bibinfo{year}{2008}\natexlab{}.
\newblock \showarticletitle{Sorec: social recommendation using probabilistic
  matrix factorization}. In \bibinfo{booktitle}{\emph{Proceedings of the 17th
  ACM conference on Information and knowledge management}}.
  \bibinfo{pages}{931--940}.
\newblock


\bibitem[\protect\citeauthoryear{Ma, Zhou, Liu, Lyu, and King}{Ma
  et~al\mbox{.}}{2011}]%
        {ma2011recommender}
\bibfield{author}{\bibinfo{person}{Hao Ma}, \bibinfo{person}{Dengyong Zhou},
  \bibinfo{person}{Chao Liu}, \bibinfo{person}{Michael~R Lyu}, {and}
  \bibinfo{person}{Irwin King}.} \bibinfo{year}{2011}\natexlab{}.
\newblock \showarticletitle{Recommender systems with social regularization}. In
  \bibinfo{booktitle}{\emph{Proceedings of the fourth ACM international
  conference on Web search and data mining}}. \bibinfo{pages}{287--296}.
\newblock


\bibitem[\protect\citeauthoryear{McMahan, Moore, Ramage, Hampson, and
  y~Arcas}{McMahan et~al\mbox{.}}{2017}]%
        {mcmahan2017communication}
\bibfield{author}{\bibinfo{person}{Brendan McMahan}, \bibinfo{person}{Eider
  Moore}, \bibinfo{person}{Daniel Ramage}, \bibinfo{person}{Seth Hampson},
  {and} \bibinfo{person}{Blaise~Aguera y Arcas}.}
  \bibinfo{year}{2017}\natexlab{}.
\newblock \showarticletitle{Communication-efficient learning of deep networks
  from decentralized data}. In \bibinfo{booktitle}{\emph{Artificial
  Intelligence and Statistics}}. PMLR, \bibinfo{pages}{1273--1282}.
\newblock


\bibitem[\protect\citeauthoryear{McSherry and Mironov}{McSherry and
  Mironov}{2009}]%
        {mcsherry2009differentially}
\bibfield{author}{\bibinfo{person}{Frank McSherry} {and} \bibinfo{person}{Ilya
  Mironov}.} \bibinfo{year}{2009}\natexlab{}.
\newblock \showarticletitle{Differentially private recommender systems:
  Building privacy into the netflix prize contenders}. In
  \bibinfo{booktitle}{\emph{Proceedings of the 15th ACM SIGKDD international
  conference on Knowledge discovery and data mining}}.
  \bibinfo{pages}{627--636}.
\newblock


\bibitem[\protect\citeauthoryear{Mu, Zha, He, and Tang}{Mu
  et~al\mbox{.}}{2019}]%
        {mu2019graph}
\bibfield{author}{\bibinfo{person}{Nan Mu}, \bibinfo{person}{Daren Zha},
  \bibinfo{person}{Yuanye He}, {and} \bibinfo{person}{Zhihao Tang}.}
  \bibinfo{year}{2019}\natexlab{}.
\newblock \showarticletitle{Graph Attention Networks for Neural Social
  Recommendation}. In \bibinfo{booktitle}{\emph{31st {IEEE} International
  Conference on Tools with Artificial Intelligence, {ICTAI} 2019, Portland, OR,
  USA, November 4-6, 2019}}. IEEE, \bibinfo{pages}{1320--1327}.
\newblock


\bibitem[\protect\citeauthoryear{Peng, Li, Song, Yang, Ranjan, Yu, and He}{Peng
  et~al\mbox{.}}{2021a}]%
        {peng2021streaming}
\bibfield{author}{\bibinfo{person}{Hao Peng}, \bibinfo{person}{Jianxin Li},
  \bibinfo{person}{Yangqiu Song}, \bibinfo{person}{Renyu Yang},
  \bibinfo{person}{Rajiv Ranjan}, \bibinfo{person}{Philip~S Yu}, {and}
  \bibinfo{person}{Lifang He}.} \bibinfo{year}{2021}\natexlab{a}.
\newblock \showarticletitle{Streaming Social Event Detection and Evolution
  Discovery in Heterogeneous Information Networks}.
\newblock \bibinfo{journal}{\emph{ACM Transactions on Knowledge Discovery from
  Data (TKDD)}} \bibinfo{volume}{15}, \bibinfo{number}{5}
  (\bibinfo{year}{2021}), \bibinfo{pages}{1--33}.
\newblock


\bibitem[\protect\citeauthoryear{Peng, Yang, Wang, Li, He, Yu, Zomaya, and
  Ranjan}{Peng et~al\mbox{.}}{2021b}]%
        {peng2021lime}
\bibfield{author}{\bibinfo{person}{Hao Peng}, \bibinfo{person}{Renyu Yang},
  \bibinfo{person}{Zheng Wang}, \bibinfo{person}{Jianxin Li},
  \bibinfo{person}{Lifang He}, \bibinfo{person}{Philip Yu},
  \bibinfo{person}{Albert Zomaya}, {and} \bibinfo{person}{Raj Ranjan}.}
  \bibinfo{year}{2021}\natexlab{b}.
\newblock \showarticletitle{Lime: Low-cost incremental learning for dynamic
  heterogeneous information networks}.
\newblock \bibinfo{journal}{\emph{IEEE Trans. Comput.}} (\bibinfo{year}{2021}).
\newblock


\bibitem[\protect\citeauthoryear{Qi, Wu, Wu, Huang, and Xie}{Qi
  et~al\mbox{.}}{2020}]%
        {qi2020privacy}
\bibfield{author}{\bibinfo{person}{Tao Qi}, \bibinfo{person}{Fangzhao Wu},
  \bibinfo{person}{Chuhan Wu}, \bibinfo{person}{Yongfeng Huang}, {and}
  \bibinfo{person}{Xing Xie}.} \bibinfo{year}{2020}\natexlab{}.
\newblock \showarticletitle{Privacy-Preserving News Recommendation Model
  Learning}. In \bibinfo{booktitle}{\emph{Proceedings of the 2020 Conference on
  Empirical Methods in Natural Language Processing: Findings}}.
  \bibinfo{pages}{1423--1432}.
\newblock


\bibitem[\protect\citeauthoryear{Rendle, Freudenthaler, Gantner, and
  Schmidt-Thieme}{Rendle et~al\mbox{.}}{2009}]%
        {rendle2009bpr}
\bibfield{author}{\bibinfo{person}{Steffen Rendle}, \bibinfo{person}{Christoph
  Freudenthaler}, \bibinfo{person}{Zeno Gantner}, {and} \bibinfo{person}{Lars
  Schmidt-Thieme}.} \bibinfo{year}{2009}\natexlab{}.
\newblock \showarticletitle{BPR: Bayesian personalized ranking from implicit
  feedback}. In \bibinfo{booktitle}{\emph{UAI}}. \bibinfo{pages}{452--461}.
\newblock


\bibitem[\protect\citeauthoryear{Ribero, Henderson, Williamson, and
  Vikalo}{Ribero et~al\mbox{.}}{2020}]%
        {ribero2020federating}
\bibfield{author}{\bibinfo{person}{M{\'o}nica Ribero}, \bibinfo{person}{Jette
  Henderson}, \bibinfo{person}{Sinead Williamson}, {and} \bibinfo{person}{Haris
  Vikalo}.} \bibinfo{year}{2020}\natexlab{}.
\newblock \showarticletitle{Federating recommendations using differentially
  private prototypes}.
\newblock \bibinfo{journal}{\emph{arXiv preprint arXiv:2003.00602}}
  (\bibinfo{year}{2020}).
\newblock


\bibitem[\protect\citeauthoryear{Shen and Jin}{Shen and Jin}{2012}]%
        {shen2012learning}
\bibfield{author}{\bibinfo{person}{Yelong Shen} {and} \bibinfo{person}{Ruoming
  Jin}.} \bibinfo{year}{2012}\natexlab{}.
\newblock \showarticletitle{Learning personal+ social latent factor model for
  social recommendation}. In \bibinfo{booktitle}{\emph{Proceedings of the 18th
  ACM SIGKDD international conference on Knowledge discovery and data mining}}.
  \bibinfo{pages}{1303--1311}.
\newblock


\bibitem[\protect\citeauthoryear{Singh and Gordon}{Singh and Gordon}{2008}]%
        {singh2008relational}
\bibfield{author}{\bibinfo{person}{Ajit~P Singh} {and}
  \bibinfo{person}{Geoffrey~J Gordon}.} \bibinfo{year}{2008}\natexlab{}.
\newblock \showarticletitle{Relational learning via collective matrix
  factorization}. In \bibinfo{booktitle}{\emph{Proceedings of the 14th ACM
  SIGKDD international conference on Knowledge discovery and data mining}}.
  \bibinfo{pages}{650--658}.
\newblock


\bibitem[\protect\citeauthoryear{Song, Xiao, Wang, Charlin, Zhang, and
  Tang}{Song et~al\mbox{.}}{2019}]%
        {song2019session}
\bibfield{author}{\bibinfo{person}{Weiping Song}, \bibinfo{person}{Zhiping
  Xiao}, \bibinfo{person}{Yifan Wang}, \bibinfo{person}{Laurent Charlin},
  \bibinfo{person}{Ming Zhang}, {and} \bibinfo{person}{Jian Tang}.}
  \bibinfo{year}{2019}\natexlab{}.
\newblock \showarticletitle{Session-based social recommendation via dynamic
  graph attention networks}. In \bibinfo{booktitle}{\emph{Proceedings of the
  Twelfth ACM International Conference on Web Search and Data Mining}}.
  \bibinfo{pages}{555--563}.
\newblock


\bibitem[\protect\citeauthoryear{Tang, Gao, Hu, and Liu}{Tang
  et~al\mbox{.}}{2013}]%
        {tang2013exploiting}
\bibfield{author}{\bibinfo{person}{Jiliang Tang}, \bibinfo{person}{Huiji Gao},
  \bibinfo{person}{Xia Hu}, {and} \bibinfo{person}{Huan Liu}.}
  \bibinfo{year}{2013}\natexlab{}.
\newblock \showarticletitle{Exploiting homophily effect for trust prediction}.
  In \bibinfo{booktitle}{\emph{Sixth {ACM} International Conference on Web
  Search and Data Mining, {WSDM} 2013, Rome, Italy, February 4-8, 2013}}.
  \bibinfo{publisher}{{ACM}}, \bibinfo{pages}{53--62}.
\newblock


\bibitem[\protect\citeauthoryear{Tang, Gao, and Liu}{Tang
  et~al\mbox{.}}{2012a}]%
        {tang2012mtrust}
\bibfield{author}{\bibinfo{person}{Jiliang Tang}, \bibinfo{person}{Huiji Gao},
  {and} \bibinfo{person}{Huan Liu}.} \bibinfo{year}{2012}\natexlab{a}.
\newblock \showarticletitle{mTrust: Discerning multi-faceted trust in a
  connected world}. In \bibinfo{booktitle}{\emph{Proceedings of the Fifth
  International Conference on Web Search and Web Data Mining, {WSDM} 2012,
  Seattle, WA, USA, February 8-12, 2012}}. \bibinfo{publisher}{{ACM}},
  \bibinfo{pages}{93--102}.
\newblock


\bibitem[\protect\citeauthoryear{Tang, Gao, Liu, and Das~Sarma}{Tang
  et~al\mbox{.}}{2012b}]%
        {tang2012etrust}
\bibfield{author}{\bibinfo{person}{Jiliang Tang}, \bibinfo{person}{Huiji Gao},
  \bibinfo{person}{Huan Liu}, {and} \bibinfo{person}{Atish Das~Sarma}.}
  \bibinfo{year}{2012}\natexlab{b}.
\newblock \showarticletitle{eTrust: Understanding trust evolution in an online
  world}. In \bibinfo{booktitle}{\emph{The 18th {ACM} {SIGKDD} International
  Conference on Knowledge Discovery and Data Mining, {KDD} '12, Beijing, China,
  August 12-16, 2012}}. \bibinfo{publisher}{{ACM}}, \bibinfo{pages}{253--261}.
\newblock


\bibitem[\protect\citeauthoryear{Vaswani, Shazeer, Parmar, Uszkoreit, Jones,
  Gomez, Kaiser, and Polosukhin}{Vaswani et~al\mbox{.}}{2017}]%
        {vaswani2017attention}
\bibfield{author}{\bibinfo{person}{Ashish Vaswani}, \bibinfo{person}{Noam
  Shazeer}, \bibinfo{person}{Niki Parmar}, \bibinfo{person}{Jakob Uszkoreit},
  \bibinfo{person}{Llion Jones}, \bibinfo{person}{Aidan~N Gomez},
  \bibinfo{person}{Lukasz Kaiser}, {and} \bibinfo{person}{Illia Polosukhin}.}
  \bibinfo{year}{2017}\natexlab{}.
\newblock \showarticletitle{Attention is all you need}.
\newblock \bibinfo{journal}{\emph{arXiv preprint arXiv:1706.03762}}
  (\bibinfo{year}{2017}).
\newblock


\bibitem[\protect\citeauthoryear{Veli{\v{c}}kovi{\'c}, Cucurull, Casanova,
  Romero, Lio, and Bengio}{Veli{\v{c}}kovi{\'c} et~al\mbox{.}}{2017}]%
        {velivckovic2017graph}
\bibfield{author}{\bibinfo{person}{Petar Veli{\v{c}}kovi{\'c}},
  \bibinfo{person}{Guillem Cucurull}, \bibinfo{person}{Arantxa Casanova},
  \bibinfo{person}{Adriana Romero}, \bibinfo{person}{Pietro Lio}, {and}
  \bibinfo{person}{Yoshua Bengio}.} \bibinfo{year}{2017}\natexlab{}.
\newblock \showarticletitle{Graph attention networks}.
\newblock \bibinfo{journal}{\emph{arXiv preprint arXiv:1710.10903}}
  (\bibinfo{year}{2017}).
\newblock


\bibitem[\protect\citeauthoryear{Wang, Liang, Liu, Zhang, and Yu}{Wang
  et~al\mbox{.}}{2021}]%
        {wang2021pre}
\bibfield{author}{\bibinfo{person}{Chen Wang}, \bibinfo{person}{Yueqing Liang},
  \bibinfo{person}{Zhiwei Liu}, \bibinfo{person}{Tao Zhang}, {and}
  \bibinfo{person}{Philip~S Yu}.} \bibinfo{year}{2021}\natexlab{}.
\newblock \showarticletitle{Pre-training Graph Neural Network for Cross Domain
  Recommendation}.
\newblock \bibinfo{journal}{\emph{arXiv preprint arXiv:2111.08268}}
  (\bibinfo{year}{2021}).
\newblock


\bibitem[\protect\citeauthoryear{Wang, He, Wang, Feng, and Chua}{Wang
  et~al\mbox{.}}{2019}]%
        {wang19neural}
\bibfield{author}{\bibinfo{person}{Xiang Wang}, \bibinfo{person}{Xiangnan He},
  \bibinfo{person}{Meng Wang}, \bibinfo{person}{Fuli Feng}, {and}
  \bibinfo{person}{Tat{-}Seng Chua}.} \bibinfo{year}{2019}\natexlab{}.
\newblock \showarticletitle{Neural Graph Collaborative Filtering}. In
  \bibinfo{booktitle}{\emph{SIGIR}}. \bibinfo{pages}{165--174}.
\newblock


\bibitem[\protect\citeauthoryear{Wang, Pan, and Xu}{Wang et~al\mbox{.}}{2014}]%
        {wang2014hgmf}
\bibfield{author}{\bibinfo{person}{Xin Wang}, \bibinfo{person}{Weike Pan},
  {and} \bibinfo{person}{Congfu Xu}.} \bibinfo{year}{2014}\natexlab{}.
\newblock \showarticletitle{Hgmf: Hierarchical group matrix factorization for
  collaborative recommendation}. In \bibinfo{booktitle}{\emph{Proceedings of
  the 23rd ACM International Conference on Conference on Information and
  Knowledge Management}}. \bibinfo{pages}{769--778}.
\newblock


\bibitem[\protect\citeauthoryear{Wu, Wu, Cao, Huang, and Xie}{Wu
  et~al\mbox{.}}{2021b}]%
        {wu2021fedgnn}
\bibfield{author}{\bibinfo{person}{Chuhan Wu}, \bibinfo{person}{Fangzhao Wu},
  \bibinfo{person}{Yang Cao}, \bibinfo{person}{Yongfeng Huang}, {and}
  \bibinfo{person}{Xing Xie}.} \bibinfo{year}{2021}\natexlab{b}.
\newblock \showarticletitle{Fedgnn: Federated graph neural network for
  privacy-preserving recommendation}.
\newblock \bibinfo{journal}{\emph{arXiv preprint arXiv:2102.04925}}
  (\bibinfo{year}{2021}).
\newblock


\bibitem[\protect\citeauthoryear{Wu, Li, Sun, Hong, Ge, and Wang}{Wu
  et~al\mbox{.}}{2020}]%
        {wu2020diffnet++}
\bibfield{author}{\bibinfo{person}{Le Wu}, \bibinfo{person}{Junwei Li},
  \bibinfo{person}{Peijie Sun}, \bibinfo{person}{Richang Hong},
  \bibinfo{person}{Yong Ge}, {and} \bibinfo{person}{Meng Wang}.}
  \bibinfo{year}{2020}\natexlab{}.
\newblock \showarticletitle{DiffNet++: A Neural Influence and Interest
  Diffusion Network for Social Recommendation}.
\newblock \bibinfo{journal}{\emph{IEEE Transactions on Knowledge and Data
  Engineering}} (\bibinfo{year}{2020}).
\newblock


\bibitem[\protect\citeauthoryear{Wu, Sun, Fu, Hong, Wang, and Wang}{Wu
  et~al\mbox{.}}{2019a}]%
        {wu2019neural}
\bibfield{author}{\bibinfo{person}{Le Wu}, \bibinfo{person}{Peijie Sun},
  \bibinfo{person}{Yanjie Fu}, \bibinfo{person}{Richang Hong},
  \bibinfo{person}{Xiting Wang}, {and} \bibinfo{person}{Meng Wang}.}
  \bibinfo{year}{2019}\natexlab{a}.
\newblock \showarticletitle{A neural influence diffusion model for social
  recommendation}. In \bibinfo{booktitle}{\emph{Proceedings of the 42nd
  International {ACM} {SIGIR} Conference on Research and Development in
  Information Retrieval, {SIGIR} 2019, Paris, France, July 21-25, 2019}}.
  \bibinfo{publisher}{{ACM}}, \bibinfo{pages}{235--244}.
\newblock


\bibitem[\protect\citeauthoryear{Wu, Sun, Hong, Fu, Wang, and Wang}{Wu
  et~al\mbox{.}}{2018}]%
        {wu2018socialgcn}
\bibfield{author}{\bibinfo{person}{Le Wu}, \bibinfo{person}{Peijie Sun},
  \bibinfo{person}{Richang Hong}, \bibinfo{person}{Yanjie Fu},
  \bibinfo{person}{Xiting Wang}, {and} \bibinfo{person}{Meng Wang}.}
  \bibinfo{year}{2018}\natexlab{}.
\newblock \showarticletitle{SocialGCN: An efficient graph convolutional network
  based model for social recommendation}.
\newblock \bibinfo{journal}{\emph{arXiv preprint arXiv:1811.02815}}
  (\bibinfo{year}{2018}).
\newblock


\bibitem[\protect\citeauthoryear{Wu, Sun, Hong, Ge, and Wang}{Wu
  et~al\mbox{.}}{2021a}]%
        {wu2018collaborative}
\bibfield{author}{\bibinfo{person}{Le Wu}, \bibinfo{person}{Peijie Sun},
  \bibinfo{person}{Richang Hong}, \bibinfo{person}{Yong Ge}, {and}
  \bibinfo{person}{Meng Wang}.} \bibinfo{year}{2021}\natexlab{a}.
\newblock \showarticletitle{Collaborative neural social recommendation}.
\newblock \bibinfo{journal}{\emph{IEEE Transactions on Systems, Man, and
  Cybernetics: Systems}} \bibinfo{volume}{51}, \bibinfo{number}{1}
  (\bibinfo{year}{2021}), \bibinfo{pages}{464--476}.
\newblock


\bibitem[\protect\citeauthoryear{Wu, Zhang, Gao, He, Weng, Gao, and Chen}{Wu
  et~al\mbox{.}}{2019b}]%
        {wu2019dual}
\bibfield{author}{\bibinfo{person}{Qitian Wu}, \bibinfo{person}{Hengrui Zhang},
  \bibinfo{person}{Xiaofeng Gao}, \bibinfo{person}{Peng He},
  \bibinfo{person}{Paul Weng}, \bibinfo{person}{Han Gao}, {and}
  \bibinfo{person}{Guihai Chen}.} \bibinfo{year}{2019}\natexlab{b}.
\newblock \showarticletitle{Dual graph attention networks for deep latent
  representation of multifaceted social effects in recommender systems}. In
  \bibinfo{booktitle}{\emph{The World Wide Web Conference, {WWW} 2019, San
  Francisco, CA, USA, May 13-17, 2019}}. \bibinfo{publisher}{{ACM}},
  \bibinfo{pages}{2091--2102}.
\newblock


\bibitem[\protect\citeauthoryear{Xia, Xiong, Philip, and Socher}{Xia
  et~al\mbox{.}}{2020}]%
        {xia2020composed}
\bibfield{author}{\bibinfo{person}{Congying Xia}, \bibinfo{person}{Caiming
  Xiong}, \bibinfo{person}{S~Yu Philip}, {and} \bibinfo{person}{Richard
  Socher}.} \bibinfo{year}{2020}\natexlab{}.
\newblock \showarticletitle{Composed Variational Natural Language Generation
  for Few-shot Intents}. In \bibinfo{booktitle}{\emph{Proceedings of the 2020
  Conference on Empirical Methods in Natural Language Processing: Findings}}.
  \bibinfo{pages}{3379--3388}.
\newblock


\bibitem[\protect\citeauthoryear{Yang, Wang, Xu, Wang, Bian, and Liu}{Yang
  et~al\mbox{.}}{2020}]%
        {yang2020heterogeneity}
\bibfield{author}{\bibinfo{person}{Chengxu Yang}, \bibinfo{person}{QiPeng
  Wang}, \bibinfo{person}{Mengwei Xu}, \bibinfo{person}{Shangguang Wang},
  \bibinfo{person}{Kaigui Bian}, {and} \bibinfo{person}{Xuanzhe Liu}.}
  \bibinfo{year}{2020}\natexlab{}.
\newblock \showarticletitle{Heterogeneity-aware federated learning}.
\newblock \bibinfo{journal}{\emph{arXiv preprint arXiv:2006.06983}}
  (\bibinfo{year}{2020}).
\newblock


\bibitem[\protect\citeauthoryear{Yang, Liu, Dou, Ma, and Yu}{Yang
  et~al\mbox{.}}{2021}]%
        {yang2021consisrec}
\bibfield{author}{\bibinfo{person}{Liangwei Yang}, \bibinfo{person}{Zhiwei
  Liu}, \bibinfo{person}{Yingtong Dou}, \bibinfo{person}{Jing Ma}, {and}
  \bibinfo{person}{Philip~S. Yu}.} \bibinfo{year}{2021}\natexlab{}.
\newblock \showarticletitle{ConsisRec: Enhancing GNN for Social Recommendation
  via Consistent Neighbor Aggregation}.
\newblock \bibinfo{journal}{\emph{Proceedings of the 44th international ACM
  SIGIR conference on Research and development in information retrieval}}.
\newblock


\bibitem[\protect\citeauthoryear{Yang, Liu, Chen, and Tong}{Yang
  et~al\mbox{.}}{2019}]%
        {yang2019federated}
\bibfield{author}{\bibinfo{person}{Qiang Yang}, \bibinfo{person}{Yang Liu},
  \bibinfo{person}{Tianjian Chen}, {and} \bibinfo{person}{Yongxin Tong}.}
  \bibinfo{year}{2019}\natexlab{}.
\newblock \showarticletitle{Federated machine learning: Concept and
  applications}.
\newblock \bibinfo{journal}{\emph{ACM Transactions on Intelligent Systems and
  Technology (TIST)}} \bibinfo{volume}{10}, \bibinfo{number}{2}
  (\bibinfo{year}{2019}), \bibinfo{pages}{1--19}.
\newblock


\bibitem[\protect\citeauthoryear{Ying, He, Chen, Eksombatchai, Hamilton, and
  Leskovec}{Ying et~al\mbox{.}}{2018}]%
        {pinsage2018ying}
\bibfield{author}{\bibinfo{person}{Rex Ying}, \bibinfo{person}{Ruining He},
  \bibinfo{person}{Kaifeng Chen}, \bibinfo{person}{Pong Eksombatchai},
  \bibinfo{person}{William~L. Hamilton}, {and} \bibinfo{person}{Jure
  Leskovec}.} \bibinfo{year}{2018}\natexlab{}.
\newblock \showarticletitle{Graph Convolutional Neural Networks for Web-Scale
  Recommender Systems}. In \bibinfo{booktitle}{\emph{SIGKDD}},
  \bibfield{editor}{\bibinfo{person}{Yike Guo} {and} \bibinfo{person}{Faisal
  Farooq}} (Eds.). \bibinfo{pages}{974--983}.
\newblock


\bibitem[\protect\citeauthoryear{Zhang, Yu, Wang, Shah, and Zhang}{Zhang
  et~al\mbox{.}}{2017}]%
        {CUNE}
\bibfield{author}{\bibinfo{person}{Chuxu Zhang}, \bibinfo{person}{Lu Yu},
  \bibinfo{person}{Yan Wang}, \bibinfo{person}{Chirag Shah}, {and}
  \bibinfo{person}{Xiangliang Zhang}.} \bibinfo{year}{2017}\natexlab{}.
\newblock \showarticletitle{Collaborative User Network Embedding for Social
  Recommender Systems}. In \bibinfo{booktitle}{\emph{Proceedings of the 2017
  {SIAM} International Conference on Data Mining, Houston, Texas, USA, April
  27-29, 2017}}, \bibfield{editor}{\bibinfo{person}{Nitesh~V. Chawla} {and}
  \bibinfo{person}{Wei Wang}} (Eds.). \bibinfo{publisher}{{SIAM}},
  \bibinfo{pages}{381--389}.
\newblock


\bibitem[\protect\citeauthoryear{Zhou, Wang, He, and Wang}{Zhou
  et~al\mbox{.}}{2021a}]%
        {zhou2021intrinsic}
\bibfield{author}{\bibinfo{person}{Yao Zhou}, \bibinfo{person}{Haonan Wang},
  \bibinfo{person}{Jingrui He}, {and} \bibinfo{person}{Haixun Wang}.}
  \bibinfo{year}{2021}\natexlab{a}.
\newblock \showarticletitle{From Intrinsic to Counterfactual: On the
  Explainability of Contextualized Recommender Systems}.
\newblock \bibinfo{journal}{\emph{arXiv preprint arXiv:2110.14844}}
  (\bibinfo{year}{2021}).
\newblock


\bibitem[\protect\citeauthoryear{Zhou, Xu, Wu, Taghavi, Korpeoglu, Achan, and
  He}{Zhou et~al\mbox{.}}{2021b}]%
        {zhou2021pure}
\bibfield{author}{\bibinfo{person}{Yao Zhou}, \bibinfo{person}{Jianpeng Xu},
  \bibinfo{person}{Jun Wu}, \bibinfo{person}{Zeinab Taghavi},
  \bibinfo{person}{Evren Korpeoglu}, \bibinfo{person}{Kannan Achan}, {and}
  \bibinfo{person}{Jingrui He}.} \bibinfo{year}{2021}\natexlab{b}.
\newblock \showarticletitle{PURE: Positive-Unlabeled Recommendation with
  Generative Adversarial Network}. In \bibinfo{booktitle}{\emph{Proceedings of
  the 27th ACM SIGKDD Conference on Knowledge Discovery \& Data Mining}}.
  \bibinfo{pages}{2409--2419}.
\newblock


\end{thebibliography}

%%
%% If your work has an appendix, this is the place to put it.
% \appendix

\end{document}